\documentclass[11pt]{article}

\usepackage{graphicx} 
\usepackage[utf8]{inputenc}
\usepackage[margin=1in]{geometry}
\usepackage{amsmath,amsthm,amssymb}
\usepackage{xcolor}
\usepackage{hyperref}
\usepackage{float}
\usepackage[ruled,vlined,linesnumbered]{algorithm2e}
\usepackage[shortlabels]{enumitem}
\usepackage[obeyFinal,textsize=scriptsize,shadow]{todonotes}
\usepackage{multirow}
\usepackage{siunitx}
\usepackage{diffcoeff}
\usepackage{listings}
\usepackage{microtype}
\usepackage{xstring}

\usepackage{tikz}
\usetikzlibrary{plotmarks,positioning,decorations.pathreplacing,calc}
\usetikzlibrary{fit,positioning,shapes.geometric, arrows.meta,backgrounds}

\usepackage{pgfplots}
\pgfplotsset{compat=1.18}

\usepackage{amsmath,amsthm,amssymb}
\usepackage{subcaption}
\usepackage{mathtools}
\usepackage{thmtools}

\usepackage{aliascnt}

\usepackage[nameinlink,capitalise]{cleveref}

\newtheorem{theorem}{Theorem}
\newtheorem{lemma}[theorem]{Lemma}
\newtheorem{claim}[theorem]{Claim}

\newtheorem{corollary}[theorem]{Corollary}

\theoremstyle{definition}
\newtheorem{definition}[theorem]{Definition}

\newtheorem*{remark}{Remark}

\crefname{lemma}{lemma}{lemmas}
\Crefname{lemma}{Lemma}{Lemmas}
\crefname{claim}{claim}{claims}
\Crefname{claim}{Claim}{Claims}

\newcommand{\ov}[1]{{\color{cyan}OV: #1}\marginpar{$\star\star$}}

\newcommand{\eps}{\varepsilon}

        {
    \end{proof}}

\newcommand{\dpalert}[1]{{\color{red} DP: #1}}

\newcommand{\eat}[1]{}

\newcommand{\resort}{{\sc Re-Sort}\xspace}

\newcommand{\delete}{{\sc Delete}\xspace}

\newcommand{\insertadst}{{\sc Insert}\xspace}
\newcommand{\routehole}{{\sc Any2End}\xspace}
\newcommand{\downhole}{{\sc JumpDown}\xspace}
\newcommand{\rebuild}{{\sc ReBuild}\xspace}
\newcommand{\addmax}{{\sc InsertMax}\xspace}
\newcommand{\swaphole}{{\sc SwapHole}\xspace}
\newcommand{\slide}{{\sc Start2End}\xspace}
\newcommand{\simpleslide}{{\sc SimpleStart2End}\xspace}
\newcommand\opt{\mathsf{opt}\xspace}
\let\polylog\relax
\DeclareMathOperator*{\polylog}{polylog}

\newcommand{\dst}{{\sc DST}\xspace}
\newcommand{\dsts}{{\sc DST}s\xspace}
\newcommand{\adst}{{\sc A-DST}\xspace}
\newcommand{\adsts}{{\sc A-DST}s\xspace}

\usetikzlibrary{decorations.pathreplacing}

\begin{document}

\title{Online and Fully-Dynamic Sorting}

\author{Yossi Azar\thanks{Department of Computer Science, Tel-Aviv University. Email: \texttt{azar@tauex.tau.ac.il}. Supported in part by the Israel Science Foundation grant No. 1728/26.}
\quad 
Sujoy Bhore\thanks{Department of Computer Science and Engineering, IIT Bombay. Email: \texttt{sujoy@cse.iitb.ac.in}. Supported in part by by ANRF ARG-MATRICS, Grant 002465.}
\quad
Devdan Dey\thanks{Department of Computer Science and Engineering, IIT Bombay. Email: \texttt{mnz.devdan@gmail.com}.} 
\quad
Debmalya Panigrahi\thanks{Department of Computer Science, Duke University. Email: \texttt{debmalya@cs.duke.edu}. Supported in part by NSF grant CCF-329230.} 
\quad
Or Vardi\thanks{Department of Computer Science, Tel-Aviv University. Email: \texttt{orvardi@mail.tau.ac.il}.}
\quad
Ian Zhang\thanks{Duke University. Email: \texttt{ian.zhang@duke.edu}.}}
\date{}

\maketitle

\pagenumbering{gobble}

\begin{abstract}

Online sorting asks how to place $n$ sequentially arriving elements in an array to minimize the sum of absolute differences between consecutive elements. Fully-dynamic sorting further extends the problem to allow both element insertions and deletions. Without recourse, these problems face strong lower bounds on the competitive ratio: $\Omega(\sqrt{n})$ for online sorting and $\Omega(n)$ for fully-dynamic sorting. In sharp contrast, we show that allowing only poly-logarithmic recourse yields at least exponential improvements in the competitive ratio for both problems:
\begin{enumerate}
\item
Our main algorithmic result is a deterministic fully-dynamic sorting algorithm that is $O(\log^2 n)$-competitive and has $O(\log^4 n)$ amortized recourse per update. We complement this with a lower bound showing that any deterministic algorithm with any $O(\polylog(n))$ competitive ratio has amortized recourse at least $\Omega(\sqrt{\log n})$.
\item
For the insertion-only setting, we establish a family of approximation--recourse tradeoffs, including $O(\log\log n)$ competitive ratio with $O(\log\log n)$ amortized recourse. Given the lower bound in the fully-dynamic case, this result shows that the insertion-only problem yields exponentially better bounds than the fully-dynamic one. 

\item
Finally, we show that the stronger approximation-recourse tradeoff for the insertion-only case extends to the fully-dynamic setting when the array has $(1+\varepsilon)n$ space.
\end{enumerate}

\end{abstract}

\clearpage

\tableofcontents

\clearpage

\pagenumbering{arabic}

\section{Introduction}\label{sec:introduction}Maintaining ordered data under online updates is a fundamental algorithmic task. Consider an array of size $n$ that must store a sequence of elements drawn from a fixed range, say $[0, 1]$. Elements arrive online, one at a time, and upon arrival must be assigned positions in the array. How should we arrange these elements so as to preserve as much order as possible while performing as little rearrangement as possible?

This question is elegantly captured in the {\em online sorting} problem, recently introduced by \cite{AABK23}, and further studied by \cite{AbrahamsenB0K024,NirjhorW25,AzarPV26,improved-stochastic-hu,kalavas26,Hermansen26,PlatanosTolias26,FanKang26,BerceaBIKP26, AamandABKKT27}. Here, the objective is to insert the elements in the array with the goal of minimizing the sum of absolute differences between consecutive elements in the array. Recent work has developed a rich theory of online sorting, establishing upper and lower bounds on the value of the objective that can be maintained without element rearrangement. The central result is that for element insertions, a competitive ratio of $O(\sqrt{n})$ can be achieved~\cite{AABK23}, and this is tight~\cite{AABK23,AbrahamsenB0K024}. But, what if we have {\em both element insertions and deletions}, i.e., we are in the {\em fully dynamic} setting? In this case, it is easy to establish a lower bound of $\Omega(n)$, which matches the trivial upper bound of $O(n)$ where the elements are re-sorted after each insertion/deletion. So, we ask: 
\begin{quote}
{\em Using limited recourse, can we achieve non-trivial bounds on the competitive ratio of the fully-dynamic sorting problem? Furthermore, using limited recourse, can we also improve on the $O(\sqrt{n})$ bound for online (i.e., insertion-only) sorting?} 
\end{quote}

\subsection{Our Results}

\paragraph{Fully-dynamic Sorting.} 
First, we consider the more general problem of 
fully-dynamic sorting. Our main result is a deterministic algorithm
with a $O(\polylog(n))$-competitive ratio, while incurring 
$O(\polylog(n))$ amortized recourse. 

\begin{restatable}{theorem}{dynamicthm}
\label{thm:dynamic}
There is a deterministic algorithm for the fully dynamic
sorting problem that maintains an array of size $n$
under arbitrary insertions and deletions, with competitive
ratio $O(\log^2 n)$ and $O(\log^4 n)$ amortized recourse
per update.
\end{restatable}

We complement this upper bound with a lower bound 
showing that any deterministic algorithm with 
an $O(\polylog(n))$ competitive ratio for this problem
must incur $\Omega(\sqrt{\log n})$ amortized recourse. So, 
our upper bound is tight up to the exact exponents in
the poly-logs.

\begin{restatable}{theorem}{lowerthm}
\label{thm:dynamic-lower}
Let $1 \le C \le O(n) $. Any deterministic $C$-competitive
algorithm for fully dynamic sorting with amortized
recourse $\rho$ satisfies
\[
    \rho =
    \Omega\!\left(
        \frac{\log n}
        {\log C+\sqrt{\log n}}
    \right).
\]
\end{restatable}


\paragraph{Insertion-only Sorting.}
Next, we consider the online (i.e., insertion-only) sorting problem.
We show that in this case, we can obtain an exponential improvement 
over the lower bound in \Cref{thm:dynamic-lower} by giving a deterministic
$O(\log\log n)$-approximation using $O(\log\log n)$ amortized recourse.

\begin{restatable}{theorem}{onlinethm}
\label{thm:insertion_only}
There is a deterministic algorithm for the online sorting
problem with competitive ratio $O(\log\log n)$ and
$O(\log\log n)$ amortized recourse per insertion.
\end{restatable}

More generally, we obtain a family of tradeoffs between
approximation and amortized recourse in the insertion-only
setting, which we present in \Cref{sec:insertion-only}.
In addition to \Cref{thm:insertion_only},
this tradeoff also yields approximations of $O(\log n)$
and $O(1)$ with $O(1)$ and $O(\log n)$ amortized recourse respectively. These approximation
bounds are at least exponentially better than the 
lower bound of $\Omega(\sqrt n)$
that can be obtained without recourse.

Finally, a popular line of work within online sorting
explores the problem when the array has extra space. 
The works of 
\cite{AABK23,NirjhorW25,AzarPV26} have shown that 
with an array of size $(1+\eps)n$, one can achieve exponential 
improvements on the competitive ratio of online 
sorting. We show that this phenomenon is also true for 
fully-dynamic sorting. Namely, we give a black-box 
reduction showing that in an array of size $(1+\eps)n$, 
fully-dynamic sorting reduces to online
(i.e., insertion only) sorting
with $O(1)$ overhead in cost and additive
$O(1/\eps)$ overhead in recourse. 

\begin{restatable}{theorem}{spacethm}
\label{thm:extraspace}
Suppose the array has size $(1+\varepsilon)n$ for any
fixed $\varepsilon>0$. Then every deterministic online
sorting algorithm with competitive ratio $c$ and
amortized recourse $r$ can be transformed into a
deterministic fully dynamic sorting algorithm with
competitive ratio $2c+1$ and amortized recourse
$r+O(1/\varepsilon)$ per update.
\end{restatable}

In particular, \Cref{thm:insertion_only,thm:extraspace}
imply that with $(1+\varepsilon)n$ space, fully dynamic
sorting admits an $O(\log\log n)$ competitive ratio
with $O(\log\log n+1/\varepsilon)$ amortized recourse.

\subsection{Our Techniques}

\paragraph{\bf Fully-Dynamic Upper Bound (\Cref{thm:dynamic}).}
The high-level idea in fully-dynamic sorting is to move any holes 
created by element deletions to the end of the array, while keeping the
array approximately sorted. Then, insertions can happen only at the end
of the array, with local re-sorting and periodic merger of the newly
inserted elements into the rest of the array. 

To implement the operation of moving a hole efficiently through the
array, we recursively organize the array into a data structure that 
we call a \emph{Dynamic Sorting Tree} (\dst). 
The challenge is to avoid shifting all
intervening elements while ensuring that successive deletions 
don't accumulate
too much disorder. The \dst organizes the array recursively at
geometrically decreasing size scales, where a level-$i$ \dst
comprises blocks that are themselves composed of sections, 
each of which is a level-$(i-1)$ \dst. The sections are 
labeled either as {\em original} or {\em transfer} sections, 
the intuition being the original sections are the ones created
when the \dst is initialized in sorted order, while the 
transfer sections allow limited disorder and are created in the 
process of moving holes through the \dst.


Crucially, a transfer section is allowed to contain elements larger than
those in the original section that follows it. When a hole reaches the
beginning of the original section, we first switch it to the transfer
section by changing their shared boundary. No element moves, but the
hole is now at the end of the transfer section. We fill this hole with a
carefully chosen minimum element from an original section further to
the right and in a lower recursive level. 
The ordering constraints ensure that this element is larger
than the receiving transfer section's existing elements, and so, accepting it
requires no shift of its contents. The resulting hole is now in 
the original section that lost an element, but since we have moved 
one level down in the recursion, this process can be continued 
down the recursion tree. 

In addition to these moves, we also need to initially move the hole
from the position of the deleted element (which can be arbitrary)
to the beginning of an original section. This requires further recursive moves 
within the section containing the deleted element, which are morally
similar to moving the hole across the \dst albeit at a lower recursive
level. Combining these two recursively defined operations, 
the one that initially moves the hole to the beginning to an 
original section, and the other that moves it all the way
to the end of the array, yields a recourse bound of $O(i^3)$
for an $i$-level \dst. 
Moving elements and changing section boundaries gradually alters
section sizes, which is fixed by periodic calls to a rebuilding operation 
that reconstitutes a \dst that violates its size invariants.  Rebuilding 
is expensive, but it only done occasionally, thereby allowing 
us to amortize the resulting recourse to a sequence of element 
deletions preceding it. 
Finally, the 
approximation bound is controlled by ordering invariants that 
we maintain between the different sections in the \dst.
For a level-$i$ \dst, this yields a cost bound of $O(i)$ times 
the overall range of values stored in the \dst.


To support insertions as well as deletions, we arrange independent
\dst components of decreasing depths into an \emph{Augmented Dynamic
Sorting Tree} (\adst), with no ordering constraints between components.
A deletion first routes its hole to the end of its component and then
through the subsequent components to the small terminal component,
where new elements are inserted. Periodic suffix rebuilding
consolidates a suffix into a clean \dst and empties its smaller
components, maintaining the geometric size bounds. Each such
reconstruction is amortized over a number of updates proportional to
the suffix's size bound, contributing only $O(d)$ amortized recourse
for a $d$-level structure. The component cost bounds sum to $O(d^2)$,
while hole routing together with structural rebuilding incurs
$O(d^4)$ amortized recourse per update. Choosing $d=\Theta(\log n)$
therefore gives an $O(\log^2 n)$ competitive ratio with
$O(\log^4 n)$ amortized recourse.

\paragraph{Fully-Dynamic Lower Bound (\Cref{thm:dynamic-lower}).}

Fix a deterministic $C$-competitive algorithm. We show
that maintaining sorting cost at most $C\cdot \operatorname{OPT}$ requires large recourse. 
The adversary works in rounds. Each round starts with the array
containing $n$ equally spaced valued in $[0,1]$, and has two 
phases. During the
\emph{forward phase}, the adversary repeatedly deletes an existing element
and inserts a copy of $1$. At some stage, the adversary switches to 
the \emph{backward phase}, where it removes the $1$s it had inserted
and restores the original elements. This marks the start of the 
next round. Since all values lie in
$[0,1]$, we always have $ \operatorname{OPT}\le 1$, so the algorithm must
maintain sorting cost at most $C$.

The challenge is to choose the deletions so that the algorithm
must either move many elements or incur a large cost.
The adversary organizes the original values into a hierarchy
of intervals. For each interval $I_v$, some child intervals
are labeled \emph{targets}, that can be further explored for reaching a leaf for deletion,
and others \emph{buffers}, that cannot.
The labels are chosen using the algorithm's current arrangement.
The adversary divides the elements of the middle children
intervals, in array order, into consecutive groups called
\emph{windows}. It chooses the labels for the children so that a constant
fraction of these windows are \emph{balanced}, meaning that
each contains many elements of both types.

If the algorithm moves enough buffer elements, we call the
corresponding interval $I_v$ \emph{paid}. These movements
contribute to the recourse lower bound, and so the adversary
\emph{releases} its buffer children and allows their elements
to be further explored towards a leaf. This lets the adversary continue removing
original elements, rather than stopping at buffers for which
the algorithm has already incurred the required recourse.
The forward phase ends once the adversary cannot 
remove any additional element using legal search moves.

To bound the cost,
we define, for any interval $I_v$, the cost {\em through} the interval
as the sum of lengths of intersects between $I_v$ and the interval spanned by each pair of consecutive array elements. The importance of this parameter is that the total cost of the array is lower bounded by the sum of costs through any set of disjoint intervals. Our main lemma states that for unpaid intervals $I_v$, either many original elements in $I_v$ that belonged to target children were not changed to $1$, or the cost through $I_v$ is large compared to the length of $I_v$. 
We now use the main lemma recursively to conclude
that at most $n/4$ original elements remain at the end of the forward phase
using an appropriate choice of parameters. This implies that there are many paid intervals.

An interval that never becomes paid retains at least half of
its original elements. Since few original elements remain,
the paid intervals at each depth must together have held
\(\Omega(n)\) elements at the start of the round.
Using the definition of a paid interval, this implies 
that the algorithm must have
moved \(\Omega(n)\) original buffer elements in total at each depth.
We show that these movements can be chosen so that
none is counted at two different depths.
Summing over the \(d\) depths gives \(\Omega(nd)\) recourse
in a round of \(O(n)\) updates, and hence \(\Omega(d)\)
amortized recourse. Choosing the branching factor and depth
appropriately then yields the stated tradeoff with cost $C$.

\paragraph{\bf Online (Insertion-Only) Sorting (Theorem~\ref{thm:insertion_only}).} 
Our insertion-only algorithm proceeds in two stages. First, we partition the insertion sequence into geometrically decreasing batches and use the recent online sorting algorithm of~\cite{AzarPV26} to transform each batch into a contiguous sorted block while incurring only $O(1)$ amortized recourse per element and constant cost. This reduces the problem to maintaining a sequence of sorted blocks in an approximately sorted array. The remaining challenge is to balance two competing objectives: keeping few sorted regions, which leads to a low competitive ratio, and minimizing the number of times each block is re-sorted, which leads to low recourse.

To resolve this tension, we introduce two complementary block gadgets. The \emph{Low Recourse Block Gadget} organizes the blocks as the leaves of a virtual $b$-ary tree and re-sorts the blocks corresponding to completed subtrees. Since every block belongs to only one subtree at each height, it participates in just $O(\log_b k)$ re-sorts, giving $O(\log_b k)$ amortized recourse, at the cost of maintaining up to $O(b)$ active subtrees per level and hence a competitive ratio of $O(b\log_b k)$. The \emph{High Recourse Block Gadget} instead maintains a hierarchy of sorted super-blocks of geometrically increasing capacities, merging a super-block into its parent whenever it becomes full. This ensures that there is only one sorted super-block per level, yielding a competitive ratio of $O(\log_b k)$, but each block may be merged $O(b)$ times per level, resulting in $O(b\log_b k)$ amortized recourse. By choosing the branching factor $b$, the two gadgets yield complementary approximation--recourse tradeoffs, including $O(\log n)$ approximation with $O(1)$ recourse and $O(1)$ approximation with $O(\log n)$ recourse.

\paragraph{\bf Fully-Dynamic Sorting with $(1+\varepsilon)$-Space (Theorem~\ref{thm:extraspace}).} 
For an array with $(1+\varepsilon)n$ space, we reduce fully dynamic sorting to online sorting, establishing their equivalence in this setting. The reduction periodically rebuilds the array, leaving an empty buffer of size $\varepsilon n$; insertions are handled online, while deletions are deferred until the next rebuild. Rebuilding every $\varepsilon n$ updates adds only $O(1/\varepsilon)$ amortized recourse.
We note that, at a high level, our periodic rebuilding strategy is similar in spirit to the reduction from fully-dynamic to online list labeling with extra space in \cite{bender2024nearly}, although the technical details of the two reductions are different.





\subsection{Related Work}

\paragraph{Online Sorting.}
The online sorting problem was introduced by Aamand
{\em et al.}~\cite{AABK23}, who established tight
deterministic bounds of $\Theta(\sqrt n)$ for $m=n$,
later extended to randomized algorithms by Abrahamsen
{\em et al.}~\cite{AbrahamsenB0K024}.
For $m=(1+\varepsilon)n$, Aamand {\em et al.} ~\cite{AABK23}
established a deterministic lower bound of
$\Omega(\log n/\log\log n)$ and a subpolynomial upper
bound. Subsequent works~\cite{NirjhorW25,AzarPV26}
obtained polylogarithmic upper bounds, with Azar
{\em et al.}~\cite{AzarPV26} achieving
$O(\log^2 n/\varepsilon)$ competitiveness and
$O(1)$ competitiveness with $O(n\log^2 n)$ space
when the value range is known.
Azar {\em et al.}~\cite{AzarPV26-icalp} extended these
results to online metric TSP, which was introduced by \cite{AbrahamsenB0K024}, obtaining an
$O(\log^3 n/\varepsilon)$-competitive algorithm using
$m=(1+\varepsilon)n$ space. They also showed that
when $\opt$ is unknown, any deterministic algorithm
using $m=n^{1+\delta}$ space has competitive ratio
$\Omega(1/\delta)$, even for online sorting.

Several variants of online sorting have also been
studied. Under stochastic and random-order arrivals,
recent works established tight logarithmic
bounds~\cite{AbrahamsenB0K024,improved-stochastic-hu,
kalavas26,Hermansen26,PlatanosTolias26,FanKang26}.
Bercea {\em et al.}~\cite{BerceaBIKP26} also studied the problem in the learning-augmented setting, establishing tight consistency robustness tradeoffs.
Finally, online sorting techniques have been applied
to geometric problems, initially for lower bounds ~\cite{AABK23}.
While recently in \cite{AamandABKKT27} several upper bounds for geometric problems are shown, the major one being an
$O(\log^2 n)$-competitive algorithm for online
translational packing of convex polygons.

\paragraph{\bf Recourse in Online and Fully-Dynamic Algorithms.}
Allowing online algorithms to revise earlier decisions while bounding
the number of such changes, known as \emph{recourse}, has proved
powerful across a wide range of problems.
This includes matching, scheduling, load balancing, and network
flows~\cite{GuptaKS14, Krishnaswamy0S23},
Steiner tree and forest~\cite{GuG016, LongMST26},
and minimum spanning tree and TSP~\cite{MegowSVW16}.
These works demonstrate that even limited recourse can substantially
improve the achievable competitive ratio compared with
irrevocable online algorithms.

A closely related line of research studies recourse in the
fully-dynamic setting, where elements or requests may both arrive
and depart. Bounded-recourse algorithms have been developed for
load balancing~\cite{FoussoulGK26, BenderKSS26},
Steiner tree~\cite{GuptaK14},
graph spanners~\cite{BhattacharyaSS22},
set cover and submodular cover~\cite{
GuptaK0P17, GuptaL20, SolomonU23, SolomonU26,
BhattacharyaCP26},
facility location and clustering~\cite{
Cohen-AddadHPSS19, LackiHGJR24, ForsterS25,
BhattacharyaCGL24, BhattacharyaCF25, BhattacharyaCFJJL26},
and bin packing~\cite{FeldkordFGGKRW18}.
More general frameworks for controlling recourse in dynamic
settings have also been developed~\cite{BhattacharyaBLS23}.

\paragraph{Online List-Labeling.}
The \emph{online list labeling} (also known as \emph{file maintenance}) problem introduced by \cite{itai1981} asks a complementary question for the same setting of element insertions: what is the minimum recourse of an algorithm that is required to keep the array precisely sorted? 
For deterministic algorithms, tight bounds of
$\Theta(\log^2 n)$ and $\Theta(\log n)$ amortized recourse
are known for linear and polynomial space, respectively
~\cite{bulanek2012}.
Recent randomized algorithms have improved the linear-space
upper bound to $O(\log n\cdot\mathrm{polyloglog}\,n)$
~\cite{bender2024siam,bender2024layered,bender2024nearly}.
Our work can be interpreted as relaxing the requirement that the array stays perfectly sorted, while allowing bounded recourse.



\subsection{Organization of the Paper}

We begin in \Cref{sec:two-level-int} by giving a simple $2$-level data structure that presents some of our ideas in fully-dynamic sorting without the technical complications of the full result. We then introduce the Dynamic Sorting Tree (\dst) in \Cref{sec:new-dst} and build on it to construct the Augmented Dynamic Sorting Tree (\adst) in \Cref{sec:fully-dynamic-lowcost}; the \adst structure realizes \Cref{thm:dynamic}. In \Cref{sec:polylog-lower-bound}, we establish our lower bound for this setting.  \Cref{sec:insertion-only} considers the insertion-only setting (online sorting), where we design two new data structures, the Low Recourse Block Gadget and the High Recourse Block Gadget, and use them to yield \Cref{thm:insertion_only}. Finally, in \Cref{fully-dynamic-extraspace}, we extend these tradeoffs to the fully dynamic setting with $(1+\eps)n$ space, establishing \Cref{thm:extraspace}. All proofs that are omitted from the main body of the paper appear in the appendix.


\section{Preliminaries}\label{sec:preliminaries}We consider an array of size $m\ge n$ storing real-valued elements that arrive online, where $n$ is an upper bound on the number of elements present at any time. Each element occupies a distinct cell, so the array has at least $m-n$ empty cells.

Let $S=(x_1,\ldots,x_k)$ denote the elements currently stored in the array, ordered from left to right and ignoring empty cells. The cost of the array is
\[
    \operatorname{cost}(S)=\sum_{j=1}^{k-1}|x_{j+1}-x_j|.
\]
An optimal offline arrangement sorts the elements and therefore has cost
$\operatorname{OPT}(S)=\max S-\min S$.
For sequences containing at most one element, both costs are defined to be zero.

It will be useful to express the cost in terms of threshold crossings. For a threshold $z\in\mathbb{R}$, let $\operatorname{cross}_z(S)$ denote the number of adjacent pairs in $S$ with one element at most $z$ and the other greater than $z$. Formally,
\[
    \operatorname{cross}_z(S)=\bigl|\{j\in[k-1]:(x_j>z)\ne(x_{j+1}>z)\}\bigr|.
\]
For sequences containing at most one element, this number is zero. Since each adjacent pair crosses precisely the thresholds between its two values, summing over all pairs gives
\begin{equation}
\label{eq:cost-crossing}
    \operatorname{cost}(S)=\int_{\mathbb R}\operatorname{cross}_z(S)\,dz.
\end{equation}

The \emph{recourse} of an update is the number of previously stored elements that change positions during that update. We measure amortized recourse as the total recourse over a sequence of updates divided by the number of updates.

\paragraph{Insertion-Only and Fully Dynamic Updates.}
We consider two variants of online sorting.
In the \emph{insertion-only} setting, elements arrive
one at a time and remain in the array permanently.
Upon each arrival, the algorithm must assign the
new element to an empty cell, possibly rearranging
previously inserted elements.

In the \emph{fully dynamic} setting, the algorithm
must also support deletions of previously inserted
elements. A deletion frees the cell occupied by
the removed element, and the algorithm may rearrange
the remaining elements to maintain a low-cost array.
We assume that at most $n$ elements are present
at any time.

\paragraph{Initialization and Reduction to Deletion-Insertion Pairs.}
We describe our fully dynamic algorithm under the
simplifying assumption that updates consist of
deletion-insertion pairs. To handle arbitrary
update sequences, we initialize an \adst\ containing
$n$ virtual elements, all with value $0$ and marked
as inactive. These elements are initially sorted,
and the \adst\ is constructed with all transfer
sections empty. Since no active elements are present,
this initialization incurs no recourse.

When an active element is deleted, we simply mark
it as inactive without changing the array.
When a new element $x$ arrives, we choose an arbitrary
inactive element $y$, delete it from the \adst,
and insert $x$ using a deletion-insertion pair.
Such an inactive element always exists, since
immediately before an insertion there are at most
$n-1$ active elements. Thus, the \adst\ maintains
exactly $n$ elements throughout the execution,
some of which may be inactive. Inactive elements
correspond to empty cells in the actual array,
but their values are retained internally for
comparisons.

The active elements, in their relative order,
form a subsequence of the sequence maintained
by the \adst. By the following lemma, a bound
on the number of crossings at every threshold
for the maintained sequence also holds for the
active elements, preserving the competitive
guarantee. Moreover, each actual insertion
requires only one deletion-insertion pair,
while actual deletions incur no recourse.
Consequently, the amortized recourse bound for
deletion-insertion pairs also applies to
arbitrary update sequences.

\begin{lemma}
\label{lem:subsequence-crossings}
Let $S'$ be a subsequence of a sequence $S$. Then, for every threshold $z$,
$\operatorname{cross}_z(S')\le \operatorname{cross}_z(S)$.
In particular, if $\operatorname{cross}_z(S)\le C$ for every $z$, then $\operatorname{cost}(S')\le C\cdot\operatorname{OPT}(S')$.
\end{lemma}


Thus, an algorithm that guarantees at most $C$ crossings at every threshold remains $C$-competitive when inactive elements are ignored. For algorithms satisfying this crossing bound, it therefore suffices to analyze deletion--insertion pairs.

\section{Intuition: A $2$-Level Data Structure}\label{sec:two-level-int}

We first describe a $2$-level data structure that 
captures some of the intuition of our eventual data structures,
\dst and \adst,
while keeping the technical details much simpler.
We then give the actual construction of our data structures 
in the next two sections, and 
establish \Cref{thm:dynamic}.

By the previous section, we imagine, w.l.o.g., 
that we start with a sorted array and then handle an 
alternating sequence of element deletions and insertions.
Deleting an element from a sorted array creates a hole.
A na\"ive idea is to
move this hole to the end while preserving sorted order,
then insert the next element in this hole, and move this
element by shifting elements to its position in the 
sorted order. While this keeps the array precisely sorted,
it incurs $\Theta(n)$ recourse in each step.
(Throughout this section, $n$ denotes the size of the array.)
Instead, we allow a limited amount of disorder
within the array, so that a hole can pass through the 
array more efficiently.

To illustrate the idea, consider partitioning the array
into approximately $\sqrt{n}$ blocks, each containing
approximately $\sqrt{n}$ elements.
We keep the blocks ordered by value, i.e., for two 
consecutive blocks, all elements in the right block 
are larger than all elements in the left block.
Initially, we also sort the elements within each block.

Now, suppose there is a hole inside a block.
With $O(\sqrt{n})$ element movements, we can move the 
hole to the end of the block.
Filling the hole with the minimum element of the next block,
the hole moves to the beginning of the next block. 
But, how do we now move the hole 
to the end of the next block? Doing this na\"ively would 
incur $O(\sqrt{n})$ movement in the next block as well, 
which would create $\Theta(n)$ movement overall 
across the $O(\sqrt{n})$ blocks. 

\begin{figure}[t]
\centering
\resizebox{.70\linewidth}{!}{%
\begin{tikzpicture}[
    x=1cm,y=1cm,
    every node/.style={font=\scriptsize},
    blk/.style={draw=black!70,very thick,rounded corners=3pt},
    firstb/.style={draw=orange!85!black,very thick,rounded corners=3pt},
    lastb/.style={draw=green!50!black,very thick,rounded corners=3pt},
    tsec/.style={fill=blue!15},
    osec/.style={fill=gray!15},
    cell/.style={
        draw=black!40,minimum width=.60cm,
        minimum height=.55cm,inner sep=0pt
    },
    hole/.style={
        draw=red!85!black,line width=1.2pt,
        fill=white,minimum width=.60cm,
        minimum height=.55cm,inner sep=0pt
    },
    stage/.style={anchor=west,inner sep=0pt},
    actiontext/.style={anchor=center,fill=white,inner sep=1pt},
    transfer/.style={
        -{Latex[length=1.4mm,width=.9mm]},
        densely dotted,line cap=round,
        line width=.6pt,draw=black!65
    }
]

\newcommand{\twoBlockRow}[1]{%
    \draw[firstb] (0,{#1-.36}) rectangle (1.62,{#1+.36});
    \draw[blk] (1.97,{#1-.36}) rectangle (5.75,{#1+.36});
    \draw[blk] (6.10,{#1-.36}) rectangle (9.16,{#1+.36});
    \draw[lastb] (9.51,{#1-.36}) rectangle (11.13,{#1+.36});%
}

\newcommand{\twoOrigCell}[3]{%
    \node[cell,osec] at (#1,#2) {$#3$};%
}
\newcommand{\twoTransferCell}[3]{%
    \node[cell,tsec] at (#1,#2) {$#3$};%
}
\newcommand{\twoHoleCell}[2]{%
    \node[hole] at (#1,#2) {$\circ$};%
}
\newcommand{\twoFixedFirst}[1]{%
    \twoOrigCell{.45}{#1}{1}\twoOrigCell{1.17}{#1}{2}%
}
\newcommand{\twoState}[2]{%
    \node[stage] at (0,{#1+.64}) {#2};
    \twoFixedFirst{#1}%
}

\newcommand{\twoDonorMove}[3]{%
    \pgfmathsetmacro{\twoMoveDepth}{min(.30,.17*abs(#2-#1))}%
    \draw[transfer]
        (#1,{#3-.42})
        .. controls ({#1+.25*(#2-#1)},{#3-.42-\twoMoveDepth})
        and ({#1+.75*(#2-#1)},{#3-.42-\twoMoveDepth})
        .. (#2,{#3-.42});%
}

\node[stage] at (0,12.00) {(a) Before deletion};

\foreach \x/\name in {.81/B_1,3.86/B_2,7.63/B_3,10.32/B_4}{
    \node at (\x,11.55) {$\name$};
}
\node[text=blue!65!black] at (2.42,11.10) {$T_2$};
\node[text=black!70] at (4.20,11.10) {$O_2$};
\node[text=blue!65!black] at (6.55,11.10) {$T_3$};
\node[text=black!70] at (8.00,11.10) {$O_3$};

\twoFixedFirst{10.45}
\twoTransferCell{2.42}{10.45}{8}
\twoOrigCell{3.14}{10.45}{3}\twoOrigCell{3.86}{10.45}{4}
\twoOrigCell{4.58}{10.45}{5}\twoOrigCell{5.30}{10.45}{6}
\twoTransferCell{6.55}{10.45}{12}
\twoOrigCell{7.27}{10.45}{9}
\twoOrigCell{7.99}{10.45}{10}\twoOrigCell{8.71}{10.45}{11}
\twoOrigCell{9.96}{10.45}{13}\twoOrigCell{10.68}{10.45}{14}
\twoBlockRow{10.45}

\node[actiontext] at (5.565,9.57)
    {Delete $4$; shift $5,6$ to the left};

\twoState{8.60}{(b) Hole at end of $O_2$}
\twoTransferCell{2.42}{8.60}{8}
\twoOrigCell{3.14}{8.60}{3}\twoOrigCell{3.86}{8.60}{5}
\twoOrigCell{4.58}{8.60}{6}\twoHoleCell{5.30}{8.60}
\twoTransferCell{6.55}{8.60}{12}
\twoOrigCell{7.27}{8.60}{9}
\twoOrigCell{7.99}{8.60}{10}\twoOrigCell{8.71}{8.60}{11}
\twoOrigCell{9.96}{8.60}{13}\twoOrigCell{10.68}{8.60}{14}
\twoBlockRow{8.60}
\twoDonorMove{2.42}{5.30}{8.60}

\node[actiontext] at (5.565,7.72)
    {Move $8=\min T_2$ into $O_2$};

\twoState{6.75}{(c) Hole in $T_2$}
\twoHoleCell{2.42}{6.75}
\twoOrigCell{3.14}{6.75}{3}\twoOrigCell{3.86}{6.75}{5}
\twoOrigCell{4.58}{6.75}{6}\twoOrigCell{5.30}{6.75}{8}
\twoTransferCell{6.55}{6.75}{12}
\twoOrigCell{7.27}{6.75}{9}
\twoOrigCell{7.99}{6.75}{10}\twoOrigCell{8.71}{6.75}{11}
\twoOrigCell{9.96}{6.75}{13}\twoOrigCell{10.68}{6.75}{14}
\twoBlockRow{6.75}
\twoDonorMove{7.27}{2.42}{6.75}

\node[actiontext] at (5.565,5.87)
    {Move $9$ into $T_2$; apply \swaphole$(O_3,T_3)$};

\twoState{4.90}{(d) Hole in $T_3$}
\twoTransferCell{2.42}{4.90}{9}
\twoOrigCell{3.14}{4.90}{3}\twoOrigCell{3.86}{4.90}{5}
\twoOrigCell{4.58}{4.90}{6}\twoOrigCell{5.30}{4.90}{8}
\twoTransferCell{6.55}{4.90}{12}\twoHoleCell{7.27}{4.90}
\twoOrigCell{7.99}{4.90}{10}\twoOrigCell{8.71}{4.90}{11}
\twoOrigCell{9.96}{4.90}{13}\twoOrigCell{10.68}{4.90}{14}
\twoBlockRow{4.90}
\twoDonorMove{9.96}{7.27}{4.90}

\node[actiontext] at (5.565,4.02)
    {Move $13=\min B_4$ into $T_3$};

\twoState{3.05}{(e) Hole in $B_4$}
\twoTransferCell{2.42}{3.05}{9}
\twoOrigCell{3.14}{3.05}{3}\twoOrigCell{3.86}{3.05}{5}
\twoOrigCell{4.58}{3.05}{6}\twoOrigCell{5.30}{3.05}{8}
\twoTransferCell{6.55}{3.05}{12}
\twoTransferCell{7.27}{3.05}{13}
\twoOrigCell{7.99}{3.05}{10}\twoOrigCell{8.71}{3.05}{11}
\twoHoleCell{9.96}{3.05}\twoOrigCell{10.68}{3.05}{14}
\twoBlockRow{3.05}
\twoDonorMove{10.68}{9.96}{3.05}

\node[actiontext] at (5.565,2.17)
    {Shift $14$ one position to the left};

\twoState{1.20}{(f) Final hole}
\twoTransferCell{2.42}{1.20}{9}
\twoOrigCell{3.14}{1.20}{3}\twoOrigCell{3.86}{1.20}{5}
\twoOrigCell{4.58}{1.20}{6}\twoOrigCell{5.30}{1.20}{8}
\twoTransferCell{6.55}{1.20}{12}
\twoTransferCell{7.27}{1.20}{13}
\twoOrigCell{7.99}{1.20}{10}\twoOrigCell{8.71}{1.20}{11}
\twoOrigCell{9.96}{1.20}{14}\twoHoleCell{10.68}{1.20}
\twoBlockRow{1.20}

\end{tikzpicture}%
}
\caption{Deletion of $4$ in the simplified two-level structure. After deleting $4$, the hole is moved to the end of $O_2$. The subsequent transfers move $8$ into $O_2$, $9$ into $T_2$, and $13$ into $T_3$. The last two transitions are shown separately: first borrowing $13$ from $B_4$, then shifting $14$ to leave the hole in the final cell. Gray and blue cells belong to original and transfer sections; the red outline marks the hole. Dotted arrows indicate element movements.}
\label{fig:dst-example}
\end{figure}
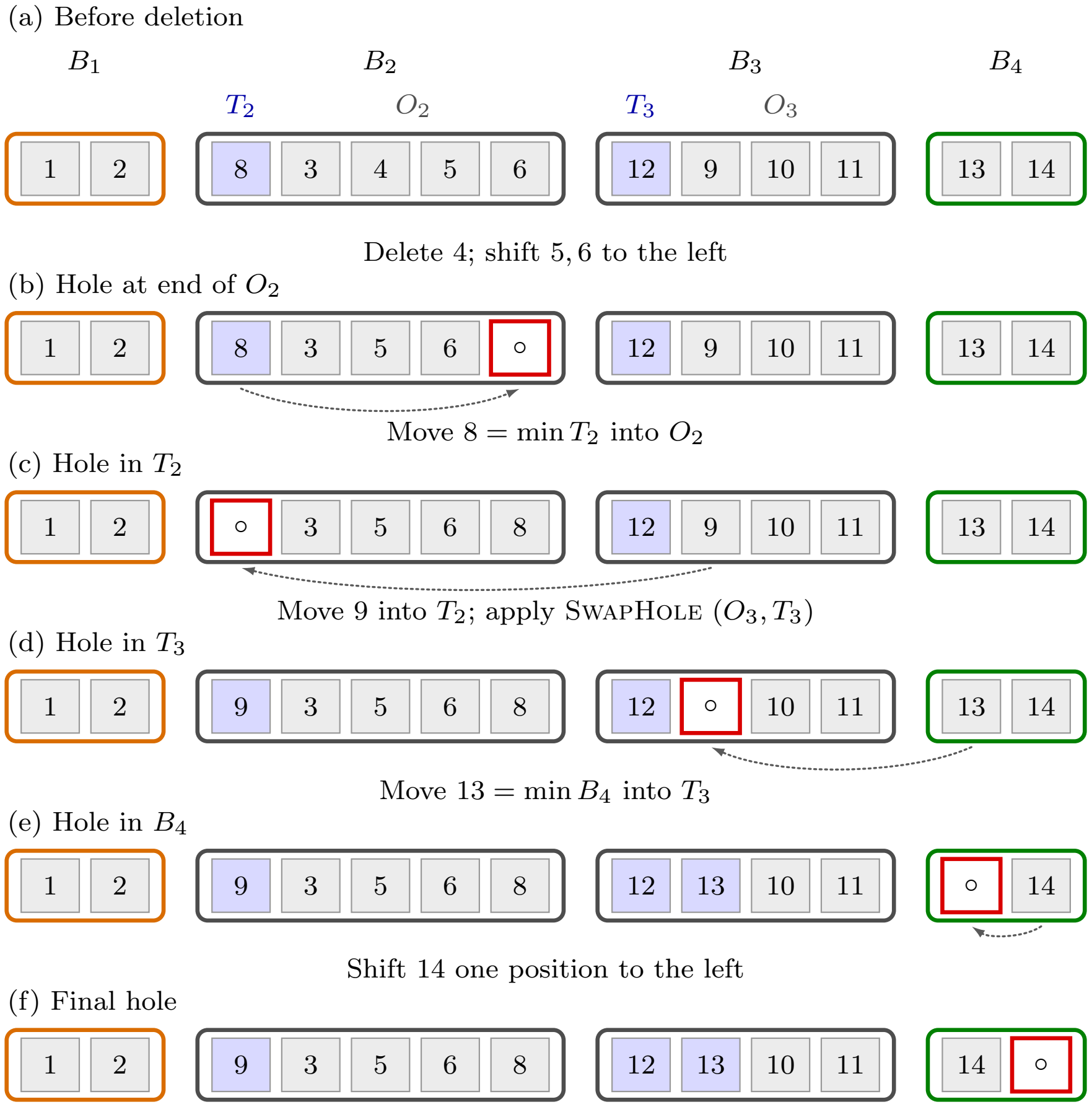

To make this more efficient, we divide each block
into a \emph{transfer section} $T$ followed by an
\emph{original section} $O$. (Initially, $T$
is empty and the block only has $O$.) Each section is 
individually sorted, but every element of $T$ is larger
than every element of $O$. 
In other words, the two sections are 
opposite to the sorted order. (For example,
see the third block in the final row of \Cref{fig:dst-example}.)
This means that the minimum element
of a block is at the beginning of $O$,
which is adjacent to the maximum element of the block,
which is at the end of $T$. 
This adjacency is crucial. Suppose the hole is 
at the end of $T$ in a block.
We fill this hole with the minimum element of the {\em next} block,
which is at the beginning of {\em its} original section.
Since the hole now lies immediately after the transfer
section of the next block, 
we can assign this empty cell to $T$
by changing the boundary between the two sections.
This boundary change moves no elements.
This shifts the hole
to the end of the transfer section
of the next block, thereby 
recreating our starting configuration in the 
next block.

Cascading these
steps across the $O(\sqrt{n})$ blocks, 
we can move a hole at the end of the transfer
section in {\em any} block
to the end of the transfer section in a later block,
with recourse proportional to the total number of blocks.
This efficient method of moving a hole
through the array, while only creating limited
disorder, lies at the heart of our construction.

Finally, to complete the picture, we need
to describe the initial steps that brings
the hole to the end of $T$ in a block, and 
the final movement that brings it to the end 
of the last block. For the two-level data 
structure, this can be done by simply shifting 
elements in the block originally containing 
the hole and in the last block, 
incurring $O(\sqrt{n})$ moves in 
each step. However, to introduce the structure used later, we describe a slightly more elaborate procedure that is closer to our eventual construction.

We make the first and last block special in 
that they only contain an original section,
and no transfer section. This ensures, for
instance, that the minimum and maximum array 
elements are always in the first and last position
respectively, a property that will be important
later in the recursive data structure. 
So, the hole is moved to the end of $T$ in 
the penultimate block, and then filled by the 
minimum element of the last block. Finally, 
the hole is moved from the beginning to the 
end of the last block by na\"ively shifting 
$O(\sqrt{n})$ elements. 

We use a more elaborate process for the initial
positioning of the hole at the end of a transfer 
block. If the hole is in an original section,
we first move it to the end of its section by 
shifting elements. Next, we fill this hole 
with the minimum element of the transfer section,
which moves the hole to the beginning of the transfer 
section. Finally, the hole is moved by shifting
elements to the end of the transfer section,
at which point the cascade of moves through the 
sequence of blocks can happen. (We also need 
some special cases when the transfer section is 
empty, or if the hole is in the first block,
but we omit these from this description because
the two-level data structure only serves an 
intuitive purpose for us.)
An illustration of how a hole moves through the array 
is shown in \Cref{fig:dst-example}. 


When a deletion occurs in one of the original blocks,
we route the resulting hole to the end of the last block
and transfer its cell to a separate sorted suffix.
New elements are inserted into this suffix, which we
maintain in sorted order. Deletions within the suffix
are handled using a simple shift to the end of the suffix. After every
$\sqrt{n}$ updates, we sort
the entire array, resetting this suffix to empty.
Note, that the last original block is initially given enough
cells to ensure that it remains nonempty until
the next time we sort the array.
 Clearly,
each insertion or deletion only incurs $O(\sqrt{n})$ 
recourse, and the re-sorts also incur $O(\sqrt{n})$
amortized recourse per step.
Importantly, we can bound the cost due to 
relaxation of sorted order.

The original blocks remain ordered by value, while
each intermediate block consists of two sorted sections
in reverse value order. Within each block, this
introduces an additional cost of at most a constant
multiple of its value range. Since the value ranges
of different original blocks are disjoint, their
total length is at most the value range of the entire
array. The sorted suffix and its boundary with the
preceding block contribute at most another constant
of the overall value range. Thus, the
sorting cost remains within a constant factor of
the cost of a sorted array.


Our goal, in \Cref{thm:dynamic} is to adopt this idea
but improve the recourse bound from $O(\sqrt{n})$
to $O(\polylog(n))$, while incurring $O({\rm polylog}(n))$
cost. A natural idea to do this is to generalize
the $2$-level structure to a $d$-level structure,
where $d = O(\log n)$. 
We organize the array into recursively defined 
sections containing either one or
three blocks, rather than using many blocks at
a level. The main complication is that whereas
a hole was moving across a sequence of blocks 
previously, it will now need to be moved through
the last block by
descending through its recursive structure.
While this still uses the high-level idea of the 
$2$-level operation, making this recursive process
efficient is technically complicated. In particular,
the resulting ordering constraints between the 
different sections at different levels of 
recursion are more delicate; in fact, the 
transfer section of the middle block need no
longer be smaller than every element of the last block.
The next two sections of the paper are devoted to describing 
a general $d$-level data structure that adopts 
some ideas from the $2$-level data structure and 
eventually establishes \Cref{thm:dynamic}.

\section{Dynamic Sorting Trees (\dst)}\label{sec:new-dst}For online sorting with fully dynamic updates, we propose
a new data structure called an
\emph{Augmented Dynamic Sorting Tree} (\adst).
The \adst\ determines how elements are moved within
the array in response to insertions and deletions.
To describe it, we first introduce a simpler
data structure called a \emph{Dynamic Sorting Tree}
(\dst), which will serve as its main building block.

A \dst\ is a recursive organization of the array
that keeps its elements approximately sorted while
supporting the routing of holes. A \emph{hole} is
an empty cell created by deleting an
element. Our goal is to move a hole to the last
cell of the \dst\ using few element movements,
rather than shifting all the elements that follow it.

We define two operations for routing holes.
The \routehole\ operation moves a hole from an
arbitrary position to the last cell of a \dst.
The \slide\ operation handles the special case
where the hole is initially in the first cell,
and does so more efficiently.
We show that these operations require
$O(d^3)$ and $O(d^2)$ element movements,
respectively, apart from the occasional
rebuilding needed to maintain the recursive
structure. The recourse incurred in rebuilding
will be accounted for in the amortized analysis
of the \adst. 

We next describe the recursive structure of
the \dst\ and the operations used to maintain it.

\subsection{Description of a $d$-Level \dst}\begin{figure}[!htbp]
\centering
\resizebox{.61\linewidth}{!}{%
\begin{tikzpicture}[
    x=1cm,y=1cm,
    every node/.style={font=\scriptsize},
    firstb/.style={draw=orange!85!black,very thick,rounded corners=2pt},
    stdb/.style={draw=black!70,very thick,rounded corners=2pt},
    lastb/.style={draw=green!50!black,very thick,rounded corners=2pt},
    tsec/.style={fill=blue!15},
    osec/.style={fill=gray!15}
]

\node at (4.75,3.85) {$d$-level \dst};

\draw[dashed,black!40] (4.70,2.10) -- (3.95,.75);
\draw[dashed,black!40] (9.50,2.10) -- (10.25,.75);

\fill[osec] (.08,2.18) rectangle (1.42,2.77);
\draw[firstb] (0,2.10) rectangle (1.50,2.85);
\node at (.75,2.475) {$O_1$};
\node[above=3pt] at (.75,2.85) {$B_1$ (first)};

\fill[tsec] (1.78,2.18) rectangle (2.78,2.77);
\fill[osec] (2.82,2.18) rectangle (4.42,2.77);
\draw[stdb] (1.70,2.10) rectangle (4.50,2.85);
\node at (2.28,2.475) {$T_2$};
\node at (3.62,2.475) {$O_2$};
\node[above=3pt] at (3.10,2.85) {$B_2$ (middle)};

\fill[osec] (4.78,2.18) rectangle (9.42,2.77);
\draw[lastb] (4.70,2.10) rectangle (9.50,2.85);
\node at (7.10,2.475) {$O_3$};
\node[above=3pt] at (7.10,2.85) {$B_3$ (last)};

\fill[osec] (4.03,.08) rectangle (5.37,.67);
\draw[firstb] (3.95,0) rectangle (5.45,.75);
\node at (4.70,.375) {$O'_1$};

\fill[tsec] (5.73,.08) rectangle (6.85,.67);
\fill[osec] (6.89,.08) rectangle (8.47,.67);
\draw[stdb] (5.65,0) rectangle (8.55,.75);
\node at (6.29,.375) {$T'_2$};
\node at (7.68,.375) {$O'_2$};

\fill[osec] (8.83,.08) rectangle (10.17,.67);
\draw[lastb] (8.75,0) rectangle (10.25,.75);
\node at (9.50,.375) {$O'_3$};

\node[below=6pt] at (7.10,0) {$O_3$ as a $(d-1)$-level \dst};

\end{tikzpicture}%
}
\caption{A three-block $d$-level \dst\ for $d\ge3$. Original sections are gray and transfer sections blue. The lower row expands $O_3$ into its three-block structure at level $d-1$.}
\label{fig:tdst-layout}
\end{figure}
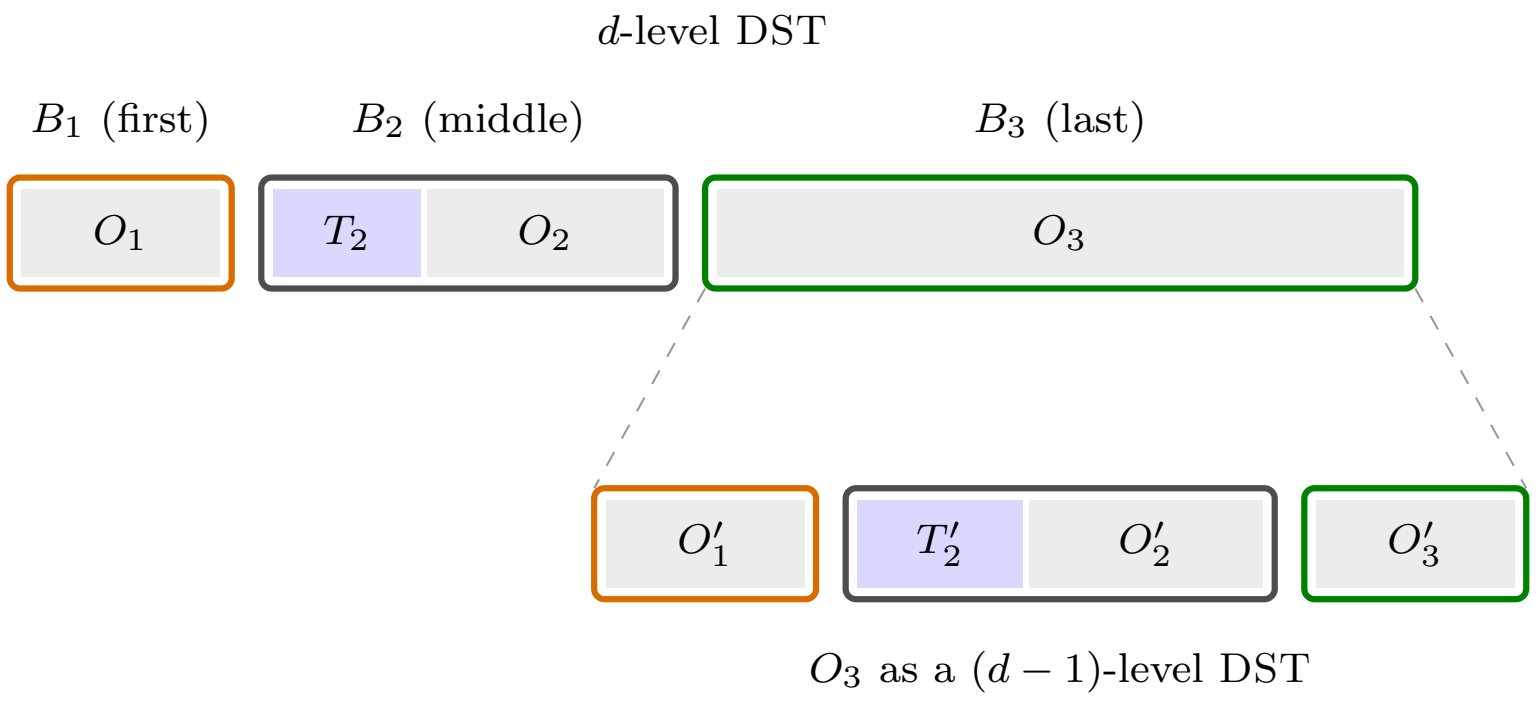

A dynamic sorting tree (\dst) is a recursively
partitioned array designed to support the routing
of holes while keeping its elements approximately
sorted. We define its structure recursively:
for $i>1$, an $i$-level \dst\ is partitioned
into contiguous blocks, each consisting of
sections that are themselves $(i-1)$-level \dsts.
The recursion terminates at level $1$,
where a \dst\ is simply a sorted array.

\paragraph{Blocks and sections.}
For $d\ge2$, a $d$-level \dst\ consists of either one or three consecutive level-$(d-1)$ \emph{blocks}. Every block contains an \emph{original section}, and a middle block also contains a \emph{transfer section}, which may be empty. Each nonempty section is itself a $(d-1)$-level \dst. Elements are called \emph{original} or \emph{transfer} according to the section in which they reside.
If the \dst\ consists of a single block, that block contains only an original section, and its size may grow or shrink over time. Otherwise, the three blocks are called the \emph{first}, \emph{middle}, and \emph{last} blocks, in left-to-right order.

\begin{enumerate}
    \item \textbf{Middle block.} The transfer section precedes the original section and is empty when the block is built. Between rebuilds, the original section can only shrink and the transfer section can only grow. Their combined size, and hence the size of the block, remains fixed.

    \item \textbf{First and last blocks.} Each consists only of an original section and has no transfer section. The sizes of these blocks may grow or shrink over time.
\end{enumerate}

Thus, a three-block section has the form $O_1\mid T_2\,O_2\mid O_3$, as illustrated in \cref{fig:tdst-layout}.

\paragraph{Holes, levels, and the tree view.}
A \dst\ contains no holes in its steady state,
but may temporarily contain a single hole while
executing a hole-routing operation.

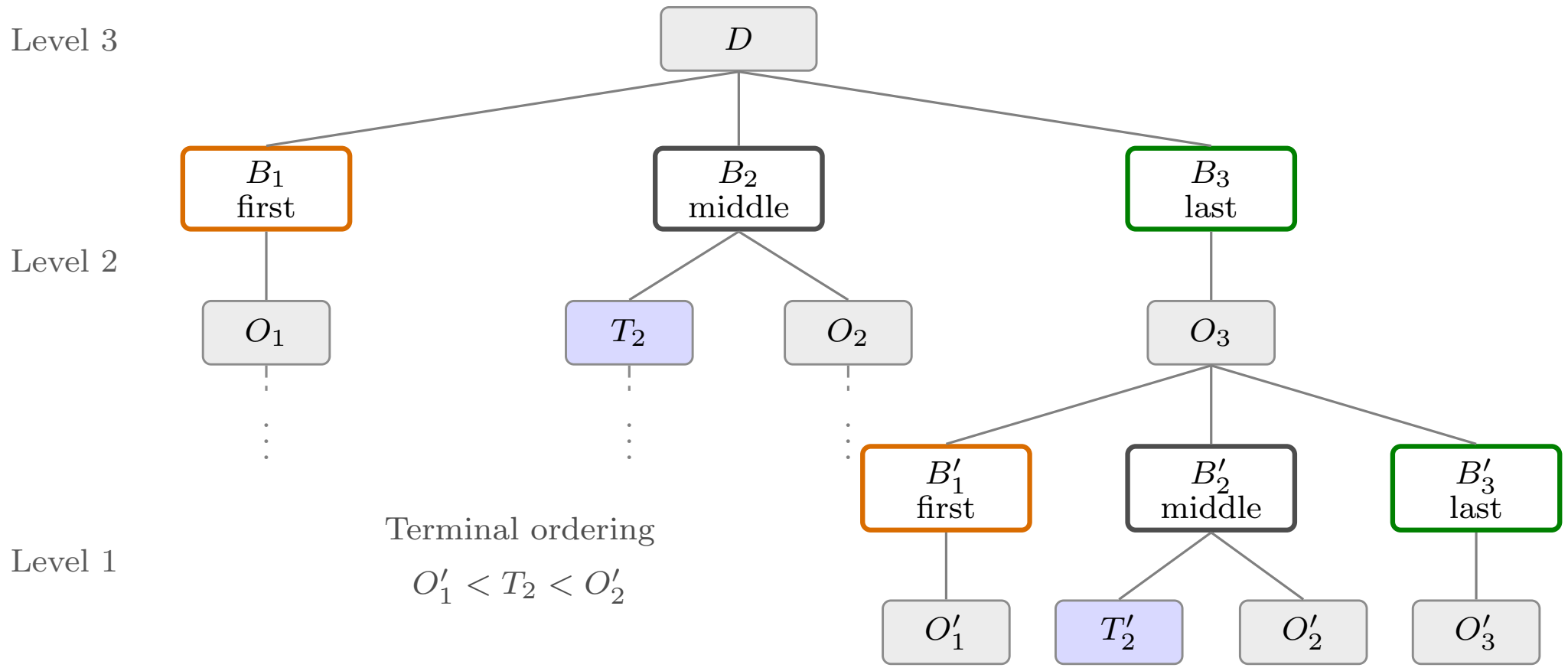
\begin{figure}[t]
\centering
\resizebox{.82\linewidth}{!}{%
\begin{tikzpicture}[
    x=1cm,y=1cm,
    every node/.style={font=\scriptsize},
    block/.style={
        draw=black!70,very thick,rounded corners=2pt,
        fill=white,minimum width=1.45cm,minimum height=.70cm,
        align=center,inner sep=3pt
    },
    firstb/.style={block,draw=orange!85!black},
    stdb/.style={block,draw=black!70},
    lastb/.style={block,draw=green!50!black},
    section/.style={
        draw=black!45,line width=.55pt,rounded corners=2pt,
        minimum width=1.10cm,minimum height=.55cm,inner sep=2pt
    },
    orig/.style={section,fill=gray!15},
    trans/.style={section,fill=blue!15},
    edge/.style={draw=black!50,line width=.6pt},
    omitted/.style={draw=black!45,densely dashed,line width=.6pt},
    leveltext/.style={anchor=east,text=black!65}
]

\node[orig,minimum width=1.35cm] (D) at (.15,0) {$D$};

\node[firstb] (B1) at (-3.95,-1.30) {$B_1$\\[-1pt]first};
\node[stdb]   (B2) at (.15,-1.30) {$B_2$\\[-1pt]middle};
\node[lastb]  (B3) at (4.25,-1.30) {$B_3$\\[-1pt]last};

\node[orig]  (O1) at (-3.95,-2.55) {$O_1$};
\node[trans] (T2) at (-.80,-2.55) {$T_2$};
\node[orig]  (O2) at (1.10,-2.55) {$O_2$};
\node[orig]  (O3) at (4.25,-2.55) {$O_3$};

\foreach \c in {B1,B2,B3} {
    \draw[edge] (D.south) -- (\c.north);
}
\draw[edge] (B1.south) -- (O1.north);
\draw[edge] (B2.south) -- (T2.north);
\draw[edge] (B2.south) -- (O2.north);
\draw[edge] (B3.south) -- (O3.north);

\node[firstb] (C1) at (1.95,-3.90) {$B'_1$\\[-1pt]first};
\node[stdb]   (C2) at (4.25,-3.90) {$B'_2$\\[-1pt]middle};
\node[lastb]  (C3) at (6.55,-3.90) {$B'_3$\\[-1pt]last};

\node[orig]  (P1) at (1.95,-5.15) {$O'_1$};
\node[trans] (Q2) at (3.45,-5.15) {$T'_2$};
\node[orig]  (P2) at (5.05,-5.15) {$O'_2$};
\node[orig]  (P3) at (6.55,-5.15) {$O'_3$};

\foreach \c in {C1,C2,C3} {
    \draw[edge] (O3.south) -- (\c.north);
}
\draw[edge] (C1.south) -- (P1.north);
\draw[edge] (C2.south) -- (Q2.north);
\draw[edge] (C2.south) -- (P2.north);
\draw[edge] (C3.south) -- (P3.north);

\foreach \s in {O1,T2,O2} {
    \draw[omitted] (\s.south) -- ++(0,-.22);
    \node[text=black!50,anchor=north,inner sep=1pt]
        at ($ (\s.south)+(0,-.25) $) {$\vdots$};
}

\node[leveltext] at (-5.10,0) {Level $3$};
\node[leveltext] at (-5.10,-1.925) {Level $2$};
\node[leveltext] at (-5.10,-4.525) {Level $1$};

\node[align=center,text=black!70] at (-1.75,-4.525)
    {Terminal ordering\\[4pt]$O'_1<T_2<O'_2$};

\end{tikzpicture}%
}
\caption{Tree view of a three-level \dst, with the subtree rooted at $O_3$ expanded. Original sections are gray and transfer sections blue. The ordering $O'_1<T_2<O'_2$ relates sections at different levels. Other subtrees are omitted.}
\label{fig:tree_DST}
\end{figure}

We view an $i$-level \dst\ as a level-$i$ section.
Unrolling the recursive decomposition gives a rooted
tree whose nodes alternate between sections and blocks.
The root is the level-$d$ section corresponding to the
entire \dst\ (see \cref{fig:tree_DST}).

For $i\ge2$, each nonempty level-$i$ section has
either one or three level-$(i-1)$ blocks as children,
ordered from left to right. Each level-$i$ block,
in turn, has its constituent level-$i$ sections as
children. A middle block has two children: its transfer
section followed by its original section. The transfer
child is retained even when it is empty. A first,
last, or singleton block has a single child,
corresponding to its original section.
The level-$1$ sections are sorted arrays. Arranged
from left to right, these arrays constitute the
underlying array.
We use this tree to describe relationships between
blocks and sections. The parent of a level-$i$ block
is the level-$(i+1)$ section containing it, while the
parent of a nonroot level-$i$ section is the level-$i$
block containing it. Two nodes with the same parent
are called \emph{siblings}, and their order is
inherited from the underlying array.

\paragraph{Roadmap for the remainder of the section.}
Having described the recursive structure of a \dst,
we next present its structural invariants in
\Cref{subsec:dst-invariants} and use them to establish
the cost bound in \Cref{cl:dst-cost}.

In \Cref{subsec:operations}, we introduce the
operations supported by a \dst, specifying their
behavior and dependencies without yet describing
their implementations. We then describe the basic
operations \addmax, \rebuild, and \swaphole\ in
\Cref{subsec:dst-basic-operations}.

Finally, in \Cref{subsec:dst-moving}, we give the
implementations of the hole-routing operations,
starting with \downhole\ and \simpleslide, followed
by \slide\ and \routehole. We establish their
correctness and recourse bounds alongside their
implementations. The recourse incurred by \rebuild\
is accounted for separately in the amortized
analysis of the \adst\ in
\Cref{sec:fully-dynamic-lowcost}.

\subsection{The Structural Invariants}
\label{subsec:dst-invariants}

The \dst\ satisfies the following structural invariants between updates. We assume without loss of generality that all element values are distinct, using infinitesimal perturbations to break ties.

For two sets of elements $A$ and $B$, we write $A<B$ if every element of $A$ is smaller than every element of $B$. We use the same notation for sections, comparing the elements they contain. Comparisons involving an empty set are vacuous.

\begin{enumerate}
    \item \textbf{Number and Structure of Blocks.}
    Every level-$i$ section with $i\ge2$ consists of either one or three level-$(i-1)$ blocks. A single block contains only an original section, denoted by $O$.

    In a three-block section, the blocks are, from left to right, the first, middle, and last blocks. The first and last blocks contain only original sections, while the middle block contains a transfer section followed by an original section. Thus, a three-block section has the form $O_1\mid T_2\,O_2\mid O_3$. The transfer section may be empty, but the original section of every middle block contains at least one element.

    \item \textbf{Base-Level Sorting.}
    Every level-$1$ section is sorted in increasing order from left to right.

    \item \textbf{Block Sizes.}
    For a section consisting of level-$i$ blocks, the size bounds are as follows:
    \begin{itemize}
        \item If the section consists of a single block, its size is at most $12\cdot 2^i$.
        \item If the section consists of three blocks, the middle block has size exactly $2^i$, the first block has size in $[2^i,12\cdot 2^i)$, and the last block has size in $[7\cdot 2^i,12\cdot 2^i)$.
    \end{itemize}

    These bounds imply that, in a three-block section, the original section of the last block also consists of three blocks unless it is at level~$1$. Indeed, a last level-$i$ block with $i>1$ owns at least $7\cdot 2^i$ cells. If its original section consisted of a single level-$(i-1)$ block, it would own at most
    \[
        12\cdot 2^{i-1}=6\cdot 2^i<7\cdot 2^i
    \]
    cells, a contradiction.

    \item \textbf{Block Ordering.}
    Consider a three-block section $O_1\mid T_2 \mid O_2\mid O_3$. We require $O_1<O_2<T_2$ and $O_2<O_3$.

    If $O_3$ is a level-$1$ section, we additionally require $T_2<O_3$. Otherwise, the block-size bounds ensure that $O_3$ consists of three blocks. Writing $O_3=O'_1\mid T'_2\,O'_2\mid O'_3$, we require $O'_1<T_2<O'_2$. Since $O'_2$ is nonempty, the ordering conditions within $O_3$ also give
    \[
        O'_1<T_2<(T'_2\cup O'_2\cup O'_3).
    \]
\end{enumerate}




The ordering invariants give the following property of the minimum element.

\begin{claim}
\label{cl:new-dst-minimum}
In every nonempty $i$-level \dst, the minimum element occupies the leftmost cell.
\end{claim}

They also give the corresponding property for the maximum element.

\begin{claim}
\label{cl:new-dst-maximum}
In every nonempty $i$-level \dst, the maximum element occupies the rightmost cell.
\end{claim}

\paragraph{Sorting-cost bound.}
The ordering invariants also bound the sorting cost of a \dst.
\begin{lemma}
\label{cl:dst-cost}
Let $S$ be an $i$-level \dst. Then, for every
threshold $z$, we have $\operatorname{cross}_z(S)\le 4i$.
Consequently, if all elements of $S$ lie in an
interval $I$, then we have
    $\operatorname{cost}(S)\le 4i\cdot |I|$.
\end{lemma}

\subsection{Overview of Operations and Inter-dependencies}\label{subsec:operations}The structural invariants guarantee the sorting-cost bound. It remains
to implement the operations while preserving these invariants.
We first specify the interface that a \dst\ provides to the surrounding
\adst\ (Section~\ref{sec:fully-dynamic-lowcost}), and then describe
the implementation.
\paragraph{Interface.}

A \dst\ supports two operations that are invoked by the surrounding \adst:

\begin{itemize}

    \item \textbf{\routehole}:
    Given a \dst\ with a hole in an arbitrary
    position, this operation moves the hole to its last position. An \adst\
    uses this operation to move the hole to the end of a \dst where a deletion occurs in.

    \item \textbf{\slide}:
    Given a \dst\ with a hole in its first
    position, this operation moves the hole to its last position. An \adst\
    uses this operation to pass a hole through a sequence of \dsts.

\end{itemize}

Note that \routehole is more general than \slide, which 
only moves a hole from the beginning of a \dst to its end.
Indeed, \routehole will use \slide in its implementation.

\paragraph{Auxiliary Operations and Inter-dependencies.}
The two main operations that move a hole within a \dst,
\slide and \routehole, use two auxiliary operations 
internal to a \dst, which we describe next.

First, we distinguish a special case of \slide that is
used internally by \slide, and is more efficient than 
a generic \slide operation:
\begin{itemize}    
    \item \simpleslide: Like \slide, this operation also moves a 
    hole from the first to the last position of a \dst, but it only 
    applies when the \dst is not the unique original section of a
    last block. 
\end{itemize}
Finally, all the previous operations need to move a hole
within the unique original section of the last block.
 
We describe this next:
\begin{itemize}    
    \item \downhole: This operation moves a hole from the
    end of the transfer section in the middle block of a \dst
    all the way to the end of the \dst by descending
    through the recursive structure of the last block.
\end{itemize}



\begin{figure}[t]
\centering
\resizebox{.72\linewidth}{!}{%
\begin{tikzpicture}[
    x=1cm,y=1cm,
    every node/.style={font=\scriptsize},
    op/.style={
        draw=orange!75!black,very thick,rounded corners=2pt,
        fill=none,minimum width=2cm,minimum height=.60cm,
        inner xsep=5pt,inner ysep=3pt,align=center
    },
    entry/.style={
        op,draw=green!45!black
    },
    arrow/.style={
        -{Latex[length=1.8mm,width=1.2mm]},
        draw=black!65,line width=.65pt
    }
]

\node[entry] (route) at (0,0) {\routehole};
\node[entry,right=6mm of route] (slide) {\slide};
\node[op,minimum width=2.8cm,right=6mm of slide] (simple) {\simpleslide};
\node[op,right=6mm of simple] (down) {\downhole};

\draw[arrow] (route.east) -- (slide.west);
\draw[arrow] (slide.east) -- (simple.west);
\draw[arrow] (simple.east) -- (down.west);

\draw[arrow] (slide.north)
    -- ++(0,.70) -| (down.north);

\draw[arrow] ([xshift=4pt]route.south)
    -- ++(0,-.50) -| (simple.south);
\draw[arrow] ([xshift=-4pt]route.south)
    -- ++(0,-.95) -| (down.south);

\end{tikzpicture}%
}
\caption{The interface operations of a \dst\ and their auxiliary dependencies. The interface operations \routehole\ and \slide\ have green borders. Recursive calls to the same operation are not shown.}
\label{fig:dst-call-structure}
\end{figure}
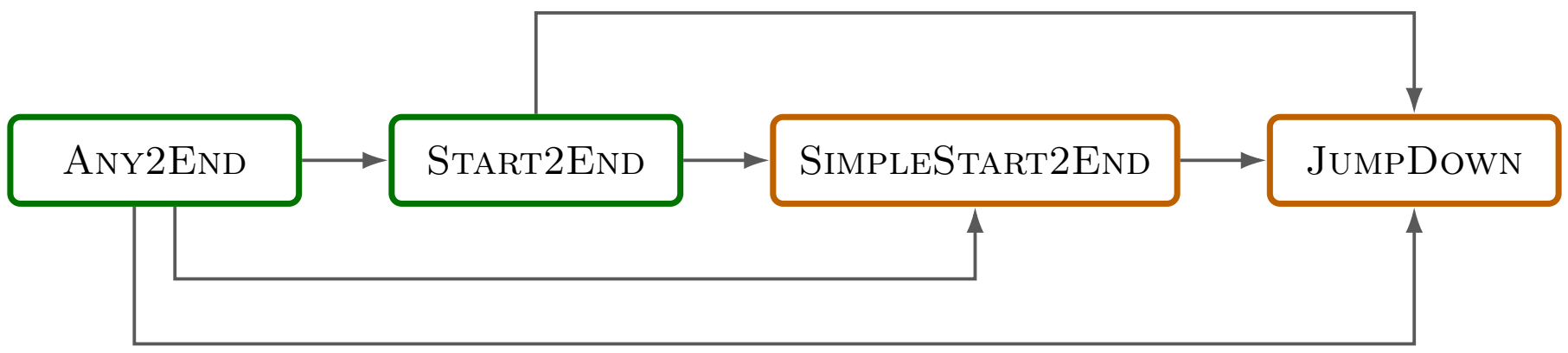

In \Cref{fig:dst-call-structure}, we show the structure of auxiliary calls between these operations. Each arrow indicates that the operation at the head calls the operation at the tail of the arrow. We note that there are no cyclic sequence of calls, which means that in order to ensure that each operation terminates, we need to only ensure that the recursion within each operation is well-defined.

For each operation, all {\em recursive calls} issued for a level-$i$ \dst are to level-$j$ \dsts for $j < i$. Note that this is not to say that all auxiliary operations called within an operation are to \dsts at a lower level. In fact, several operations will call other operations at the same recursive level. But, this does not affect finite termination because each call either moves forward on the DAG in \Cref{fig:dst-call-structure} or reduces the level of the \dst. Eventually, the only base cases are for level-$1$ \dsts, which are sorted arrays.

Given the somewhat complicated structure of the recursion, we sometimes explicitly mention the level of the \dst on which an operation is being called. While the input to the operation is simply a \dst, explicitly writing the level makes it clear to the reader that the recursion satisfies the condition of decreasing the level in each recursive call.

\subsection{Basic Operations and Rebuilding}\label{subsec:dst-basic-operations}

We start our detailed descriptions of the implementation of the \dst operations 
by first describing some basic operations that will be used by all the operations
specified above that actually move a hole within a \dst.

\paragraph{The \addmax operation.}

Given a \dst\ whose last position is a hole,
and an element $y$ larger than every element currently
stored in the \dst, this operation places $y$ in the hole.


\begin{claim}
\label{cl:add-max-invariants}
The \addmax operation maintains all structural
invariants of the \dst.
\end{claim}

The operation moves only $y$ into the hole. All other elements remain in place.

\begin{claim}
\label{cl:add-max-recourse}
The recourse of \addmax\ is $1$.
\end{claim}

\paragraph{The \rebuild operation.}

Moving a hole across the boundary between two
adjacent \dsts\ changes their sizes. Such changes
may violate the Block Sizes invariant of
\Cref{subsec:dst-invariants}. To repair these
violations, we introduce the \rebuild\ operation,
which reorganizes the affected part of the structure.

Given a \dst $D$ requiring rebuilding, \rebuild\
selects a \dst to reconstruct according to the rules
below. This may be a larger enclosing \dst when
necessary to preserve the ordering invariants.
The operation sorts the stored elements and reconstructs
the recursive structure, with all transfer sections empty.
If the selected \dst contains a hole, it is placed
at the end of the \dst. Rebuilding preserves the total
number of cells, and the hole counts toward the block
sizes just like any occupied cell.
The \rebuild\ operation is defined as follows.

\begin{itemize}
    \item For a level-$1$ \dst, we simply sort its
    stored elements.

    \item For a level-$(i+1)$ DST $D$, where $i\ge1$, we first
    consider the sizes of its constituent blocks. If they
    satisfy the Block Sizes invariant, we preserve the number
    and sizes of these blocks, but not necessarily their
    contents. Otherwise, we choose a new block division as
    follows: if $|D|\le11\cdot2^i$, we create a single block.
    Otherwise, we create three blocks, where the middle
    contains exactly $2^i$ cells, the last contains
    \[
    \max\left\{
        8\cdot2^i,
        \left\lceil\frac{|D|-2^i}{2}\right\rceil
    \right\}
    \]
    cells, and the first contains the remaining cells.
    
    In either case, we sort all the elements of $D$ and
    distribute them among the original sections of the blocks
    in increasing order, according to the designated block
    sizes. The transfer section of the middle block, if
    present, is made empty. If $D$ contains a hole, we treat
    it as the largest element, so that it occupies the last
    cell of $D$. As before, the hole counts toward block sizes.
    
    We then apply \rebuild\ recursively to each original
    section, rebuilding its constituent DST all the way down
    to level $1$. This recursion is performed regardless of
    whether the block division was changed.

\end{itemize}

\begin{claim}
\label{cl:rebuild-invariants}
The \rebuild operation maintains all structural
invariants of the \dst.
\end{claim}

Rebuilding moves elements only within the selected \dst, which contains $O(2^i)$ cells at level $i$.

\begin{claim}
\label{cl:rebuild-recourse}
The worst-case recourse of \rebuild\ on a
level-$i$ \dst is $O(2^i)$.
\end{claim}

\paragraph{The \swaphole operation.}

To move a hole through the recursive structure of
a \dst, we sometimes need to transfer an empty cell
between adjacent sections. In particular, this
occurs between the transfer and original sections
of a middle block, which are represented by
adjacent \dsts.

The \swaphole\ operation performs this transfer
by changing which section owns the empty cell,
without moving any stored element.
However, changing the sizes of the sections may
violate the structural invariants, which we restore
using \rebuild.

The operation is implemented as follows.

\begin{enumerate}
    \item Let $D$ and $D'$ denote the \dsts supplying
    and receiving the hole, respectively.
    Suppose $D$ is immediately to the left of $D'$,
    and its last cell $h$ contains the hole.
    We remove $h$ from $D$ and make it the first
    cell of $D'$.

    This is done recursively along the chains of
    last original sections of $D$ and first original
    sections of $D'$, down to level $1$.
    A transfer in the opposite direction is symmetric.

    \item After transferring the hole, we examine
    the affected sections and their ancestors,
    from highest level to lowest.
    A section requires repair if a Block Size
    invariant is violated or if it has three blocks
    and its middle original section becomes empty.

    Let $S$ be the highest-level section requiring
    repair. We apply \rebuild\ to $S$, unless $S$
    is the last original section of a three-block
    parent $P$, in which case we rebuild $P$ instead.
    This empties the transfer section of $P$,
    removing the ordering constraint that rebuilding
    $S$ might violate. Since $P$ was examined before
    $S$, its block sizes are valid and are preserved
    by \rebuild.

    We repeat until all affected sections satisfy
    the structural invariants.
\end{enumerate}

\begin{claim}
\label{cl:swap-hole-invariants}
The \swaphole\ operation maintains all structural
invariants of the \dsts.
\end{claim}

\subsection{Implementation Details of Operations that Move a Hole}\label{subsec:dst-moving}

Now, we give detailed implementations of the operations that move 
a hole within a \dst. These operations use the previously 
defined basic operations, \addmax and \swaphole, along with
calling each other according to \Cref{fig:dst-call-structure}
(and recursively calling themselves at lower levels). To keep this description 
organized, we first describe the two simpler operations, 
\downhole and \simpleslide, and then the more complex ones,
\slide and \routehole.

\begin{remark}
    Several of the operations we describe below will call \swaphole,
    and therefore might trigger \rebuild. If this happens,
    \rebuild may change the recursive decomposition and
    move the hole to the end of the \dst it was called from, 
    or even to the end of a higher-level \dst containing it.
    In this case, all the operation calls for which the hole
    has already moved to the end of its \dst are truncated
    since they do not need to execute their remaining steps.
    (Note that the purpose of all the operations moving a hole 
    are to bring the hole at the end of the \dst they are called from.)
\end{remark}

\subsubsection{The Simpler Operations: \downhole and \simpleslide}

We begin with the two operations for moving holes that
are already at the boundaries of recursive sections.
The main idea is to move a hole through a section
by transferring only a few elements, rather than
shifting the entire section.

\paragraph{The \downhole\ operation.}
Consider a three-block section
    $S=O_1\mid T_2\mid O_2\mid O_3$.
Suppose the hole is at the end of $T_2$.
    We would like to fill it with an element from
    $O_3$, but $\min O_3$ may be smaller than the
    elements of $T_2$.
    Instead, consider the three-block structure
    of $O_3$, given by
        $O_3=O'_1\mid T'_2\mid O'_2\mid O'_3$.
    The ordering invariant $O'_1<T_2<O'_2$
    allows us to move $\min O'_2$ into the hole.
    We then use \swaphole\ to move the resulting
    hole to the end of $T'_2$, without moving
    any additional elements.
    This brings the hole to the analogous position
    of its starting location, but in a \dst at a lower
    level. So, we can now recurse on $T'_2$.

Formally, the \downhole operation is implemented recursively as follows:

\begin{enumerate}
    \item 
    If $O_3$ is a level-$1$ \dst, move
    $\min O_3$ into the hole in $T_2$ using \addmax,
    and shift the elements of $O_3$ one
    position to the left to move the hole to
    the end of $O_3$.

    \item Otherwise, we perform the following steps:
    \begin{enumerate}
        \item Move $\min O'_2$ into the hole in
        $T_2$ using \addmax.

        \item Apply \swaphole$(O'_2,T'_2)$ from
        right to left, transferring the hole to
        the end of $T'_2$.

        \item Invoke \downhole$(i-1)$ on $O_3$,
        moving the hole to the end of $S$.
    \end{enumerate}
\end{enumerate}

As in all operations, if rebuilding
has already moved the hole to the end of $S$ or an
enclosing section, the corresponding unfinished
operations terminate.

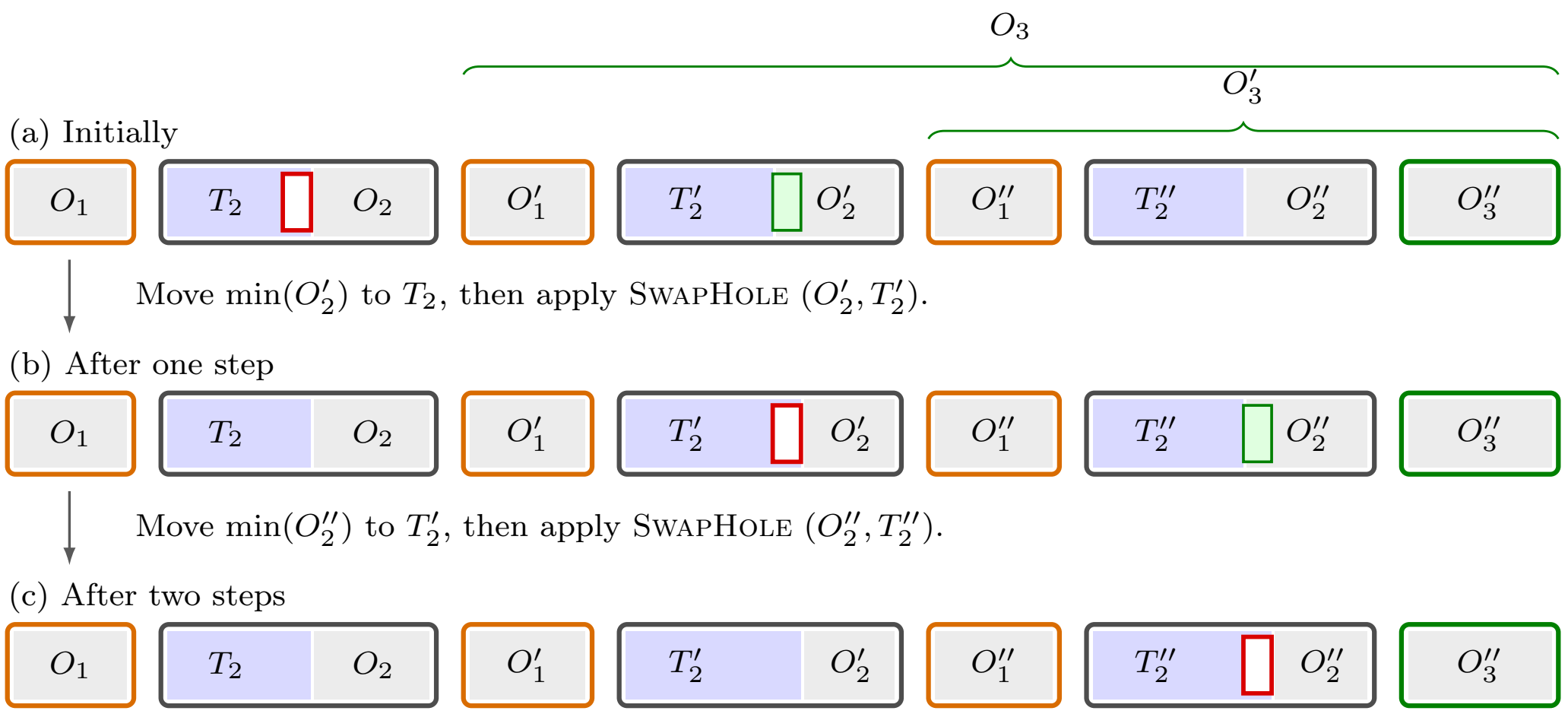
\begin{figure}[t]
\centering
\resizebox{.82\linewidth}{!}{%
\begin{tikzpicture}[
    x=1cm,y=1cm,
    every node/.style={font=\scriptsize},
    firstb/.style={draw=orange!85!black,very thick,rounded corners=2pt},
    stdb/.style={draw=black!70,very thick,rounded corners=2pt},
    lastb/.style={draw=green!50!black,very thick,rounded corners=2pt},
    osec/.style={fill=gray!15},
    tsec/.style={fill=blue!15},
    hole/.style={draw=red!85!black,fill=white,line width=1.1pt},
    donor/.style={draw=green!50!black,fill=green!12,line width=.7pt},
    move/.style={-{Latex[length=1.6mm,width=1.1mm]},
        draw=black!65,line width=.65pt},
    brc/.style={draw=green!50!black,line width=.55pt,decorate,
        decoration={brace,amplitude=4pt}},
    state/.style={anchor=west,inner sep=0pt}
]

\newcommand{\dstDownState}[6]{%
    \fill[osec] (0.06,{#1+.06}) rectangle (1.06,{#1+.66});
    \draw[firstb] (0,#1) rectangle (1.12,{#1+.72});

    \fill[tsec] (1.42,{#1+.06}) rectangle (2.69,{#1+.66});
    \fill[osec] (2.72,{#1+.06}) rectangle (3.74,{#1+.66});
    \draw[stdb] (1.36,#1) rectangle (3.80,{#1+.72});

    \fill[osec] (4.10,{#1+.06}) rectangle (5.12,{#1+.66});
    \draw[firstb] (4.04,#1) rectangle (5.18,{#1+.72});

    \fill[tsec] (5.48,{#1+.06}) rectangle (#3,{#1+.66});
    \fill[osec] ({#3+.03},{#1+.06}) rectangle (7.86,{#1+.66});
    \draw[stdb] (5.42,#1) rectangle (7.92,{#1+.72});

    \fill[osec] (8.22,{#1+.06}) rectangle (9.26,{#1+.66});
    \draw[firstb] (8.16,#1) rectangle (9.32,{#1+.72});

    \fill[tsec] (9.62,{#1+.06}) rectangle (#4,{#1+.66});
    \fill[osec] ({#4+.03},{#1+.06}) rectangle (12.05,{#1+.66});
    \draw[stdb] (9.56,#1) rectangle (12.11,{#1+.72});

    \fill[osec] (12.41,{#1+.06}) rectangle (13.68,{#1+.66});
    \draw[lastb] (12.35,#1) rectangle (13.74,{#1+.72});

    \node at (.56,{#1+.36}) {$O_1$};
    \node at (1.94,{#1+.36}) {$T_2$};
    \node at (3.245,{#1+.36}) {$O_2$};
    \node at (4.61,{#1+.36}) {$O'_1$};
    \node at (6.02,{#1+.36}) {$T'_2$};
    \node at ({(#3+7.92)/2},{#1+.36}) {$O'_2$};
    \node at (8.74,{#1+.36}) {$O''_1$};
    \node at (10.18,{#1+.36}) {$T''_2$};
    \node at ({(#4+12.11)/2},{#1+.36}) {$O''_2$};
    \node at (13.045,{#1+.36}) {$O''_3$};

    \draw[hole] (#2,{#1+.11})
        rectangle ({#2+.25},{#1+.61});
    \ifnum#6=1
        \draw[donor] (#5,{#1+.11})
            rectangle ({#5+.25},{#1+.61});
    \fi
}

\dstDownState{4.10}{2.44}{6.78}{10.95}{6.78}{1}
\dstDownState{2.05}{6.78}{7.03}{10.95}{10.95}{1}
\dstDownState{0.00}{10.95}{7.03}{11.20}{0}{0}

\draw[brc] (4.04,5.59) -- (13.74,5.59)
    node[midway,above=5pt,text=black] {$O_3$};
\draw[brc] (8.16,5.02) -- (13.74,5.02)
    node[midway,above=5pt,text=black] {$O'_3$};

\node[state] at (0,5.06) {(a) Initially};
\node[state] at (0,3.01) {(b) After one step};
\node[state] at (0,.96) {(c) After two steps};

\draw[move] (.55,3.95) -- (.55,3.29);
\node[anchor=west] at (1.00,3.62)
    {Move $\min(O'_2)$ to $T_2$, then apply \swaphole$(O'_2,T'_2)$.};

\draw[move] (.55,1.90) -- (.55,1.24);
\node[anchor=west] at (1.00,1.57)
    {Move $\min(O''_2)$ to $T'_2$, then apply \swaphole$(O''_2,T''_2)$.};

\end{tikzpicture}%
}
\caption{Two steps of \downhole$(i)$ through the expanded last sections $O_3$ and $O'_3$. Red outlines mark holes and green cells mark the next elements to move. After each transfer, \swaphole\ assigns the vacated cell to the preceding transfer section. The hole passes from $T_2$ to $T'_2$ and then to $T''_2$.}
\label{fig:downhole}
\end{figure}

\begin{claim}
\label{cl:down-hole-invariants}
The \downhole\ operation preserves all structural invariants
and moves the hole to the last position of the \dst.
\end{claim}

The hole descends through at most $i$ levels, with $O(1)$ recourse per level apart from rebuilding.

\begin{claim}
\label{cl:down-hole-recourse}
Excluding recourse incurred by any \rebuild calls, 
\downhole$(i)$ incurs $O(i)$ recourse.
\end{claim}

\begin{proof}
We prove the claim by induction on $i$.
The base case shifts elements in a level-$2$ \dst, since $O_3$ is a level-$1$ \dst.
Since a level-$2$ \dst has $O(1)$ elements this incurs $O(1)$ recourse.
In the inductive case, we move one element from $O'_2$ into 
$T_2$ using \addmax (recourse of $1$ by \Cref{cl:add-max-recourse})
and then
apply \swaphole which only incurs recourse if it
calls \rebuild; the latter recourse
is explicitly omitted from this claim.
Finally, we recursively call \downhole$(i-1)$.
Thus, excluding \rebuild, the recourse satisfies
$R(i)\le R(i-1)+O(1)$,
    where $R(1)=O(1)$. The claim follows.
\end{proof}

\paragraph{The \simpleslide\ operation.}
    To move a hole from the beginning to the end
    of $S$, we first move it recursively through
    $O_1$. We then fill it with $\min O_2$,
    leaving a hole at the beginning of $O_2$.
    Using \swaphole, we move this hole to the
    end of $T_2$ without moving any elements.
    We can now apply \downhole\ to complete
    the movement to the end of $S$.

Formally, the operation is implemented recursively as follows.

\begin{figure}[t]
\centering
\resizebox{.45\linewidth}{!}{%
\begin{tikzpicture}[
    x=1cm,y=1cm,
    every node/.style={font=\scriptsize},
    firstb/.style={draw=orange!85!black,very thick,rounded corners=2pt},
    stdb/.style={draw=black!70,very thick,rounded corners=2pt},
    lastb/.style={draw=green!50!black,very thick,rounded corners=2pt},
    osec/.style={fill=gray!15},
    tsec/.style={fill=blue!15},
    hole/.style={draw=red!85!black,fill=white,line width=1.1pt},
    operation/.style={-{Latex[length=1.5mm,width=1mm]},
        draw=black!65,line width=.65pt}
]

\newcommand{\figsixstate}[3]{%
    \fill[osec] (.08,{#1+.08}) rectangle (1.42,{#1+.52});
    \fill[tsec] (1.78,{#1+.08}) rectangle (#3,{#1+.52});
    \fill[osec] (#3,{#1+.08}) rectangle (4.82,{#1+.52});
    \fill[osec] (5.18,{#1+.08}) rectangle (7.42,{#1+.52});

    \draw[firstb] (0,#1) rectangle (1.50,{#1+.60});
    \draw[stdb] (1.70,#1) rectangle (4.90,{#1+.60});
    \draw[lastb] (5.10,#1) rectangle (7.50,{#1+.60});

    \node at (.75,{#1+.30}) {$O_1$};
    \node at ({(1.78+#3)/2},{#1+.30}) {$T_2$};
    \node at ({(#3+4.82)/2},{#1+.30}) {$O_2$};
    \node at (6.30,{#1+.30}) {$O_3$};

    \draw[hole] (#2,{#1+.08})
        rectangle ({#2+.26},{#1+.52});
}

\figsixstate{6.40}{0.08}{3.30}
\figsixstate{4.80}{1.16}{3.30}
\figsixstate{3.20}{3.30}{3.30}
\figsixstate{1.60}{3.30}{3.56}
\figsixstate{0.00}{7.16}{3.56}

\draw[operation] (.65,6.23) -- (.65,5.57);
\node[anchor=west] at (1.10,5.90)
    {\simpleslide$(i-1)$ on $O_1$};

\draw[operation] (.65,4.63) -- (.65,3.97);
\node[anchor=west] at (1.10,4.30)
    {\addmax: $\min(O_2)\to O_1$};

\draw[operation] (.65,3.03) -- (.65,2.37);
\node[anchor=west] at (1.10,2.70)
    {\swaphole$(O_2,T_2)$};

\draw[operation] (.65,1.43) -- (.65,.77);
\node[anchor=west] at (1.10,1.10)
    {\downhole$(i)$ on $S$};

\end{tikzpicture}%
}
\caption{\simpleslide$(i)$ on $S=O_1\mid T_2\,O_2\mid O_3$, without rebuilding. After the hole reaches the end of $O_1$, \addmax\ fills it with $\min(O_2)$. The vacated cell becomes part of $T_2$ under \swaphole, and \downhole\ moves the hole to the end of $S$.}
\label{fig:simpleslide}
\end{figure}
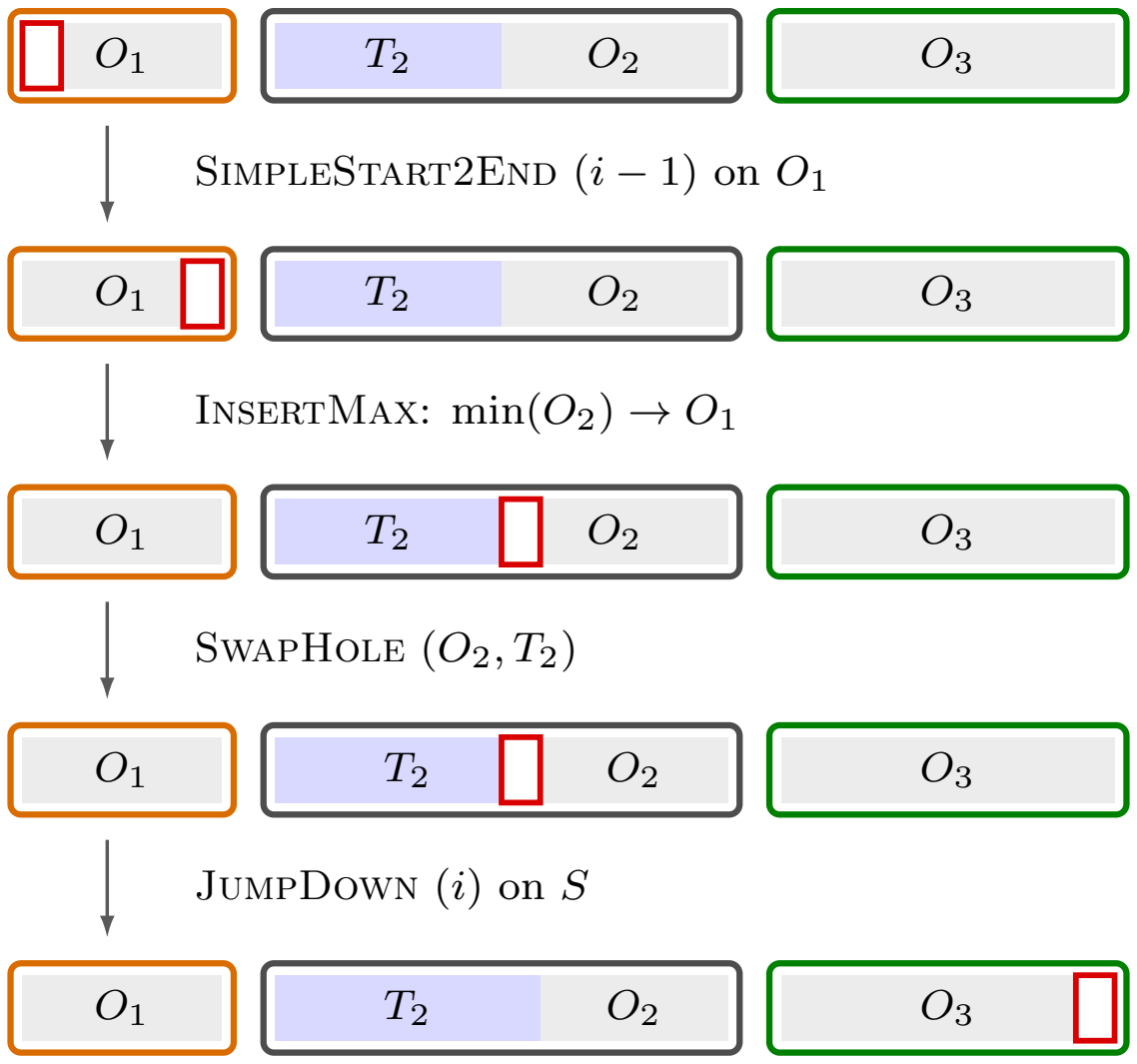

\begin{enumerate}
    \item If $i=1$, shift the elements of the sorted
    array to move the hole to its last position.

    \item If $S$ contains a single block
    with original section $O$, invoke
    \simpleslide($i-1$) on $O$.

    \item Otherwise, write
    \(
        S=O_1\mid T_2\mid O_2\mid O_3
    \)
    and perform the following steps:

    \begin{enumerate}
        \item Apply \simpleslide($i-1$) to $O_1$, moving
        the hole to the end of $O_1$.

        \item Move $\min O_2$ into the hole in $O_1$
        using \addmax.

        \item Apply \swaphole$(O_2,T_2)$ from right
        to left, moving the hole to the end of $T_2$.

        \item Finally, apply \downhole($i$) to move
        the hole to the end of $S$.
    \end{enumerate}
    We illustrate this operation in \Cref{fig:simpleslide}.
\end{enumerate}


\begin{claim}
\label{cl:simpleslide-invariants}
The \simpleslide operation moves the hole from
the first to the last position of the \dst,
while preserving all structural invariants.
\end{claim}

As the recursion returns through level $j$, it performs $O(1)$ local movements and calls \downhole$(j)$. Excluding rebuilding, the resulting $O(j)$ recourse per level sums to $O(i^2)$.

\begin{claim}
\label{cl:simpleslide-recourse}
Excluding recourse incurred by any \rebuild calls, 
\simpleslide($i$) incurs $O(i^2)$ recourse.
\end{claim}

\begin{proof}
The first base case is on $O(1)$-sized level-$1$
\dst incurring $O(1)$ recourse. The second base
case is dominated by the inductive case, which
incurs $O(i)$ recourse for \downhole($i$)
(by \Cref{cl:down-hole-recourse}), 
$O(1)$ recourse for \addmax (by
\Cref{cl:add-max-recourse}), and no recourse
for \swaphole excluding \rebuild calls.
Since \simpleslide($i$) makes one recursive
call to \simpleslide($i-1$), we get
$R(i)\le R(i-1)+O(i)$. The claim follows.
\end{proof}

\subsubsection{The Complex Operations: \slide and \routehole}

The operations \downhole and \simpleslide 
described above do not yet handle all situations
of moving a hole through a \dst. 
First, recall that \simpleslide does not handle 
the case when the \dst $S$ is the (unique) section 
of the last block of its parent. We will handle this explicitly
in the \slide operation, which will include all cases of 
moving a hole from the beginning to the end of a \dst. 
Second, a deletion may create a hole
at an arbitrary position of a \dst, and not at the 
beginning of the \dst. In \routehole, we will generalize 
\slide to handling the case when the hole is initially in 
an arbitrary position of the array. Later, we will use \routehole
to implement an element deletion in the \delete operation of the \adst. 
   
\paragraph{The \slide\ operation.}
    The main difficulty beyond \simpleslide\ arises
    when a \dst\ $S$ is the last original section of
    a three-block parent $\hat{S}$ with a nonempty
    transfer section.
    
    In this case, write
    \[
    S=O_1\mid T_2\mid O_2\mid O_3,
    \qquad
    \hat{S}=\hat{O}_1\mid\hat{T}_2\mid
    \hat{O}_2\mid S.
    \]
    The ordering invariants require
    \[
        O_1<\hat{T}_2<O_2.
    \]
    Thus, we cannot move $\min O_2$ directly into
    $O_1$, as in \simpleslide.
    
    Instead, we move $\min\hat{T}_2$ into $O_1$,
    route the resulting hole through $\hat{T}_2$
    using \simpleslide, and fill it with $\min O_2$.
    The remaining steps are then identical to
    \simpleslide.
    
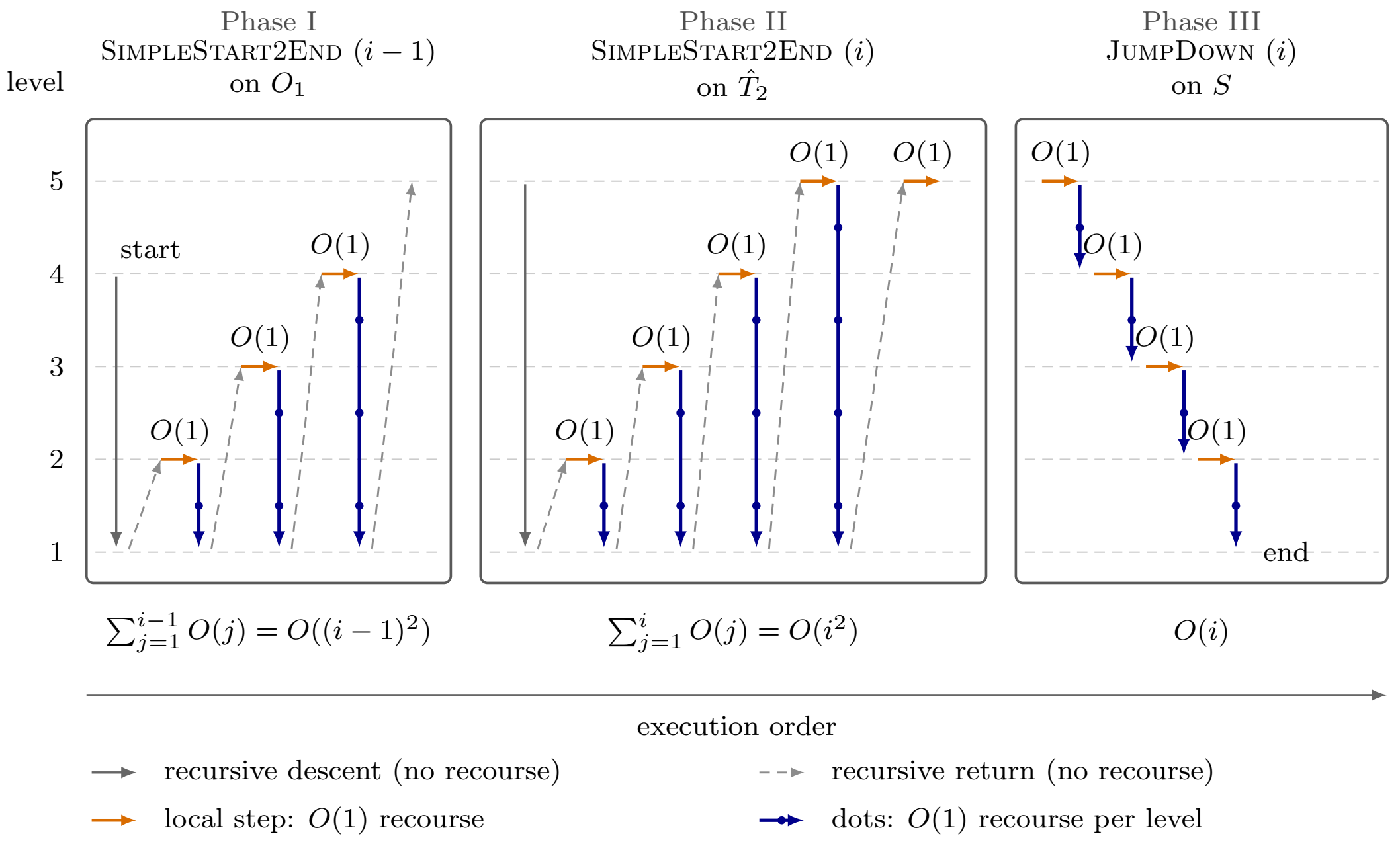
\begin{figure}[t]
\centering
\resizebox{.90\linewidth}{!}{%
\begin{tikzpicture}[
    x=0.75cm,y=0.78cm,>=Latex,
    every node/.style={font=\scriptsize},
    panel/.style={
        draw=black!65,rounded corners=2pt,line width=.7pt
    },
    guide/.style={
        black!20,densely dashed,line width=.4pt
    },
    call/.style={
        -{Latex[length=1.6mm]},draw=black!60,line width=.65pt
    },
    upward/.style={
        -{Latex[length=1.4mm]},draw=black!45,
        densely dashed,line width=.55pt
    },
    smallmove/.style={
        -{Latex[length=1.6mm]},draw=orange!85!black,line width=.85pt
    },
    down/.style={
        -{Latex[length=1.7mm]},draw=blue!55!black,line width=1pt
    },
    costlabel/.style={font=\scriptsize,inner sep=1pt},
    subtitle/.style={font=\scriptsize,text=black!70},
    summary/.style={font=\scriptsize},
    legend/.style={font=\scriptsize}
]

\begin{scope}[on background layer]
    \draw[panel] (1.05,1.20) rectangle (5.95,7.20);
    \draw[panel] (6.35,1.20) rectangle (13.15,7.20);
    \draw[panel] (13.55,1.20) rectangle (18.55,7.20);

    \foreach \Y in {1.60,2.80,4.00,5.20,6.40} {
        \draw[guide] (1.17,\Y) -- (5.83,\Y);
        \draw[guide] (6.47,\Y) -- (13.03,\Y);
        \draw[guide] (13.67,\Y) -- (18.43,\Y);
    }
\end{scope}

\node[subtitle] at (3.50,8.47) {Phase I};
\node at (3.50,8.06) {\simpleslide$(i-1)$};
\node at (3.50,7.65) {on $O_1$};

\node[subtitle] at (9.75,8.47) {Phase II};
\node at (9.75,8.06) {\simpleslide$(i)$};
\node at (9.75,7.65) {on $\hat{T}_2$};

\node[subtitle] at (16.05,8.47) {Phase III};
\node at (16.05,8.06) {\downhole$(i)$};
\node at (16.05,7.65) {on $S$};

\node[anchor=south east] at (0.93,7.40) {level};
\foreach \Y/\L in {1.60/1,2.80/2,4.00/3,5.20/4,6.40/5} {
    \node[anchor=east] at (0.93,\Y) {\L};
}


\node[anchor=south west,inner sep=1pt]
    at (1.45,5.35) {start};

\draw[call] (1.45,5.16) -- (1.45,1.65);

\draw[upward] (1.62,1.64) -- (2.05,2.80);
\draw[smallmove] (2.05,2.80) -- (2.56,2.80);
\node[costlabel,above] at (2.31,2.92) {$O(1)$};
\draw[down] (2.56,2.75) -- (2.56,1.65);
\fill[blue!55!black] (2.56,2.20) circle[radius=1.2pt];

\draw[upward] (2.73,1.64) -- (3.13,4.00);
\draw[smallmove] (3.13,4.00) -- (3.64,4.00);
\node[costlabel,above] at (3.39,4.12) {$O(1)$};
\draw[down] (3.64,3.95) -- (3.64,1.65);
\foreach \Y in {2.20,3.40} {
    \fill[blue!55!black] (3.64,\Y) circle[radius=1.2pt];
}

\draw[upward] (3.81,1.64) -- (4.21,5.20);
\draw[smallmove] (4.21,5.20) -- (4.72,5.20);
\node[costlabel,above] at (4.47,5.32) {$O(1)$};
\draw[down] (4.72,5.15) -- (4.72,1.65);
\foreach \Y in {2.20,3.40,4.60} {
    \fill[blue!55!black] (4.72,\Y) circle[radius=1.2pt];
}
\draw[upward] (4.89,1.64) -- (5.43,6.40);


\draw[call] (6.95,6.36) -- (6.95,1.65);

\draw[upward] (7.12,1.64) -- (7.50,2.80);
\draw[smallmove] (7.50,2.80) -- (8.01,2.80);
\node[costlabel,above] at (7.76,2.92) {$O(1)$};
\draw[down] (8.01,2.75) -- (8.01,1.65);
\fill[blue!55!black] (8.01,2.20) circle[radius=1.2pt];

\draw[upward] (8.18,1.64) -- (8.53,4.00);
\draw[smallmove] (8.53,4.00) -- (9.04,4.00);
\node[costlabel,above] at (8.79,4.12) {$O(1)$};
\draw[down] (9.04,3.95) -- (9.04,1.65);
\foreach \Y in {2.20,3.40} {
    \fill[blue!55!black] (9.04,\Y) circle[radius=1.2pt];
}

\draw[upward] (9.21,1.64) -- (9.55,5.20);
\draw[smallmove] (9.55,5.20) -- (10.06,5.20);
\node[costlabel,above] at (9.80,5.32) {$O(1)$};
\draw[down] (10.06,5.15) -- (10.06,1.65);
\foreach \Y in {2.20,3.40,4.60} {
    \fill[blue!55!black] (10.06,\Y) circle[radius=1.2pt];
}

\draw[upward] (10.23,1.64) -- (10.65,6.40);
\draw[smallmove] (10.65,6.40) -- (11.16,6.40);
\node[costlabel,above] at (10.91,6.51) {$O(1)$};
\draw[down] (11.16,6.35) -- (11.16,1.65);
\foreach \Y in {2.20,3.40,4.60,5.80} {
    \fill[blue!55!black] (11.16,\Y) circle[radius=1.2pt];
}

\draw[upward] (11.33,1.64) -- (12.04,6.40);
\draw[smallmove] (12.04,6.40) -- (12.55,6.40);
\node[costlabel,above] at (12.30,6.51) {$O(1)$};


\draw[smallmove] (13.90,6.40) -- (14.41,6.40);
\node[costlabel,above] at (14.16,6.52) {$O(1)$};
\draw[down] (14.41,6.35) -- (14.41,5.25);
\fill[blue!55!black] (14.41,5.80) circle[radius=1.2pt];

\draw[smallmove] (14.60,5.20) -- (15.11,5.20);
\node[costlabel,above] at (14.86,5.32) {$O(1)$};
\draw[down] (15.11,5.15) -- (15.11,4.05);
\fill[blue!55!black] (15.11,4.60) circle[radius=1.2pt];

\draw[smallmove] (15.30,4.00) -- (15.81,4.00);
\node[costlabel,above] at (15.56,4.12) {$O(1)$};
\draw[down] (15.81,3.95) -- (15.81,2.85);
\fill[blue!55!black] (15.81,3.40) circle[radius=1.2pt];

\draw[smallmove] (16.00,2.80) -- (16.51,2.80);
\node[costlabel,above] at (16.26,2.92) {$O(1)$};
\draw[down] (16.51,2.75) -- (16.51,1.65);
\fill[blue!55!black] (16.51,2.20) circle[radius=1.2pt];

\node[anchor=west] at (16.70,1.61) {end};

\node[summary] at (3.50,0.56)
    {$\sum_{j=1}^{i-1}O(j)=O((i-1)^2)$};
\node[summary] at (9.75,0.56)
    {$\sum_{j=1}^{i}O(j)=O(i^2)$};
\node[summary] at (16.05,0.56) {$O(i)$};

\draw[-{Latex[length=1.6mm]},black!60,line width=.65pt]
    (1.05,-0.25) -- (18.55,-0.25);
\node[legend] at (9.80,-0.64) {execution order};

\draw[call] (1.12,-1.25) -- (1.73,-1.25);
\node[legend,anchor=west] at (1.91,-1.25)
    {recursive descent (no recourse)};

\draw[upward] (10.10,-1.25) -- (10.71,-1.25);
\node[legend,anchor=west] at (10.89,-1.25)
    {recursive return (no recourse)};

\draw[smallmove] (1.12,-1.87) -- (1.73,-1.87);
\node[legend,anchor=west] at (1.91,-1.87)
    {local step: $O(1)$ recourse};

\draw[down] (10.10,-1.87) -- (10.71,-1.87);
\fill[blue!55!black] (10.40,-1.87) circle[radius=1.2pt];
\node[legend,anchor=west] at (10.89,-1.87)
    {dots: $O(1)$ recourse per level};

\coordinate (panelcenter) at (9.80,0);
\path let \p1=(current bounding box.south west),
          \p2=(current bounding box.north east),
          \p3=(panelcenter),
          \n1={max(\x3-\x1,\x2-\x3)}
      in (\x3-\n1,\y1) rectangle (\x3+\n1,\y2);

\end{tikzpicture}%
}
\caption{Recursive execution of \slide$(i)$ when $S$ is its parent's last original section and $\hat{T}_2$ is nonempty. Each \simpleslide\ descends to level $1$; on returning to level $j$, it performs $O(1)$ local work and calls \downhole$(j)$. These repeated descents give $O((i-1)^2)$ and $O(i^2)$ recourse in the first two phases, followed by $O(i)$ in the third. Gray arrows show recursive calls and returns; orange and dark blue arrows show local movements and \downhole, respectively.}
\label{fig:slide-zigzag}
\end{figure}

    \Cref{fig:slide-zigzag} illustrates these
    recursive movements.

Formally, the \slide operation is implemented recursively as follows:

\begin{enumerate}
    \item If $i=1$, shift the elements of the sorted
    array to move the hole to its last position.

    \item If $S$ contains a single block
    with original section $O$, call 
    \slide($i-1$) on $O$.

    \item Otherwise, $S=O_1\mid T_2\mid O_2\mid O_3$ is a
    three-block DST. If $S$ is not the last original section
    of a three-block parent, call \simpleslide($i$) on $S$.
    
    \item Otherwise, $S=\widehat O_3$ is the last original
    section of a three-block parent
    $\widehat S=\widehat O_1\mid\widehat T_2
    \mid\widehat O_2\mid S$.
    In this case, we have  the following steps:

    \begin{enumerate}
        \item Apply \simpleslide($i-1$) to $O_1$,
        moving the hole to the end of $O_1$. (Note that 
        this is valid because $O_1$ is in first block in $S$.)

        \item We have two cases:
        \begin{enumerate}
            \item If $\hat{T}_2$ is empty, we move $\min O_2$ to the hole using
            \addmax.
            \item Otherwise, $\hat{T}_2$ is non-empty. In this case, we do the following:
            \begin{enumerate}
                \item move $\min \hat{T_2}$ into $O_1$ using \addmax. 
                \item apply \simpleslide($i$) to $\hat{T_2}$.
                \item move the minimum element in $O_2$ into the resulting hole in $\hat{T_2}$, using \addmax.
            \end{enumerate}
        \end{enumerate}
        \item Apply \swaphole$(O_2,T_2)$ from right
        to left, moving the hole to the end of $T_2$.

        \item Finally, apply \downhole($i$) to move
        the hole to the end of $S$.
    \end{enumerate}
\end{enumerate}

As usual, if rebuilding has already moved the hole
to the end of $S$ or an enclosing section, the
corresponding unfinished operations terminate.

\begin{claim}
\label{cl:slide-invariants}
The \slide operation moves the hole from
the first to the last position of the \dst,
while preserving all structural invariants.
\end{claim}

Excluding rebuilding, the calls to \simpleslide\ contribute $O(i^2)$ recourse in total, while the final call to \downhole\ contributes only $O(i)$.

\begin{claim}
\label{cl:slide-recourse}
Excluding recourse incurred by any \rebuild calls, 
\slide$(i)$ incurs $O(i^2)$ recourse.
\end{claim}

\begin{proof}
For $i=1$, the operation shifts a sorted array
of constant size, incurring $O(1)$ recourse.

For $i>1$, if the \dst\ consists of a single block,
the operation recursively calls \slide($i-1$).
The next case, which just calls \simpleslide ($i$),
subsumed by the the last case. In the last case, 
there are calls to \simpleslide($i$), 
\simpleslide($i-1$), \downhole($i$),
and $O(1)$ calls to \addmax\ and \swaphole.

By \Cref{cl:simpleslide-recourse}, the calls to
\simpleslide($i-1$) and \simpleslide($i$) incur
$O((i-1)^2)$ and $O(i^2)$ recourse, respectively.
By \Cref{cl:down-hole-recourse}, the call to
\downhole($i$) incurs $O(i)$ recourse.
Finally, by \Cref{cl:add-max-recourse}, the
$O(1)$ calls to \addmax\ incur $O(1)$ recourse,
while \swaphole\ incurs no recourse aside from rebuilding.
Thus, the total recourse is
\[
    O((i-1)^2)+O(i^2)+O(i)+O(1)=O(i^2).
\]

Thus, writing $R(i)$ for the recourse of \slide($i$),
excluding rebuilding, we obtain
\[
    R(i)\le \max\left\{R(i-1),O(i^2)\right\}.
\]
Together with $R(1)=O(1)$, this gives
$R(i)=O(i^2)$.
\end{proof}

\paragraph{The \routehole\ operation.}
Finally, we are left with the most general operation
that moves a hole from an arbitrary position in a 
\dst $S$ to its end. To do this, we
    first move the hole recursively to the end of the
    section containing it within $S$ (and therefore, 
    one level lower than $S$). 
    The remaining movement
    depends on which section contains the hole
    in $S$.

    If $S$ has a single block, then we simply call 
    \routehole recursively in its unique section.
    So, let us assume that $S$ is a $3$-block \dst
    given by $S = O_1 \mid T_2 \mid O_2 \mid O_3$.
    The simplest case is when the hole is in $O_3$, 
    in which case it has already reached the end of $S$.
    Next, suppose the hole is in the middle block.
    If it is in $T_2$, 
    we use \downhole to move it to the end of $S$.
    Otherwise, it is in $O_2$, and
    we would like to fill the hole with $\min O_3$ 
    and apply \slide\ to $O_3$. There is one case
    where we cannot do this directly: when $T_2$
    is non-empty and $O_3$ is a level-$1$ \dst. 
    In this latter case, $O_3 > T_2 > O_2$; so,
    we cannot move $\min O_3$ to $O_2$ skipping
    $T_2$. In this case,
    we move $\min T_2$ to the hole in $O_2$,
    then apply \simpleslide to $T_2$ to move the 
    hole to the end of this section, and finally 
    apply \downhole to move it to the end of $S$.
    
    Finally, suppose the hole is in $O_1$. 
    Then, we use a subroutine
    similar to \slide to move the hole to the end of
    $S$. The only difference with \slide is that the 
    latter starts with the hole at the beginning of
    $O_1$ rather than its end. But, the first step of 
    \slide moves the hole to the end of $O_1$, before
    moving it further to the end of $S$. So, we simply
    copy the steps of \slide omitting the first step.

Formally, the operation is implemented recursively as follows.

\begin{enumerate}
    \item If $i=1$, shift the elements following the
    hole one position to the left.

    \item If the \dst\ contains a single block
    with original section $O$,
    call \routehole($i-1$) on $O$.

    \item Otherwise, write
    \(
        S=O_1\mid T_2\mid O_2\mid O_3. 
    \)
    Let $A$ denote the section containing the hole.

    First, call \routehole($i-1$) on $A$.
   
 We now have the following cases:
    \begin{enumerate}
        \item If $A=O_3$, we end the operation.

        \item If $A=T_2$, call \downhole($i$).

\item If $A=O_2$, we distinguish two cases.
\begin{enumerate}
    \item If $T_2$ is nonempty and $\min T_2 < \min O_3$ (may only happen when $O_3$ is a level-$1$ section),
    move $\min T_2$ into the hole using \addmax, apply
    \simpleslide($i-1$) to $T_2$, and invoke
    \downhole($i$).

    \item Otherwise, move $\min O_3$ into the hole
    and apply \slide($i-1$) to $O_3$.
\end{enumerate}
       
    \item If $A=O_1$, we proceed as follows:
    \begin{enumerate}
        \item There are two cases. Let $\hat{S}$ be the level-$(i+1)$ \dst that $S$ is contained in. 
        
        \medskip {\bf Case 1}: 
        If $S$ is the last original section of a
        three-block parent
        $\widehat S = \widehat O_1 \mid \widehat T_2
        \mid \widehat O_2 \mid S$,
        and $\widehat T_2$ is nonempty, then we:
        \begin{enumerate}
            \item Move $\min\hat{T}_2$ into the hole
            in $O_1$ using \addmax.
            \item Apply \simpleslide($i$) to $\hat{T}_2$.
            \item Move $\min O_2$ into the resulting
            hole in $\hat{T}_2$ using \addmax.
        \end{enumerate}
        
        \medskip {\bf Case 2}: Otherwise, we move $\min O_2$ into the hole
        in $O_1$ using \addmax.
    
        \item Apply \swaphole$(O_2,T_2)$ from right
        to left, moving the hole to the end of $T_2$.
    
        \item Finally, apply \downhole($i$) to move
        the hole to the end of $S$.
\end{enumerate}

    \end{enumerate}
\end{enumerate}


\begin{figure}[t]
\centering
\resizebox{.50\linewidth}{!}{%
\begin{tikzpicture}[
    x=1cm,y=1cm,
    every node/.style={font=\scriptsize},
    firstb/.style={draw=orange!85!black,very thick,rounded corners=2pt},
    stdb/.style={draw=black!70,very thick,rounded corners=2pt},
    lastb/.style={draw=green!50!black,very thick,rounded corners=2pt},
    osec/.style={fill=gray!15},
    tsec/.style={fill=blue!15},
    hole/.style={draw=red!85!black,fill=white,line width=1.1pt},
    operation/.style={-{Latex[length=1.5mm,width=1mm]},
        draw=black!65,line width=.65pt},
    brc/.style={draw=black!55,line width=.55pt,decorate,
        decoration={brace,amplitude=3pt}}
]

\newcommand{\figeightstate}[3]{%
    \fill[osec] (.08,{#1+.08}) rectangle (.77,{#1+.52});
    \fill[tsec] (1.13,{#1+.08}) rectangle (2.27,{#1+.52});
    \fill[osec] (2.27,{#1+.08}) rectangle (2.97,{#1+.52});
    \fill[osec] (3.38,{#1+.08}) rectangle (4.37,{#1+.52});
    \fill[tsec] (4.73,{#1+.08}) rectangle (#3,{#1+.52});
    \fill[osec] (#3,{#1+.08}) rectangle (6.47,{#1+.52});
    \fill[osec] (6.83,{#1+.08}) rectangle (8.37,{#1+.52});

    \draw[firstb] (0,#1) rectangle (.85,{#1+.60});
    \draw[stdb] (1.05,#1) rectangle (3.05,{#1+.60});

    \draw[firstb] (3.30,#1) rectangle (4.45,{#1+.60});
    \draw[stdb] (4.65,#1) rectangle (6.55,{#1+.60});
    \draw[lastb] (6.75,#1) rectangle (8.45,{#1+.60});

    \node at (.425,{#1+.30}) {$\widehat O_1$};
    \node at (1.70,{#1+.30}) {$\widehat T_2$};
    \node at (2.62,{#1+.30}) {$\widehat O_2$};
    \node at (3.875,{#1+.30}) {$O_1$};
    \node at ({(4.73+#3)/2},{#1+.30}) {$T_2$};
    \node at (6.165,{#1+.30}) {$O_2$};
    \node at (7.60,{#1+.30}) {$O_3$};

    \draw[hole] (#2,{#1+.08})
        rectangle ({#2+.26},{#1+.52});
}

\figeightstate{6.75}{4.11}{5.60}
\figeightstate{5.40}{1.13}{5.60}
\figeightstate{4.05}{2.01}{5.60}
\figeightstate{2.70}{5.60}{5.60}
\figeightstate{1.35}{5.60}{5.86}
\figeightstate{0.00}{8.11}{5.86}

\draw[brc] (0,8.45) -- (8.45,8.45)
    node[midway,above=6pt,font=\scriptsize]
    {$\widehat S$ {\footnotesize(level $i+1$)}};

\draw[brc,draw=blue!55!black] (1.13,7.48) -- (2.27,7.48)
    node[midway,above=3pt,text=black,font=\scriptsize]
    {$\widehat T_2$ {\footnotesize(level $i$)}};

\draw[brc,draw=green!50!black] (3.30,7.48) -- (8.45,7.48)
    node[midway,above=3pt,text=black,font=\scriptsize]
    {$S=\widehat O_3$ {\footnotesize(level $i$)}};

\draw[operation] (.55,6.63) -- (.55,6.12);
\node[anchor=west] at (.95,6.375)
    {Move $\min(\widehat T_2)$ to $O_1$};

\draw[operation] (.55,5.28) -- (.55,4.77);
\node[anchor=west] at (.95,5.025)
    {\simpleslide$(i)$ on $\widehat T_2$};

\draw[operation] (.55,3.93) -- (.55,3.42);
\node[anchor=west] at (.95,3.675)
    {Move $\min(O_2)$ to $\widehat T_2$};

\draw[operation] (.55,2.58) -- (.55,2.07);
\node[anchor=west] at (.95,2.325)
    {\swaphole$(O_2,T_2)$};

\draw[operation] (.55,1.23) -- (.55,.72);
\node[anchor=west] at (.95,.975)
    {\downhole$(i)$ on $S$};

\end{tikzpicture}%
}
\caption{The case of \routehole$(i)$ in which
$S=\widehat O_3$ is the last original section of a
three-block parent
$\widehat S=\widehat O_1\mid\widehat T_2\mid\widehat O_2\mid S$,
and $\widehat T_2$ is nonempty. Starting with the hole
at the end of $O_1$, the hole passes through
$\widehat T_2$, returns to $S$, and then moves to the
end of $S$. Red outlines mark holes; rebuilding is omitted.}
\label{fig:routehole-higher-transfer}
\end{figure}
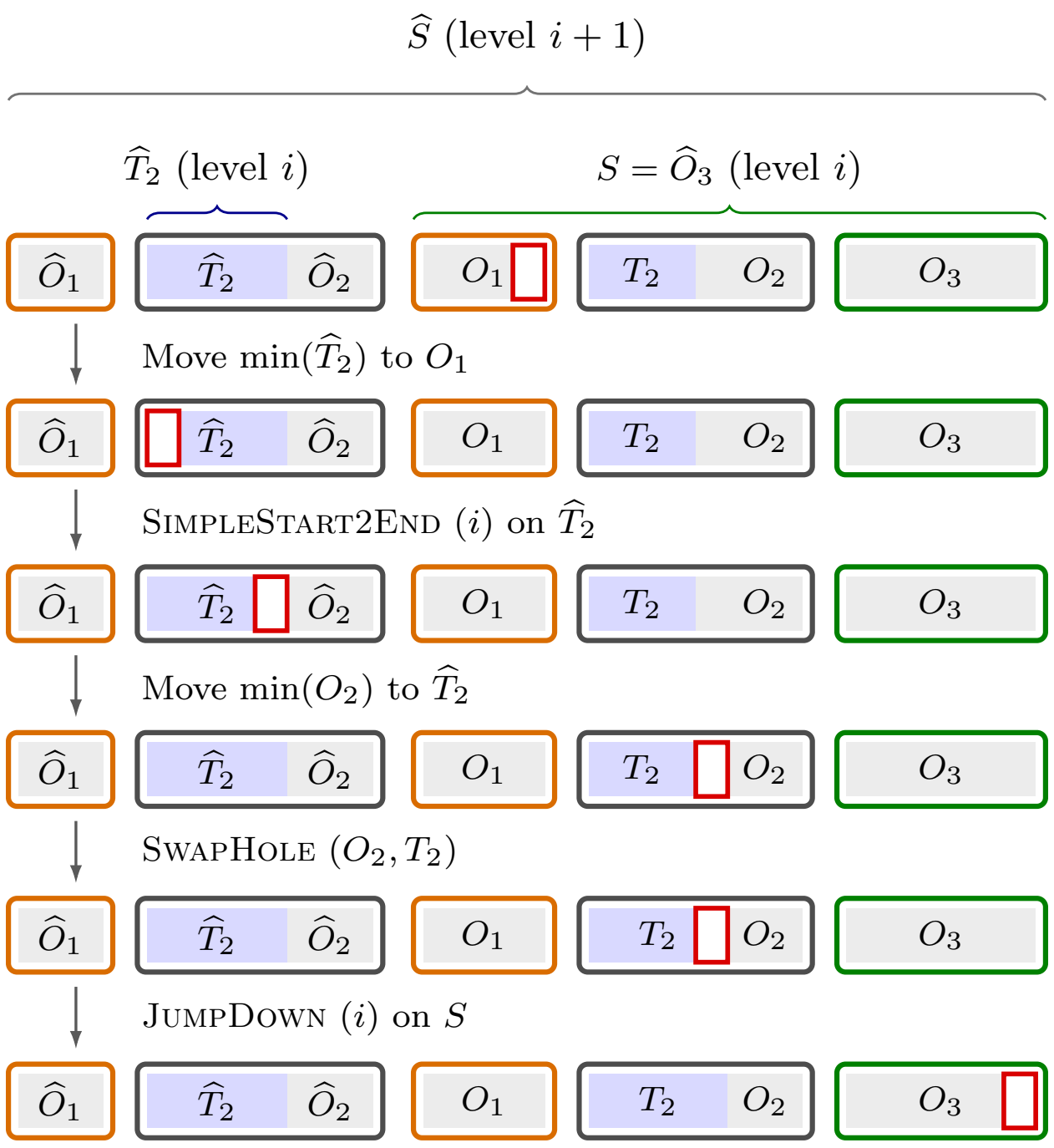

\begin{claim}
\label{cl:route-hole-invariants}
The \routehole operation moves the hole from
an arbitrary position to the last position of the \dst,
while preserving all structural invariants.
\end{claim}

Summing the $O(j^2)$ recourse at each level $j$, excluding rebuilding, gives the following bound.

\begin{claim}
\label{cl:route-hole-recourse}
Excluding recourse incurred by any \rebuild calls, 
\routehole$(i)$ incurs $O(i^3)$ recourse.
\end{claim}

\begin{proof}
We prove the claim by induction on $i$.
For $i=1$, the \dst\ is a sorted array of constant
size, so shifting the elements following the hole
incurs $O(1)$ recourse.

Now consider $i>1$. If the \dst\ consists of a
single block, the operation only calls
\routehole($i-1$), and the claim follows by induction.

Otherwise, the operation first calls
\routehole($i-1$) on the section containing the hole.
Depending on this section, the remaining steps involve
either one call to \slide, or a combination 
of one call to \simpleslide and also one call
to \downhole at level either $i$ or $i-1$, and $O(1)$ calls to
\addmax\ and \swaphole.

By \Cref{cl:slide-recourse,cl:simpleslide-recourse,cl:down-hole-recourse},
these steps incur $O(i^2)$ recourse, excluding
rebuilding. The calls to \addmax\ incur $O(1)$
recourse, while \swaphole\ incurs no recourse
apart from rebuilding.
Thus, letting $R(i)$ denote the recourse of
\routehole($i$), excluding rebuilding, we obtain
\[
    R(i)\le R(i-1)+O(i^2), \text{ with } R(1)=O(1).
\]
 Summing over the levels gives
\[
    R(i)=O\left(\sum_{j=1}^{i}j^2\right)=O(i^3).\qedhere
\]
\end{proof}








\section{Fully-Dynamic Sorting (\adst)}\label{sec:fully-dynamic-lowcost}We now use the \dst\ operations of \Cref{sec:new-dst} to construct an \emph{augmented dynamic sorting tree} (\adst) that supports element insertions and deletions. This gives an $O(\log^2 n)$-competitive algorithm with $O(\log^4 n)$ amortized recourse per update, proving \Cref{thm:dynamic}.


\subsection{Handling Insertions and Deletions: The \adst}

The \dst\ of \Cref{sec:new-dst} supports moving a hole
through its recursive structure, but does not directly
support insertions or deletions. We now use these
operations to construct a data structure that supports
both types of updates.

\begin{figure}[!htbp]
\centering
\begin{tikzpicture}[
    x=0.92cm,y=1cm,
    every node/.style={font=\scriptsize},
    comp/.style={very thick, rounded corners},
    dd/.style={comp, draw=blue!60!black, fill=blue!12},
    dm/.style={comp, draw=violet!70!black, fill=violet!12},
    dt/.style={comp, draw=orange!85!black, fill=orange!15},
    aa/.style={comp, draw=green!50!black, fill=green!15}
]
\draw[dd] (0,0) rectangle (6.4,0.8);
\node at (3.2,0.4) {$D_d$};

\draw[dm] (6.55,0) rectangle (9.35,0.8);
\node at (7.95,0.4) {$D_{d-1}$};

\node at (9.75,0.4) {\large$\cdots$};

\draw[dt] (10.15,0) rectangle (11.25,0.8);
\node at (10.7,0.4) {$D_2$};

\draw[aa] (11.4,0) rectangle (11.95,0.8);
\node at (11.675,0.4) {$A_1$};

\fill[black!3, rounded corners=2pt]
    (0,-0.85) rectangle (11.95,-0.15);

\node[anchor=west, text=black!65] at (0.12,-0.5)
    {size scale};
\node at (3.2,-0.5) {$O(2^d)$};
\node at (7.95,-0.5) {$O(2^{d-1})$};
\node at (10.7,-0.5) {$O(2^2)$};
\node at (11.675,-0.5) {$O(1)$};

\draw[->, thick, green!50!black]
    (13.0,0.4) -- (12.05,0.4);
\node[right, text=green!40!black] at (13.0,0.4)
    {insertions};

\draw[decorate,
      decoration={brace, amplitude=5pt},
      thick, black!60]
    (6.55,0.95) -- (11.95,0.95)
    node[midway, above=5pt]
    {level-$(d-1)$ \adst};
\end{tikzpicture}
\caption{
The components of a level-$d$ \adst.
A level-$j$ component has size $O(2^j)$.
Insertions enter the sorted array $A_1$, while
deletions move holes toward the right end.
Periodic reconstructions combine a component
with the smaller components to its right.
}
\label{fig:adst-sizes}
\end{figure}
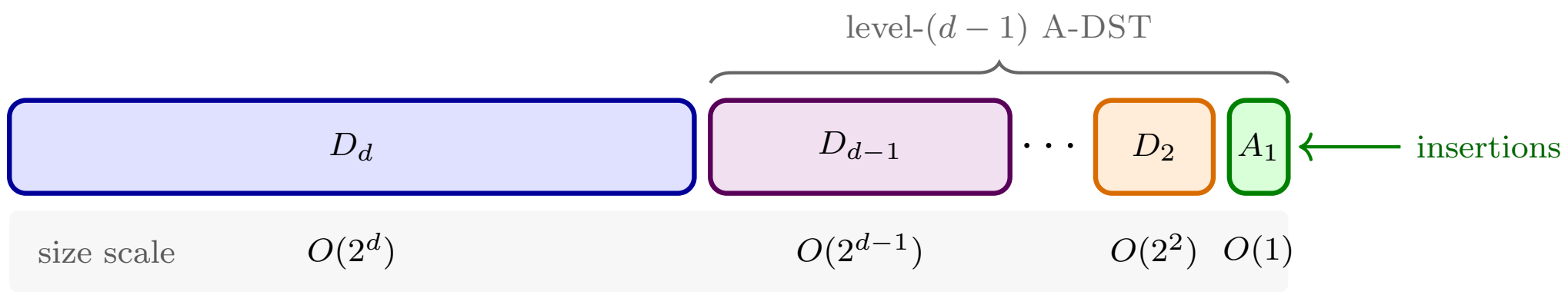

Our data structure, called an \emph{augmented dynamic
sorting tree} (\adst), consists of a sequence of
\dsts denoted $D_d,D_{d-1},\ldots,D_2$,
where $D_j$ is a level-$j$ \dst, followed by 
a sorted array, denoted $A_1$ (see \cref{fig:adst-sizes}). 
We set $d=1+\lceil\log_2 n\rceil$.
We call the constituent \dsts and the sorted 
array $A_1$ components of the overall \adst.
The components have geometrically
decreasing size bounds, with $D_j$ containing
at most $O(2^j)$ cells and $A_1$ containing $O(1)$ cells.
Any of these \dsts may be empty at any stage.

By \Cref{sec:preliminaries}, insertions and deletions
to an \adst are alternated. Before insertion, the 
\adst contains exactly one hole, which is at the 
end of $A_1$.
An insertion is performed in $A_1$, maintaining
its sorted order, shifting elements as required. 
In contrast, before deletion, the \adst
does not contain any hole. 
To delete an element, we first
remove it from its cell, creating a hole. We then
use the \dst\ operations to move this hole to the
end of its component and through the remaining
components, until it reaches the end of the array.
Finally, to prevent the smaller components from growing
too large, we periodically reconstruct a component
together with all components to its right.
Such a reconstruction sorts their elements and
builds a single \dst\ at the corresponding level,
leaving the smaller components empty.

We give details of the structure and operations of 
an \adst below.


\subsubsection{Description of a Level-$d$ \adst}

We now formally define the recursive structure of the \adst.
Its recursion depth is $d=1+\lceil\log_2 n\rceil$,
with levels indexed from $1$ to $d$, where level $d$
is the top level.
For the base case, a level-$1$ \adst\ is simply a sorted
array of length $n$. Maintaining this order requires at most
$O(n)$ recourse per update and achieves a competitive ratio
of $1$.

For $d>1$, a level-$d$ \adst\ is partitioned into two
contiguous regions: a level-$d$ \dst\ on the left and a
level-$(d-1)$ \adst\ on the right.
Unrolling the recursion, a level-$d$ \adst\ consists of
one level-$i$ \dst\ for each $i=d,d-1,\ldots,1$,
arranged in decreasing order of their levels.
This structure of an \adst is depicted in
\cref{fig:adst-sizes}.

\paragraph{Structural invariants.}
Every component $D_j$ satisfies the structural
invariants of the \dst in \Cref{subsec:dst-invariants}, and $A_1$ is sorted.
We impose no ordering between different components.
In addition, we maintain the following invariant:

\begin{itemize}
    \item \textbf{Suffix Sizes.}
    For every $1\le j<d$, the level-$j$
    \adst\ $A_j$ contains at most $2^j$ elements.
\end{itemize}

Since $D_j$ is contained in $A_j$, each component
$D_j$ contains at most $2^j$ elements.
The same holds for $D_d$, since the entire array
contains at most $n\le 2^d$ elements.
In particular, $A_1$ contains at most two elements,
so inserting into this sorted array incurs $O(1)$
recourse, excluding any rebuilding.

\paragraph{Cost bound.}We first bound the cost of an \adst\ using the
crossing bounds of its constituent \dsts.
We then extend this guarantee to the subsequence
of active elements, as required by the reduction
in the preliminaries.

\begin{claim}
\label{cl:adst-cost}
Let $A$ be a level-$i$ \adst, and let $A'$
be any subsequence of $A$. Then, for every
threshold $z$,
\[
    \operatorname{cross}_z(A')\le 4i^2.
\]
Consequently, if all elements of $A'$ lie in
an interval $I$, then
$\operatorname{cost}(A')=O(i^2)\cdot |I|$.
\end{claim}

\begin{proof}
A level-$i$ \adst\ consists of one level-$j$
\dst\ for each $j\in\{1,\ldots,i\}$.
By \Cref{cl:dst-cost}, each level-$j$ \dst\
contributes at most $4j$ crossings at any
threshold $z$. In addition, there are at most
$i-1$ boundaries between consecutive \dsts,
each contributing at most one crossing. Thus,
\[
    \operatorname{cross}_z(A)
    \le \sum_{j=1}^{i}4j+(i-1)
    =2i^2+3i-1
    \le 4i^2.
\]
Since $A'$ is a subsequence of $A$, by
\Cref{lem:subsequence-crossings},
\[
    \operatorname{cross}_z(A')
    \le \operatorname{cross}_z(A)
    \le 4i^2.
\]
Integrating over $I$ gives
$\operatorname{cost}(A')=O(i^2)\cdot |I|$.
\end{proof}


\subsection{Maintaining the Recursive Structure}

\paragraph{The \resort($i$) Operation.}

As updates are performed, cells may be transferred
from higher-level \dsts\ to lower-level \dsts,
causing the latter to grow. To maintain the Suffix
Sizes invariant, we periodically rebuild parts
of the \adst\ using \resort.

For $i\ge2$, the \resort($i$) operation collects all
elements of the level-$i$ \adst\ $A_i$, sorts them,
and constructs an level-$i$ \dst\ containing
these elements, leaving the level-$(i-1)$ \adst\ empty.

Recall that each element update comprises a deletion
followed by an insertion. For each $i$, we maintain 
a counter that counts the number of updates since the
counter was last reset to $0$. When the $i$th counter
reaches $2^{i-1}$, we say that the \adst $A_i$
is ready for rebuilding. Rebuilding is performed after 
element insertion, i.e., at the end of the current update.
If several levels
require rebuilding simultaneously, we perform
only the highest-level rebuilding and reset
its counter and those of all lower levels.
Rebuilding at level $i$ is performed by calling
\resort($i$).

\begin{claim}
\label{cl:resort-invariants}
The \resort\ operation maintains all structural invariants.
\end{claim}

Each level contributes $O(1)$ amortized rebuilding recourse per update. We obtain the following bound.

\begin{claim}
\label{cl:resort-recourse}
For a level-$i$ \adst, the amortized recourse
incurred by \resort\ is $O(i)$ per update.
\end{claim}

\begin{proof}
Consider \resort($j$), for $2\le j\le i$.
By the Suffix Sizes invariant, the level-$j$
\adst\ contains at most $2^j$ elements after
each completed update. Immediately before
rebuilding, its size may exceed this bound
by at most one. Thus, \resort($j$) incurs
$O(2^j)$ recourse.

The operation is invoked only after
$2^{j-1}$ updates since its counter was last
reset, where each update consists of a deletion
followed by an insertion. Hence its amortized
recourse per update is
\[
    \frac{O(2^j)}{2^{j-1}}=O(1).
\]

A higher-level rebuilding resets the counters
of lower levels, which can only reduce their
rebuilding frequency. Summing over the at most
$i$ levels gives $O(i)$ amortized recourse
per update.
\end{proof}

\paragraph{Transferring holes between components.}

We extend the \swaphole\ operation that we introduced 
earlier in \dsts to transfer a hole between
two adjacent components of the \adst, including
the final sorted array $A_1$.
The transfer is performed by changing the boundary
between the components, without moving any stored
element. The implementation is as follows.

\begin{enumerate}
    \item Let $D$ and $D'$ denote the components
    supplying and receiving the hole, respectively.
    Suppose $D$ is immediately to the left of $D'$,
    and its last cell $h$ contains the hole.
    We remove $h$ from $D$ and make it the first
    cell of $D'$.
    This is done recursively, removing $h$ along
    the chain of last original sections of $D$
    and adding it along the chain of first original
    sections of $D'$, down to level $1$.

    \item After transferring the hole, we examine
    the affected sections and their ancestors,
    from highest level to lowest.
    A section requires repair if a Block Sizes
    invariant is violated or if it has three blocks
    and its middle original section becomes empty.

    Let $S$ be a highest-level section requiring
    repair. We apply \rebuild\ to $S$, unless
    $S$ is the last original section of a
    three-block parent $P$, in which case we
    rebuild $P$ instead. Rebuilding $P$ empties
    its transfer section while preserving its
    top-level block sizes and the ordering
    constraints at higher levels.

    When rebuilding a section in $D'$ containing
    the hole, we treat the hole as the smallest
    element, leaving it in the first cell rather
    than the last. This convention also applies
    throughout the recursive rebuilding.

    We repeat until all affected sections satisfy
    the \dst\ invariants.
\end{enumerate}

Since $D$ and $D'$ are separate components, they
have no ordering constraints between them, and
rebuilding is performed independently within each
component. Moreover, the hole lies along the chain
of first original sections of $D'$, so the rebuilding
rule above preserves its position at the beginning
of $D'$, as required by \slide.

\begin{claim}
\label{cl:adst-swap-hole-invariants}
The \swaphole\ operation maintains all structural
invariants of the \adst.
\end{claim}

We bound the total rebuilding recourse in terms of
the number of calls to \swaphole.

\begin{claim}
\label{cl:rebuild-amortized}
At any point during the execution of a $d$-level \adst,
let $M$ denote the number of \swaphole\ calls made so far.
Then, the total recourse incurred by the resulting
\rebuild\ operations is $O(dM)$.
\end{claim}

\subsection{\adst: The interface}The interface of the \adst\ contains two operations:
\begin{itemize}
    \item \textbf{\delete}: Given an element $x$ stored
    in the \adst, deletes $x$ from the \adst.
    \item \textbf{\insertadst}: Given a new element $x$,
    inserts $x$ into the \adst.
\end{itemize}


\paragraph{Implementation of \delete($x$).}
The \delete($x$) operation removes an element $x$
from a level-$i$ \adst\ and moves the resulting
hole to its last position.

The operation is implemented as follows.

\begin{enumerate}
    \item Let $D$ be the level-$j$ component
    containing $x$, treating $A_1$ as a
    level-$1$ \dst.
    Remove $x$ from its cell, creating a hole,
    and apply \routehole($j$) to $D$ to move
    the hole to its last position.

    \item If $D=A_1$, the operation is complete.

    \item Otherwise, process the remaining
    components from left to right. For each
    such component $D'$:
    \begin{enumerate}
        \item Apply \swaphole$(D,D')$ to transfer
        ownership of the hole to $D'$.

        \item Apply \slide\ to $D'$, moving
        the hole to its last position.

        \item Set $D=D'$ and continue to the
        next component.
    \end{enumerate}
\end{enumerate}

\begin{claim}
\label{cl:adst-delete-invariants}
The \delete\ operation maintains all structural
invariants and moves the resulting hole to the last
position of the \adst.
\end{claim}

\paragraph{Implementation of \insertadst($x$).}
Recall that each deletion is followed by an insertion.
The deletion leaves a hole at the end of the \adst,
which belongs to its final level-$1$ sorted array.
We insert $x$ into its sorted position in this array,
shifting the necessary elements to fill the hole.

\begin{claim}
\label{cl:adst-insertion-invariants}
The \insertadst($x$) operation maintains all structural
invariants of the \adst.
\end{claim}

Finally, we get to proving the main lemma that shows the amortized recourse of the \adst.

\begin{lemma}
\label{cl:adst-delete-recourse}
The amortized recourse of \delete\ on a level-$i$
\adst\ is $O(i^4)$.
\end{lemma}

\begin{proof}
We first bound the recourse excluding \rebuild
calls, and then bound the recourse due to \rebuild calls 
separately.

Suppose the deleted element belongs to a level-$j$
component, where $j\le i$.
By \Cref{cl:route-hole-recourse}, moving the
resulting hole to the end of this component
incurs $O(j^3)$ recourse.
We then apply \slide\ to each of the remaining
components. By \Cref{cl:slide-recourse}, their
total recourse is
    $\sum_{k=1}^{j-1}O(k^2)=O(j^3)$.
Thus, the recourse of \delete, excluding
\rebuild calls, is $O(i^3)$.

We next account for \rebuild calls.
Consider a sequence of $K$ deletions, and let
$M$ be the total number of calls to \swaphole\
during their execution.
We distinguish between two types of \swaphole calls.
Every internal \swaphole\ call within a \dst\ is
charged to the preceding \addmax\ call
that created the hole being transferred.
Each such \addmax call corresponds to a unit of recourse,
which is charged at most once by \swaphole calls.
Hence, the number of internal \swaphole calls is 
bounded by the recourse due to \addmax calls,
which is at most $O(Ki^3)$.

The remaining \swaphole calls transfer holes between
adjacent components of the \adst.
Since there are only $i$ components in an
level-$i$ \adst, each \delete
operation makes at most $i-1$ \swaphole
calls of the second type. Cumulatively,
the $K$ element deletions make $O(Ki)$ \swaphole calls.

Therefore,
    $M=O(Ki^3)$.
By \Cref{cl:rebuild-amortized}, these calls
incur a total rebuilding recourse of
    $O(iM)=O(Ki^4)$.
Therefore, the amortized recourse
of \delete\ is $O(i^4)$, including \rebuild calls.
\end{proof}

\subsection{Cost Bound of \adst}

We use a $d$-level \adst, where
$d=1+\lceil\log_2(n)\rceil$.
As described in \Cref{sec:preliminaries},
the \adst\ always maintains $n$ elements,
some of which may be inactive.

Let $A_{\mathrm{active}}$ denote the subsequence
of active elements, and let $I$ be the interval
they span. By \Cref{cl:adst-cost},
\[
    \operatorname{cost}(A_{\mathrm{active}})
    =O(d^2)\cdot |I|.
\]
Since $\operatorname{OPT}=|I|$ and
$d=O(\log n)$, the competitive ratio is
$O(\log^2 n)$, as stated in \Cref{thm:dynamic}.

\subsection{Recourse Bound of \adst}

By \Cref{cl:adst-delete-recourse}, a \delete
operation incurs $O(d^4)$ amortized recourse.
The subsequent insertion fills the hole 
at the end of $A_1$.
By the Suffix Sizes bound, including the temporary
hole, this array has constant size. Inserting
the new element in sorted order therefore
incurs $O(1)$ recourse and does not change
the number of cells in any component.
Finally, any scheduled \resort\ is performed
after the insertion. By \Cref{cl:resort-recourse},
these \resort calls contribute $O(d)$
amortized recourse per update.

Thus, the amortized recourse per deletion-insertion
pair is
\[
    O(d^4)+O(1)+O(d)
    =O(d^4)
    =O(\log^4 n),
\]
which is the recourse bound in \Cref{thm:dynamic}.

\section{A Fully Dynamic Lower Bound}\label{sec:polylog-lower-bound}

We now establish our lower bound on the amortized
recourse required by deterministic fully dynamic
sorting algorithms.




\lowerthm*

We clarify that in \Cref{thm:dynamic-lower},
the condition $C\le O(n)$ means that
$C\le cn$ for some constant $c$.

\subsection{The instance} The initial instance contains one element of each value $0,\frac{1}{n},\frac{2}{n},\ldots,\frac{n-1}{n}$, 
and every value that will be inserted into the array will have a value of \(1\).
Hence, the optimum cost will always be at most \(1\).
We organize the elements into a hierarchy of consecutive
value ranges. The adversary will use this hierarchy to choose which
elements to replace.

Let \(B\ge 64\) and \(d\ge 1\) be integers to be chosen later,
with \(n=B^d\).
The hierarchy is represented by a complete \(B\)-ary tree
\(\mathcal{T}\) of depth \(d\).
Its leaves correspond, from left to right, to the elements
\(x_0,\ldots,x_{n-1}\).
The root represents all elements, and each internal node
divides its elements into \(B\) equal consecutive groups,
one for each child, each leaf represents a single element.

Formally, for a node \(v\), let \(X_v\) denote the set of original values
represented by its descendant leaves, and let \(N_v=|X_v|\).
A node at depth \(\ell\) therefore satisfies
\[
    N_v=B^{d-\ell}=\frac{n}{B^\ell}.
\]
If \(X_v=\{x_{a},\ldots,x_{a+N_v-1}\}\), associate with \(v\)
the value interval
\[
    I_v=\left[\frac{a}{n},\frac{a+N_v}{n}\right),
    \qquad |I_v|=\frac{N_v}{n}.
\]

\noindent
{\bf Remark.}
We assume \(n=B^d\) for simplicity.
For arbitrary \(n\), a constant-factor smaller depth and rounding
of \(B\) give \(B^d=\Theta(n)\).
Use the first \(B^d\) values for the tree and leave
the others unreplaced outside its value interval.
With population counts restricted to the tree's values,
only universal constants change in the proof.

\subsection{The adversary}
The adversary works in rounds, each consisting of a forward phase
followed by a backward phase. Each round begins with the initial
multiset of values \[
    0,\frac{1}{n},\frac{2}{n},\ldots,\frac{n-1}{n}.
\]
During the forward phase, the adversary searches the value tree
to select an element to replace by a copy of \(1\).
It restricts this search by controlling which children may be entered.

From now on we fix a single round. When an internal node is first visited, the adversary labels its
children as \emph{targets}, 
or
\emph{buffers}.
Later, the buffer children are relabeled as \emph{released}.

\paragraph{Labeling the children.} 

When the search first reaches an internal node \(v\) during the forward
phase, the adversary labels its children before continuing the search
below \(v\).
This initial labeling is chosen once per node (in each round), using
the algorithm's current arrangement.
At this point, all original values in \(X_v\) are still present,
since no element below \(v\) has yet been selected for replacement during the round.

 Let \(v_1,\ldots,v_B\) denote the children of \(v\), ordered by value.
The first and last \(B/4\) children are all labeled as buffers.
The labels of the middle \(B/2\) children are chosen according to
the positions of their elements in the array, as follows:
List the \(N_v/2\) values belonging to the middle children as
\(z_1,\ldots,z_{N_v/2}\), in the left-to-right order of the cells
containing their elements.
We partition this list into \(B/16\) consecutive groups of \(8N_v/B\) values, where consecutive refers to the order in this list, not that they occupy consecutive array cells.
We call these groups \emph{windows}.
For a given labeling of the middle children as either a target or as a buffer we call a window \emph{balanced} if it contains
at least \(N_v/B\) values from target children and at least \(N_v/B\) values from buffer children.
We choose a labeling for the children for which the number of balanced windows is $\Omega(B)$ (we show that such labeling exists). 

\paragraph{Paid intervals and released children.}
Fix a constant \(\varepsilon>0\), to be chosen later.
For each visited internal interval \(I_v\), the adversary counts
the elements in its initially labeled buffer
children that the algorithm has moved in updates after the labels
were assigned.
Each element is counted only once, even if it moves several times
or returns to the cell it occupied when the labels were assigned.

At the end of each completed update of the forward phase,
an interval \(I_v\) becomes \emph{paid} if the number of buffer elements that the algorithm moved is at least \(\varepsilon N_v\).
When this happens, all its buffer children are relabeled as
\emph{released} for the rest of the forward phase.

\paragraph{The sequence.} The adversary uses these labels to guide its search for the next element to replace.
A target child is available for search, and a buffer child becomes
available once it is relabeled as released.
Each search starts at the root.
At an internal node, the adversary first labels its children if it
has not already done so during this round.
It then searches target and released children in increasing value
order, backtracking whenever it cannot reach a remaining
element in a subtree.
Leaves whose elements have already been removed (and replaced by 1) are skipped.
When the search reaches a leaf whose element is still present,
the adversary deletes that element and then inserts a new copy of \(1\).

The forward phase ends when there is no path from the root to a leaf containing an element through target and released nodes.
Suppose the forward phase removed the values \(y_1,\ldots,y_m\),
in that order, and let \(e_j\) be the fresh copy of \(1\) inserted
after the removal of \(y_j\).
The backward phase first deletes \(e_1,\ldots,e_m\), in that order,
and then reinserts \(y_1,\ldots,y_m\), in that order.
The algorithm responds to each deletion and insertion.
This restores the initial multiset of values and completes the round.

The adversary then clears all child labels, payment statuses,
and movement counts.
The next round uses the same value tree and begins from the
algorithm's arrangement at the end of the backward phase.
The adversary repeats this procedure for a number of rounds
to be specified in the analysis.
\subsection{The analysis}
We first show that the labeling rule guarantees many balanced windows.
To prove the bound, we consider a random labeling and bound
the expected number of balanced windows.
Some fixed labeling must achieve at least this expectation,
and therefore so does the labeling chosen by the adversary.

\begin{lemma}
\label{lem:balanced-windows}
At every internal node \(v\), there exists a labeling of the children for which at least a \(7/9\)-fraction of the windows are balanced.
\end{lemma}

\begin{proof}
Consider a random labeling in which each middle child
is independently labeled as a target or a buffer, with probability
\(1/2\).

Fix a window,
for each middle child \(v_i\), let \(a_i\) be the number of its
values in \(W\).
Each child contains \(N_v/B\) values in total, while \(W\) contains
\(8N_v/B\) values, so
\[
    0\le a_i\le N_v/B,
    \qquad
    \sum_i a_i=8N_v/B.
\]
Let \(X\) be the number of values in \(W\) belonging to children
labeled as targets.
Since the child labels are independent,
\[
    \mathbb{E}[X]=\frac12\sum_i a_i=4N_v/B,
\]
and
\[
    \operatorname{Var}(X)
    =\frac14\sum_i a_i^2
    \le\frac{N_v/B}{4}\sum_i a_i
    =2(N_v/B)^2.
\]
The window is balanced precisely when \(N_v/B\le X\le7N_v/B\).
By Chebyshev's inequality,
\[
    \Pr[W\text{ is balanced}]
    \ge
    1-\frac{\operatorname{Var}(X)}{(3N_v/B)^2}
    \ge\frac79.
\]

By linearity of expectation, the expected number of balanced windows
is at least a \(7/9\)-fraction of all windows.
Therefore, some labeling achieves at least this many balanced windows.
The adversary chooses a labeling maximizing their number, so its
deterministic choice achieves the same guarantee.
\end{proof}

The balanced-window lemma provides many windows containing both
target and buffer elements.
All these elements have values in the middle half of \(I_v\).
Buffer elements that stay in their original cells can force array
cost when values outside \(I_v\) are placed between them.
To quantify this cost, we measure the portion of the array cost
lying within a value interval.
Let \(a_1,\ldots,a_q\) be the values currently present in the array,
listed from left to right.
\begin{definition}
For a value interval \(I\), let
\[
    V_I(A)
    :=\sum_{j=1}^{q-1}|[a_j,a_{j+1}]\cap I|.
\]
Here, \([x,y]\) denotes the interval with endpoints \(x\) and \(y\),
regardless of their order, and the absolute-value notation denotes
interval length.
Thus, each consecutive pair contributes the length of the portion
of the interval between its values that lies in \(I\).
\end{definition}

\begin{claim}
\label{cl:interval-cost}
For pairwise disjoint value intervals \(I_1,\ldots,I_t\),
\[
    \sum_{i=1}^t V_{I_i}(A)
    \le \operatorname{cost}(A).
\]
Moreover, for every internal node \(v\),
\[
    \sum_{w\text{ child of }v}V_{I_w}(A)=V_{I_v}(A).
\]
\end{claim}

\begin{proof}
For each consecutive pair \(a_j,a_{j+1}\), the portions of
\([a_j,a_{j+1}]\) lying in the intervals \(I_1,\ldots,I_t\)
are disjoint.
Their total length is therefore at most \(|a_j-a_{j+1}|\).
Summing over all consecutive pairs proves the first assertion.
The intervals of the children of \(v\) partition \(I_v\).
Consequently, their intersections with \([a_j,a_{j+1}]\)
partition its intersection with \(I_v\).
Summing their lengths over all consecutive pairs proves the equality.
\end{proof}

We now use the balanced windows to show that, at an unpaid node,
many target elements must remain unless the array has large cost
within the node's value interval.
The proof uses buffer elements in their original cells to bound
the cost between selected target positions.

\begin{definition}
Let \(A\) be the algorithm's array at the end of the forward phase and  \(I_v\) a visited internal interval.
\begin{itemize}
    \item \(U_v\) is the number of original elements in the target
    child intervals of \(I_v\) that are still present in the array (were not replaced by a value of $1$).
    \item \(M_v\) is the number of distinct original elements in buffer children
    that the algorithm has moved since the child intervals of
    \(I_v\) were labeled.
\end{itemize}    
\end{definition}
Note that while \(I_v\) remains unpaid, then
    \(M_v<\varepsilon N_v\), and none of its original buffer
    elements has been replaced by \(1\).
We now fix \(\varepsilon=1/1024\) in the definition of paid
intervals. With this choice, we obtain the following lemma.
\begin{lemma}
\label{lem:unpaid-forcing}
Every
visited unpaid interval \(I_v\) satisfies
\[
    U_v\ge\delta N_v
    \qquad\text{or}\qquad
    V_{I_v}(A)\ge\delta B|I_v|.
\]
for some constant \(\delta>0\).
\end{lemma}

\begin{proof}
Fix a visited unpaid interval \(I_v\).
Throughout the proof, an element's \emph{original cell} is the
cell it occupied when \(I_v\) was first visited.

The middle \(B/2\) child intervals contain \(N_v/2\) elements
in total.
These elements are partitioned into windows of \(8N_v/B\)
elements each, so the number of windows is
\[
    \frac{N_v/2}{8N_v/B}=\frac{B}{16}.
\]
By \Cref{lem:balanced-windows}, the chosen labeling makes
at least a \(7/9\) fraction of these windows balanced.
Thus, there are at least
\[
    \frac79\cdot\frac{B}{16}=\frac{7B}{144}
\]
balanced windows.
We pair consecutive balanced windows in their array order,
leaving the last window unused.
For \(B\ge64\), the bound above guarantees at least \(B/64\) pairs, we denote this number of pairs by \(r\), which is at least $1$ for \(B/64\).
From the first window of each pair, select \(N_v/B\) buffer
elements, and from the second, select \(N_v/B\) target elements.
Denote these groups, in their array order at the first visit, by
\[
    G_1,D_1,G_2,D_2,\ldots,G_r,D_r.
\]
All selected elements have values in the middle half of \(I_v\).

Call the original cells of the elements in
\(D_1,\ldots,D_r\) the \emph{selected cells}.
There are \(rN_v/B\) such cells.
We count them according to what has happened to the buffer
groups and what currently occupies the cells.

First, discard every pair whose buffer group has no unmoved
element.
Each discarded pair contains \(N_v/B\) distinct moved buffer
elements and \(N_v/B\) selected cells.
Since the buffer groups are disjoint, these pairs account for
at most \(M_v\) selected cells.
For each remaining pair, choose an unmoved element of \(G_j\).
It still occupies its original cell, since no buffer element
has been deleted.
Consider the current array segment from this fixed cell to
the rightmost original cell of \(D_j\).
This segment contains all the selected cells of \(D_j\),
even though those cells need not be consecutive.
Moreover, the segments share no array edge: all original cells
of \(D_j\) precede all original cells of \(G_{j+1}\).

Suppose such a segment contains a value outside \(I_v\).
Follow the consecutive array edges from the fixed buffer cell
to the first value outside \(I_v\).
The value intervals of all preceding edges lie inside \(I_v\);
the last edge contributes the distance from its inside endpoint
to the boundary crossed.
By the triangle inequality, their total contribution to
\(V_{I_v}(A)\) is at least the distance from the buffer's value
to that boundary.
This distance is at least \(|I_v|/4\), because the buffer's
value lies in the middle half of \(I_v\).
Since the segments share no array edge, at most
\(4V_{I_v}(A)/|I_v|\) of them can contain an outside value.
Each contains \(N_v/B\) selected cells, so these segments
account for at most
$\frac{4N_v}{B|I_v|}V_{I_v}(A)$
selected cells.

Every remaining selected cell currently contains a value
in \(I_v\).
The array is full, and the only newly inserted values are
copies of \(1\), which lies outside \(I_v\).
Its occupant is therefore an original element belonging
to a target or buffer child interval.
A target occupant is counted by \(U_v\).
A buffer occupant must have moved: at the first visit,
this cell held a target element, so the buffer element
occupied a different cell.
It is therefore counted by \(M_v\).
Distinct cells have distinct occupants, so there are at most
\(U_v+M_v\) remaining selected cells.

Combining these three bounds gives
\[
    \frac{rN_v}{B}
    \le
    M_v+\frac{4N_v}{B|I_v|}V_{I_v}(A)+(U_v+M_v).
\]
Using \(r\ge B/64\) and \(M_v<\varepsilon N_v\), we obtain
\[
    U_v+\frac{4N_v}{B|I_v|}V_{I_v}(A)
    \ge
    \left(\frac1{64}-2\varepsilon\right)N_v.
\]
With our choice \(\varepsilon=1/1024\), this gives
\[
    U_v+\frac{N_v}{B|I_v|}V_{I_v}(A)
    \ge \frac{7N_v}{2048}
    \ge \frac{N_v}{512}.
\]
At least one of the two terms must therefore be at least
\(N_v/1024\). Hence,
\[
    U_v\ge\frac{N_v}{1024}
    \qquad\text{or}\qquad
    V_{I_v}(A)\ge\frac{B|I_v|}{1024}.
\]
This proves the lemma with \(\delta=1/1024\).
\end{proof}

We next bound the number of values remaining at the end
of the forward phase in terms of the algorithm's cost.
Note that when terminating, the adversary may leave elements protected by buffer children
of unpaid nodes.
To bound this remaining population, we will use \Cref{lem:unpaid-forcing}

\begin{lemma}[Stopping lemma]
\label{lem:stopping}
For a \(C\)-competitive algorithm, the number \(S\) of
elements remaining at the end of the forward phase satisfies
$S\le\frac{2nC}{\delta^d B}$.
\end{lemma}

\begin{proof}
Let \(S_v\) be the number of original elements in \(I_v\)
remaining at the end of the forward phase.
Thus, \(S=S_{\mathrm{root}}\).
Recall that \(N_v=n|I_v|\) and \(0<\delta\le1/2\).
We call a child available if it is labeled as a target or released.
At termination, every available child of a visited interval
containing surviving original elements has been visited.
Moreover, every visited leaf has no surviving original element.

We first show how to bound \(S_v\) using either the cost through
\(I_v\) or the surviving populations of its available children.
Suppose that \(I_v\) is unpaid.
If the cost alternative of \Cref{lem:unpaid-forcing} holds, then
\[
    S_v\le N_v=n|I_v|
    \le\frac{n}{\delta B}V_{I_v}(A).
\]
Otherwise, the lemma gives \(U_v\ge\delta N_v\), and hence
\[
    S_v\le N_v\le\frac{U_v}{\delta}
    =
    \frac1\delta
    \sum_{w\text{ available child of }v}S_w.
\]
The equality holds because the available children of an unpaid
interval are precisely its target children.
Thus, when the cost is small, it suffices to bound the surviving
populations in these children.

If \(I_v\) is paid, all its children are available, so
\[
    S_v=\sum_{w\text{ available child of }v}S_w.
\]
Combining the paid and unpaid cases, and using \(\delta<1\),
every visited internal interval satisfies
\[
    S_v\le
    \frac1\delta
    \sum_{w\text{ available child of }v}S_w
    +\frac{n}{\delta B}V_{I_v}(A).
\]

We now expand this inequality from the root down to depth \(d\).
Available children that have not been visited contribute zero
surviving elements, so every nonzero population term can be
expanded. This gives
\[
    S\le
    \frac1{\delta^d}
    \sum_{\substack{v\text{ visited}\\\text{at depth }d}}S_v
    +
    \frac nB
    \sum_{j=0}^{d-1}\frac1{\delta^{j+1}}
    \sum_{\substack{v\text{ visited}\\\text{at depth }j}}
    V_{I_v}(A).
\]
The first term is zero because every visited interval at
depth \(d\) is a leaf with no surviving original element.
At each depth, the value intervals are disjoint.
By \Cref{cl:interval-cost}, their cost contributions therefore
sum to at most \(\operatorname{cost}(A)\le C\).
Consequently,
\[
    S
    \le\frac{nC}{B}\sum_{j=0}^{d-1}\delta^{-(j+1)}
    =\frac{nC}{B}\cdot\frac{\delta^{-d}-1}{1-\delta}
    \le\frac{2nC}{\delta^d B},
\]
where the last inequality uses \(\delta\le1/2\).
\end{proof}

Recall that $C \le O(n)$. Therefore the choice below is always possible
\begin{corollary}
\label{size}
By choosing \(B\le n\) to be the smallest multiple of \(16\)
satisfying \(B\ge 8C/\delta^d\) we have 
\begin{enumerate}

\item 
\(
    S\le\frac{n}{4}\ .
\)
\item
\(
    \log n
    =d\log B
    =O\bigl(d\log C+d^2\bigr)
\)
which implies that
\(
    d
    =
    \Omega\!\left(
        \frac{\log n}{\log C+\sqrt{\log n}}
    \right).
\)
\end{enumerate}

\end{corollary}
Next, we show that, at every internal depth, at least half of the original elements belong to paid nodes.
Recall that \(N_v\) counts the original elements below \(v\), and
\(S\) counts those still present at the end of the forward phase.
The following lemma uses the fact that unpaid nodes retain their
outer buffer elements to show that paid nodes account for at least
half of the original elements at each internal depth.

\begin{claim}
\label{cl:paid-mass}
If \(S \le n/4\), then $\sum_{v\text{ paid}}N_v\ge\frac{d\cdot n}{2}$.
\end{claim}

\begin{proof}
Fix an internal depth \(\ell\in\{0,\ldots,d-1\}\).
Every node \(v\) at this depth that never became paid retains
at least \(N_v/2\) of its original elements.

Indeed, if \(v\) was never visited, none of its original elements
was replaced, since replacing an element below \(v\) requires
the search to visit \(v\).
All \(N_v\) elements therefore remain.
If \(v\) was visited but remains unpaid, its outer \(B/2\) children
remain labeled as buffers.
The adversary never searches these children, so their
\(N_v/2\) original elements all remain.

The sets of original elements belonging to nodes at depth \(\ell\)
are disjoint.
Consequently,
\[
    S\ge
    \frac12
    \sum_{\substack{v\text{ at depth }\ell\\v\text{ not paid}}}N_v.
\]
which means that
\[
\sum_{\substack{v\text{ at depth }\ell\\v\text{ not paid}}}N_v
\le 2S \le \frac{n}{2}.
\]
The nodes at this depth partition the \(n\) original
elements. Hence,
\[
    \sum_{\substack{v\text{ at depth }\ell\\v\text{ paid}}}N_v
    =
    n- \sum_{\substack{v\text{ at depth }\ell\\v\text{ not paid}}}N_v
    \ge\frac{n}{2}.
\]
Summing over the \(d\) internal depths gives
\[
    \sum_{v\text{ paid}}N_v\ge\frac{d n}{2}.
\]
\end{proof}

Each paid interval \(I_v\) requires at least \(\varepsilon N_v\)
distinct buffer elements to have moved after its children were
labeled.
To use the preceding bound on \(\sum_{v\text{ paid}}N_v\),
we must account for possible overlap between these elements.
The proof below shows that the payments can be charged without
counting the same movement twice.
Here, movements of the same element in different updates
count separately toward recourse.

\begin{lemma}
\label{lem:payment-additivity}
In every round, the total recourse \(R\) during the forward
phase satisfies $R\ge\frac{\varepsilon dn}{2}$.
Consequently, repeating the rounds forces
\(\Omega(d)\) amortized recourse.
\end{lemma}

\begin{proof}
For each paid node \(v\), at least
\(\lceil\varepsilon N_v\rceil\) distinct original buffer elements
were moved after its labels were assigned and no later than
the update in which it became paid.
Choose \(\lceil\varepsilon N_v\rceil\) of these elements,
and charge \(v\) to the first movement of each chosen element
after its labels were assigned.

We show that no movement is charged to two distinct nodes.
If neither node is an ancestor of the other, their sets of
original elements are disjoint, so their charges concern
different elements.

It remains to consider two nodes \(v,w\), where \(v\) is
an ancestor of \(w\).
Suppose first that \(v\) was still unpaid when \(w\) was
first visited.
At that time, the search could descend from \(v\) only through
a target child.
All original elements below \(w\) therefore belong to a target
child of \(v\).
In particular, no element charged by \(w\) is a buffer element
of \(v\), so the two nodes cannot charge the same movement.

Suppose instead that \(v\) was already paid when \(w\) was
first visited.
Every movement charged to \(v\) occurred no later than
the update in which \(v\) became paid.
The labels at \(w\) were assigned after that update, and \(w\)
counts only movements in subsequent updates.
Thus, even if the two nodes charge movements of the same element,
those movements occur in different updates.

Each movement consequently receives at most one charge.
Since each paid node \(v\) receives
\(\lceil\varepsilon N_v\rceil\) charges, we obtain
\[
    R\ge
    \sum_{v\text{ paid}}\lceil\varepsilon N_v\rceil
    \ge\varepsilon\sum_{v\text{ paid}}N_v
    \ge\frac{\varepsilon dn}{2},
\]
where the last inequality follows from \Cref{cl:paid-mass}.
\end{proof}

We complete the proof of the lower bound plugging by the above lemma into \Cref{size}.


\section{Online (Insertion-Only) Sorting}\label{sec:insertion-only}\newcommand{\lrbg}{{\sc lrbg}\xspace}
\newcommand{\hrbg}{{\sc hrbg}\xspace}
\newcommand{\apv}{{\sc apv}\xspace}

In this section, we consider the {\em online sorting} problem—the insertion-only setting of the dynamic sorting problem studied in the previous section, which can also be looked at as \emph{online sorting with recourse}. It is known that the best competitive ratio achievable for online sorting {\em without recourse} is $\Theta(\sqrt{n})$. Our main result is to show that the competitive ratio can be exponentially improved to $O(\log n)$ using $O(1)$ amortized recourse per element. More generally, we show the following: 
\begin{theorem}\label{thm:online}
    There are deterministic algorithms for the online sorting problem that achieve the following tradeoffs:
    \begin{itemize}
        \item[-] competitive ratio of $O\left(R\cdot \log^{1/R} n\right)$ with recourse $R \in [O(1), O(\log\log n)]$, and
        \item[-] competitive ratio of $C\in [O(1), O(\log\log n)]$ with recourse $O(C \cdot \log^{1/C} n)$.
    \end{itemize}
\end{theorem}    

\begin{remark}
Our result assumes, as done in many papers on this topic, that the optimal cost of the complete
insertion sequence is known to the algorithm. For example,
this holds when all values lie in $[0,1]$ and both $0$ and $1$
appear in the sequence, so that $OPT=1$. This assumption
is required by the algorithm of Azar et al.~\cite{AzarPV26},
which we use as a subroutine.
\end{remark}

\begin{figure}
\centering

\begin{tikzpicture}[scale=0.6]

\def\a{3}

\draw[->, thick] (0,0) -- (11,0);
\draw[->, thick] (0,0) -- (0,8);

\node at (5.5,-0.8) { Recourse};
\node[rotate=90] at (-1.2,4)
    {Competitive Ratio};

\draw[blue, very thick]
    plot[smooth] coordinates {
        (0.5,7.0)
        (1.0,5.5)
        (1.7,4.3)
        (2.4,3.5)
        (\a,\a)
    };

\draw[red, very thick, domain=\a:10, samples=150]
    plot ({\x}, {\a/(1 + ln(\x/\a))});

\fill (0.5,7.0) circle (2.5pt);
\fill (\a,\a) circle (2.5pt);
\fill (10,{ \a/(1 + ln(10/\a)) }) circle (2.5pt);

\node[above right=3pt] at (0.5,7.0)
    {$\bigl(O(1),\,O(\log n)\bigr)$};

\node[above=10pt,right = 5pt] at (\a,\a)
    {$\bigl(O(\log\log n),\,O(\log\log n)\bigr)$};

\node[right=5pt] at (10,{ \a/(1 + ln(10/\a)) })
    {$\bigl(O(\log n),\,O(1)\bigr)$};

\end{tikzpicture}
\caption{The tradeoff between competitive ratio and recourse achieved by \Cref{thm:online}.}
\label{fig:tradeoff}
\end{figure}
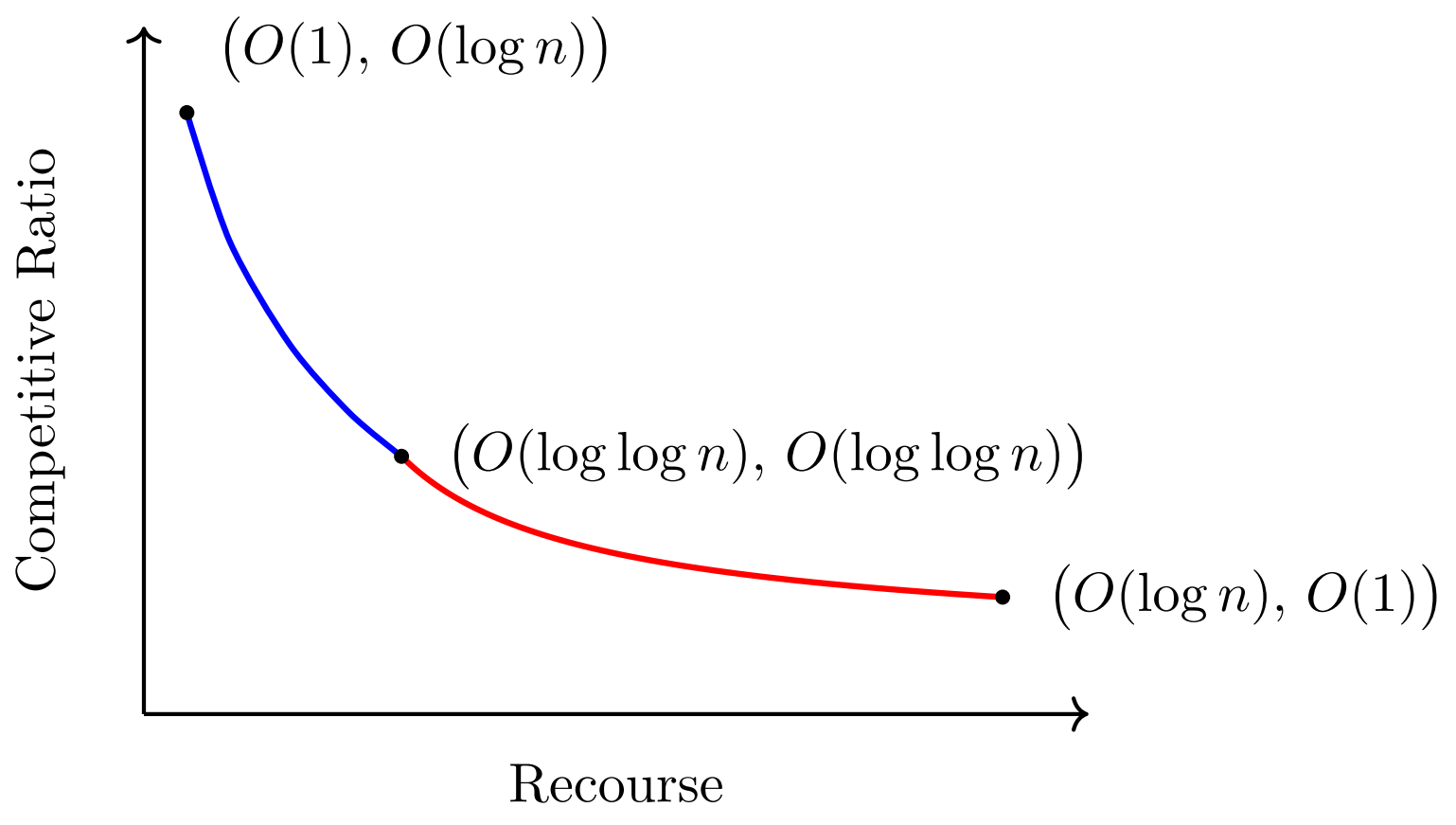

The tradeoff achieved by the above theorem is shown in \cref{fig:tradeoff}. In particular, we get the following bounds:
    \begin{itemize}
        \item[-] A competitive ratio of $O(\log n)$ with $O(1)$ amortized recourse per element. This is obtained by setting $R = O(1)$ in the first tradeoff.
        \item[-] A competitive ratio of $O(\log\log n)$ with $O(\log\log n)$ amortized recourse per element. This can be obtained by either setting $R = O(\log \log n)$ in the first tradeoff or $C = O(\log\log n)$ in the second tradeoff.
        \item[-] A competitive ratio of $O(1)$ with $O(\log n)$ amortized recourse per element. This is obtained by setting $C = O(1)$ in the second tradeoff.
    \end{itemize}

The two tradeoffs in \Cref{thm:online} are achieved by two different data structures:
\begin{itemize}
    \item[-] {\em Low-Recourse Block Gadget} (\lrbg): This data structure takes as input a sequence of $k$ sorted blocks of possibly unequal size, and returns an array of cost $O(b\cdot \log_b k)$ using $O(\log_b k)$ recourse per element, for some parameter $b$. We describe this gadget in \Cref{sec:lrbg}.
    \item[-] {\em High Recourse Block Gadget} (\hrbg): Similar to \lrbg, this data structure takes as input a sequence of $k$ sorted blocks of possibly unequal size. However, it uses more recourse to achieve a lower cost bound than \lrbg: namely, it returns an array of cost $O(\log_b k)$ using $O(b\cdot \log_b k)$ recourse per element, for some parameter $b$. We describe this gadget in \Cref{sec:hrbg}. 
\end{itemize}

The \lrbg and \hrbg data structures operate on sorted blocks. To create these blocks from a sequence of inserted elements, we use a data structure from prior work of Azar, Panigrahi, and Vardi~\cite{AzarPV26}. The \apv data structure (see Section 6 in \cite{AzarPV26}) has the following property: for any $\beta \in [O(\log n), O(\log^2 n)]$, given an input sequence comprising $n$ elements, it inserts the elements in an array of size $O(\beta n \log n)$ such that the sum of absolute differences between elements in consecutive non-empty array locations is at most $O\left(\frac{\log n}{\beta}\right)$. The data structure does not use any recourse.    

We use the \apv gadget in multiple iterations on input sequences of various lengths in our data structure. For an input sequence of length $r$, we set $\beta := O(\log r)$. Furthermore, at the end of inserting $r$ elements, we re-sort these elements and make the array compact by inserting them in contiguous locations. This last step has overall recourse of $O(r)$, i.e., $O(1)$ recourse per element. Therefore, we have the following claim:
\begin{claim}\label{cl:apv}
    Given a sequence of $r$ elements and an empty array of size $O(r\log^2 r)$, the \apv algorithm can be used to sort the elements and return them in a contiguous sorted block of size $r$, while incurring $O(1)$ recourse per element. At any intermediate stage of this process, the cost of the elements is at most $O(1)$.
\end{claim}

We now describe the use of \Cref{cl:apv} to create sorted blocks that are input to the \lrbg and \hrbg gadgets. The overall algorithm comprises a set of phases that process geometrically decreasing numbers of elements. In other words, the $i$th phase inserts $n_i := n/2^i$ elements. We further subdivide a phase comprising $n_i$ elements into $t = O(\log^2 n)$ subphases, each of which inserts $n_i/t$ elements. Each subphase is a separate invocation of the \apv gadget with $r := n_i/t$. Note that in such a subphase, the array has contiguous empty space of size at least $n_i = (n_i/t) \cdot t = \Omega(r \log^2 r)$, which satisfies the condition of \Cref{cl:apv}. Therefore, the \apv gadget returns the elements in a subphase in sorted order, incurring $O(1)$ recourse per element and at most $O(1)$ cost at any intermediate stage. 

We summarize the use of the \apv algorithm below:
\begin{claim}\label{cl:apv-summary}
    The invocations of the \apv algorithm produce a sequence of $O(\log^3 n)$ sorted blocks (of unequal size), have cost at most $O(1)$ at any intermediate stage within a block, and incur $O(1)$ amortized recourse per element. 
\end{claim}

\subsection{Low Recourse Block Gadget}\label{sec:lrbg}

We now describe the Low Recourse Block Gadget (with parameter $b$). It is convenient to first consider the special case where each sorted block is just a single element. In this case, the \lrbg data structure is defined via a virtual tree of arity $b$ with $k$ leaves that are identified with the $k$ array cells. The insertion algorithm is simple: the arriving element is inserted in the first empty cell in the array. There is only one additional operation to maintain the data structure: when all the leaves in any subtree of the virtual tree are full, the corresponding subarray is sorted at $O(1)$ recourse per element. 

To extend this data structure to sorted input blocks, we make a simple modification: each leaf in the virtual tree is no longer associated with a single array cell, but the $i$th leaf is associated with a subarray of size equal to the $i$th sorted block inserted into the data structure. The re-sorting steps are exactly as earlier, i.e., when all the leaves in a subtree are full, we sort all the elements in the corresponding subarrays at $O(1)$ recourse per element.

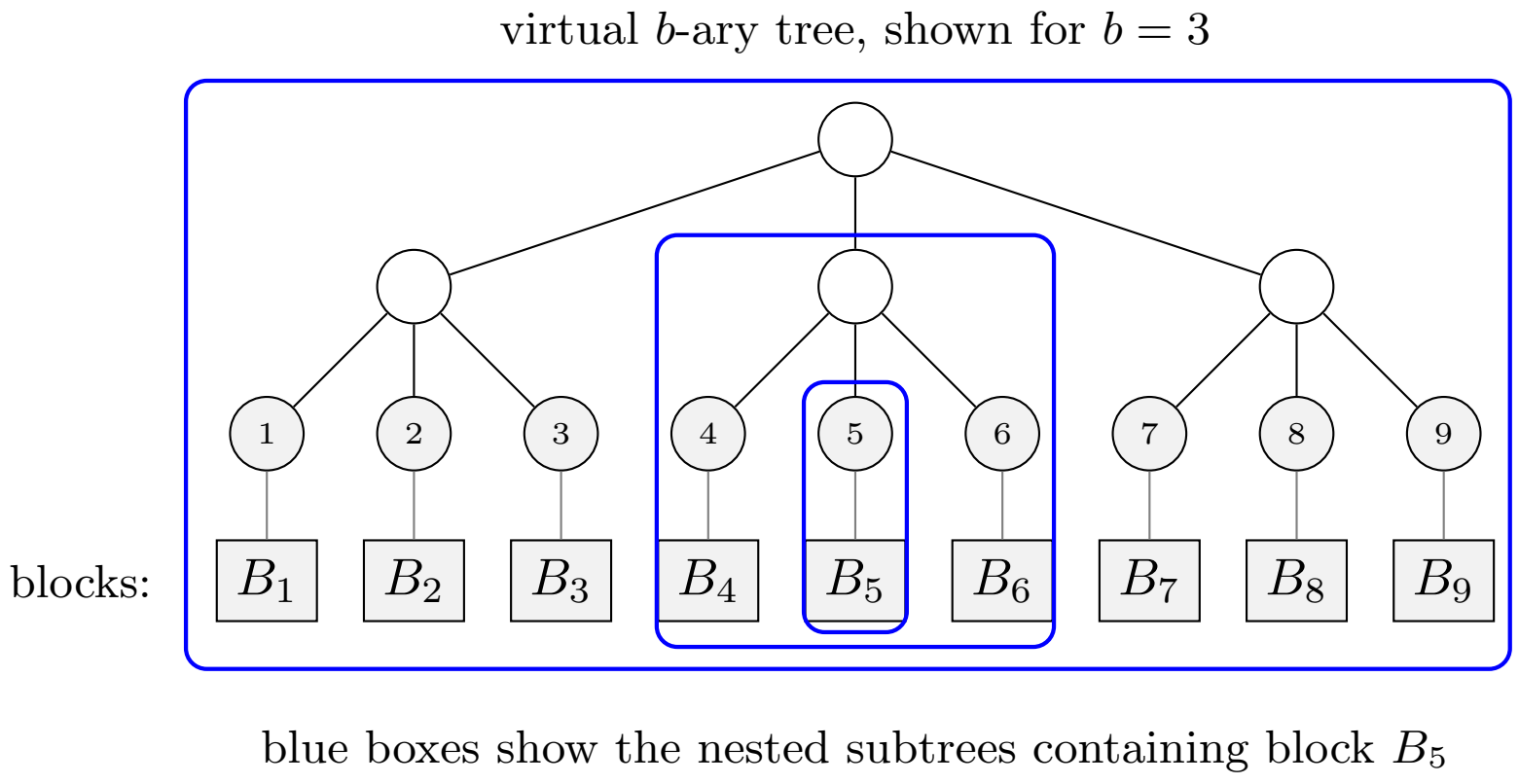
\begin{figure}[!h]
\centering
\begin{tikzpicture}[
    x=1cm,y=1cm,
    every node/.style={font=\small},
    internal/.style={circle,draw,minimum size=5mm,inner sep=0pt},
    leaf/.style={circle,draw,fill=gray!10,minimum size=5mm,inner sep=0pt},
    block/.style={draw,minimum width=6.5mm,minimum height=5mm,fill=gray!10},
    subtree/.style={draw=blue,thick,rounded corners},
    annot/.style={font=\footnotesize}
]

\node[internal] (r) at (4,3) {};
\node[internal] (a) at (1,2) {};
\node[internal] (b) at (4,2) {};
\node[internal] (c) at (7,2) {};

\draw (r)--(a);
\draw (r)--(b);
\draw (r)--(c);

\foreach \x/\i/\p in {
0/1/a, 1/2/a, 2/3/a,
3/4/b, 4/5/b, 5/6/b,
6/7/c, 7/8/c, 8/9/c}
{
    \node[leaf] (l\i) at (\x,1) {\tiny \i};
    \draw (\p)--(l\i);
    \node[block] (B\i) at (\x,0) {$B_{\i}$};
    \draw[gray] (l\i)--(B\i);
}

\node[annot,above=2mm of r] {virtual $b$-ary tree, shown for $b=3$};
\node[annot,left=3mm of B1] {blocks:};

\draw[subtree]
    ($(l5)+(-0.35,0.35)$)
    rectangle
    ($(B5)+(0.35,-0.35)$);

\draw[subtree]
    ($(b)+(-1.35,0.35)$)
    rectangle
    ($(B6)+(0.35,-0.45)$);

\draw[subtree]
    ($(r)+(-4.55,0.4)$)
    rectangle
    ($(B9)+(0.45,-0.60)$);

\node[annot,below=6mm of B5]
{blue boxes show the nested subtrees containing block $B_5$};

\end{tikzpicture}

\caption{
Low Recourse Block Gadget (\lrbg) for arity $b$, illustrated with $b=3$.
Each leaf of the virtual tree is associated with one sorted block $B_i$.
The blue boxes show the nested subtrees containing a particular block, here $B_5$:
the leaf subtree, its parent subtree, and the root subtree.
Whenever such a subtree becomes full, the corresponding contiguous interval of
blocks is re-sorted.
}
\label{fig:lrbg}
\end{figure}

We give an illustration of the \lrbg data structure in \cref{fig:lrbg}.

\begin{lemma}
\label{lem:lrbg}
    The Low Recourse Block Gadget with parameter $b$ inserts a sequence of $k$ sorted blocks (of arbitrary sizes) with $O(b \cdot \log_b k)$ cost and $O(\log_b k)$ amortized recourse per element. 
\end{lemma}
\begin{proof}
For the recourse bound, note that every element participates in at most $O(\log_b k)$ re-sorts, i.e., once for a subtree of each height. For the cost bound, call a subtree of the virtual tree active if its leaves are already full, but not all leaves in its parent subtree are full yet. Note that there are at most $b-1$ active subtrees at every height at any time. Moreover, the elements in any active subtree are sorted. Therefore, the total cost is at most $(b-1)\cdot O(\log_b k) = O(b\cdot \log_b k)$.
\end{proof}


\subsection{High Recourse Block Gadget}\label{sec:hrbg}

We now describe the High Recourse Block Gadget (with parameter $b$). As in the case of \lrbg, it will be convenient to first consider the special case of singleton input blocks. The data structure uses a parameter $L = O(\log_b k)$ to denote the number of levels. At any time, the elements inserted so far are organized into a set of sorted {\em super-blocks} that are contiguous from left to right in the array. There is exactly one super-block at each level $\ell\in \{L, L-1, \ldots, 0\}$ and these super-blocks are organized in decreasing order of levels from left to right in the array. We say that the super-block at level $\ell+1$ is the {\em parent} of the super-block at level $\ell$. The size of the super-block at level $\ell$ is not fixed: it can vary from being empty, i.e., size $0$, to having a maximum size of $k^{\ell/L}$. A super-block that is at its maximum size is said to be {\em full}. Eventually, the super-block at level $L$ is full and all other super-blocks are empty, i.e., the entire array of size $k$ is sorted.

The super-blocks are maintained using the following insertion subroutine. The incoming element is always inserted in the super-block at level $0$. Note that this super-block can hold at most $1$ element. By the subsequent operations that we will describe, this super-block will always be empty once we have completed the insertion of an element and are ready to insert the next element. Therefore, it is always feasible to insert the new element in this super-block.  We have just one additional maintenance operation: whenever a super-block becomes full, we move the elements in the super-block to its parent super-block and re-sort all the elements in the parent super-block at $O(1)$ recourse per element of the parent super-block. Note that these operations can cascade, i.e., we merge elements in the super-block at level $\ell$ into the super-block at level $\ell+1$, which makes it full and has to be merged into the super-block at level $\ell+2$, and so on. After merging the elements of a super-block into its parent, the super-block itself becomes empty. Therefore, the super-block at level $0$, which can only hold $1$ element, immediately becomes full on inserting a new element, gets merged into the super-block at level $1$ thereby becoming empty, and is ready to accept the next incoming element.

To extend this data structure to an insertion sequence comprising $k$ sorted blocks, we make only one change: the super-block at level $\ell$ can now accept at most $k^{\ell/L}$ sorted input blocks rather than $k^{\ell/L}$ elements. Extending this idea, the super-block becomes full once it has accepted $k^{\ell/L}$ sorted input blocks rather than $k^{\ell/L}$ elements. As earlier, once it becomes full, all its elements are merged into its parent super-block at level $\ell+1$ at $O(1)$ recourse per element of the parent super-block. As earlier, after the merger, the super-block at level $\ell$ becomes empty. This means that the maximum size of a super-block is no longer fixed, but this does not affect the operation of our data structure. 

\begin{figure}[!h]
\centering
\begin{tikzpicture}[
    every node/.style={font=\small},
    sb/.style={
        draw,
        rounded corners,
        minimum height=6mm,
        fill=gray!10
    },
    caplabel/.style={font=\footnotesize},
    arr/.style={->,thick}
]

\node[sb,minimum width=2.2cm] (L)   at (0,0)    {$S_L$};
\node[sb,minimum width=1.8cm] (Lm1) at (2.25,0) {$S_{L-1}$};
\node[sb,minimum width=1.3cm] (D)   at (4.00,0) {$\cdots$};
\node[sb,minimum width=1.2cm] (S1)  at (5.55,0) {$S_1$};
\node[sb,minimum width=0.8cm] (S0)  at (6.85,0) {$S_0$};

\node[caplabel] at ($(L.north)+(0,0.30)$) {$\le k$};
\node[caplabel] at ($(Lm1.north)+(0,0.30)$) {$\le k^{(L-1)/L}$};
\node[caplabel] at ($(D.north)+(0,0.30)$) {$\cdots$};
\node[caplabel] at ($(S1.north)+(0,0.30)$) {$\le k^{1/L}$};
\node[caplabel] at ($(S0.north)+(0,0.30)$) {$\le 1$};

\node[caplabel] at ($(L.north west)+(-0.75,0.30)$) {capacity};

\draw[arr]
    ($(S0.north)+(0,0.80)$)
    -- node[above,font=\footnotesize]
       {merge into parent when full}
    ($(L.north)+(0,0.80)$);

\node[draw,circle,fill=blue!10,minimum size=5mm,inner sep=0pt]
    (x) at ($(S0.south)+(0,-0.95)$) {$x$};

\draw[arr]
    (x.north) -- node[right,font=\footnotesize] {insert} (S0.south);

\draw[decorate,decoration={brace,mirror,amplitude=4pt}]
    (L.south west) -- (S0.south east)
    node[midway,below=6pt,font=\footnotesize]
    {array layout: $S_L,S_{L-1},\ldots,S_0$};

\end{tikzpicture}

\caption{
High Recourse Block Gadget (\hrbg).
The array consists of sorted super-blocks arranged from left to right in decreasing
level order. The capacity of level $\ell$ is $k^{\ell/L}$.
A new item is inserted into $S_0$.
Whenever a super-block becomes full, its contents are merged into its parent,
the parent is re-sorted, and the super-block becomes empty.
}
\label{fig:hrbg}
\end{figure}
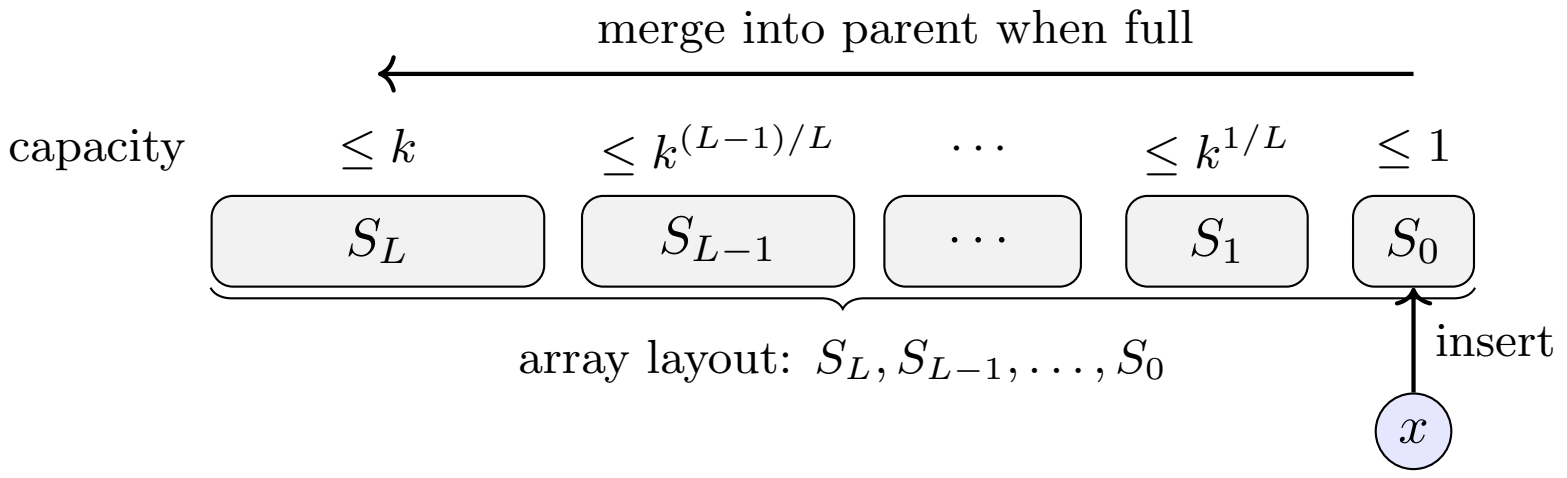

We give an illustration of the \hrbg data structure in \cref{fig:hrbg}.

\begin{lemma}
\label{lem:hrbg}
    The High Recourse Block Gadget with parameter $b$ inserts a sequence of $k$ sorted blocks (of arbitrary sizes) with $O(\log_b k)$ cost and $O(b\cdot \log_b k)$ amortized recourse per element. 
\end{lemma}
\begin{proof}
    Since the algorithm maintains $L+1 = O(\log_b k)$ sorted super-blocks at any stage (some of these can be empty), the cost is at most $O(\log_b k)$. To bound recourse, note that an element participates in a re-sort operation at any level $\ell+1$  every time the super-block at level $\ell$ becomes full. This can happen at most $k^{1/L}$ times, after which the super-block at level $\ell+1$ will be full and the element will move to the super-block at level $\ell+2$. Therefore, each element participates in at most $(L+1)\cdot k^{1/L} = O(b\cdot \log_b k)$ re-sort operations, and incurs $(L+1)\cdot k^{1/L} = O(b\cdot \log_b k)$ amortized recourse.
\end{proof}


\subsection{Putting it Together}

\begin{proof}[Proof of \Cref{thm:online}]
Our overall algorithm uses the sequence of sorted blocks produced by \apv (\Cref{cl:apv-summary}) as input to either \lrbg or \hrbg, i.e., $k = O(\log^3 n)$. For \lrbg, we set $\log_b k = 3R$, i.e.,
$b = k^{1/(3R)} = O(\log^{1/R} n)$, in
\Cref{lem:lrbg} to obtain the first tradeoff in \Cref{thm:online}, with $O(R)$ amortized recourse.
For \hrbg, we similarly set $\log_b k = 3C$, i.e.,
$b = k^{1/(3C)} = O(\log^{1/C} n)$, in
\Cref{lem:hrbg} to obtain the second tradeoff in
\Cref{thm:online}, with $O(C)$ competitiveness. These tradeoffs include the $O(1)$ cost and $O(1)$ amortized recourse  per element due to \apv as given in \Cref{cl:apv-summary}.
\end{proof}

\section{Online and Dynamic Sorting with $(1+\eps)n$-Space}\label{fully-dynamic-extraspace}In this section, we show that if the array has extra space, i.e., if the array is of size $m = (1+\eps)n$ for any fixed $\eps > 0$, and there are at most $n$ elements stored in it at any time, then the online and dynamic sorting problems are equivalent:

\begin{claim}\label{cl:extra-reduction}
    Suppose there is a (deterministic) algorithm for online sorting that inserts a sequence of $n$ elements in an array of size $n$ with cost $c$ and amortized recourse $r$ per element. Then, for any $\eps > 0$, there is a (deterministic) algorithm for dynamic sorting that maintains at most $n$ elements in an array of size $(1+\eps) n$ with cost $2c+1$ and amortized recourse $r+O(1/\eps)$ per element.
\end{claim}

\begin{remark}
We assume, as before, that the minimum and
maximum values are represented by two permanent elements
that cannot be deleted. In particular, we may assume that
all values lie in $[0,1]$ and that $0$ and $1$ remain
present throughout the execution.
\end{remark}

\begin{proof}
    The dynamic sorting algorithm re-sorts the elements after every $\eps n$ steps (element insertions or deletions), and compacts them in the array leaving an empty subarray of size $\eps n$. Between any two re-sorts, element insertions execute the online sorting algorithm on the empty subarray of size $\eps n$. This is valid because there can be at most $\eps n$ insertions between two re-sorts. All element deletions are postponed to the next re-sort, at which time the deleted elements are not included in the sorted array. In other words, although an element is actually removed from the array on deletion, the online sorting algorithm (which cannot handle element deletions) continues assuming that the deleted element is still in the array. 

    The algorithm has a cost of $c$ on the part of the array where elements are being inserted, and a cost of $1$ between the previously re-sorted subarray and this subarray where insertions are being made. For the recourse bound, note that the only additional recourse is $O(n)$ for every re-sort, which can be charged to the $\eps n$ steps between every two consecutive re-sorts.
\end{proof}

Using \Cref{cl:extra-reduction} and \Cref{thm:online}, we get:
\begin{theorem}
    For any $\eps > 0$, there are deterministic algorithms for dynamic sorting that maintain at most $n$ elements in an array of size $(1+\eps)n$ with the following tradeoffs:
        \begin{itemize}
        \item[-] competitive ratio of $O\left(R\cdot \log^{1/R} n\right)$ with recourse $R + O(1/\eps)$, where $R \in [O(1), O(\log\log n)]$, and
        \item[-] competitive ratio of $C\in [O(1), O(\log\log n)]$ with recourse $O(C \cdot \log^{1/C} n + 1/\eps)$.
    \end{itemize}
\end{theorem}

In particular, this implies that we can match the tradeoffs listed after \Cref{thm:online}---$O(\log n)$-competitiveness with $O(1)$ recourse, $O(\log\log n)$-competitiveness with $O(\log\log n)$ recourse, and $O(1)$-competitiveness with $O(\log n)$ recourse---even in the fully-dynamic setting, provided the array is of size $(1+\eps)n$. This represents a substantial improvement in the competitive ratio over the best-known bound without recourse even in the insertion-only setting, that of $O\left(\frac{\log^2 n}{\eps}\right)$ obtained by \cite{AzarPV26}.

\section{Closing Remarks}\label{sec:closing}In this paper, we studied online sorting with recourse, focusing on the fully dynamic setting where elements may be inserted and deleted without additional space. Our main result is a deterministic algorithm that achieves an $O(\log^2 n)$ competitive ratio with $O(\log^4 n)$ amortized recourse per update. 
The main ingredient is a recursive data structure that allows limited disorder to support efficient updates. We also complement this upper bound with a lower bound showing that for deterministic algorithms, the recourse bound cannot be improved beyond the exact exponent in the poly-logarithmic factor. Our results for insertion-only sorting and fully dynamic sorting with additional space provide stronger tradeoffs, thereby showing that these problems are strictly easier than the full-dynamic problem.

Several questions remain open. Most importantly, we would like to understand the optimal tradeoff between competitiveness and recourse when the array has exactly $n$ cells. In particular, can we obtain an $O(1)$ competitive ratio with poly-logarithmic amortized recourse? Our lower bound shows that $O(1)$-competitiveness requires $\Omega(\sqrt{\log n})$ amortized recourse for deterministic algorithms, but considerable gaps remain between the known upper and lower bounds.

Another direction is to obtain worst-case rather than amortized recourse guarantees. Our algorithm relies on rebuilding operations that may move many elements in a single update, and it is unclear whether these operations can be distributed over time without additional space or a significant loss in competitiveness. Finally, it would be interesting to extend our techniques to fully dynamic online metric TSP~\cite{AbrahamsenB0K024,Bertram25,AzarPV26-icalp}. Our construction relies heavily on linear ordering of the elements, and understanding which of its ideas extend to general metrics remains an interesting challenge.

\section*{AI Methodology}

We used ChatGPT to generate figures (some of which were
subsequently edited), assist with proof checking, and
improve the writing and presentation of the paper.
In terms of the results themselves, the proof of the fully dynamic lower
bound was developed through multiple rounds of interaction with ChatGPT.
The remaining results and underlying proof arguments
were developed by the authors.
The authors take full responsibility for the correctness
and contents of the paper.

\bibliographystyle{alphaurl}\bibliography{main}

\appendix

\section{Omitted Proof from \Cref{sec:preliminaries}} \label{appendix:prelim}
\begin{proof}[Proof of \Cref{lem:subsequence-crossings}]
Write $S=(x_1,\ldots,x_k)$ and
$S'=(x_{p_1},\ldots,x_{p_m})$, where
$p_1<\cdots<p_m$.

Fix a threshold $z$, and let
$b_j=\mathbf{1}[x_j>z]$.
For every pair of consecutive elements in $S'$,
\[
    |b_{p_{r+1}}-b_{p_r}|
    \le
    \sum_{j=p_r}^{p_{r+1}-1}|b_{j+1}-b_j|.
\]
Since these index ranges are disjoint, summing
over $r=1,\ldots,m-1$ gives
\[
    cross_z(S')
    \le \sum_{j=1}^{k-1}|b_{j+1}-b_j|
    =cross_z(S).
\]

Now suppose $\operatorname{cross}_z(S)\le C$ for every $z$.
By \eqref{eq:cost-crossing},
\[
    \operatorname{cost}(S')
    =\int_{\min S'}^{\max S'}\operatorname{cross}_z(S')\,dz
    \le C(\max S'-\min S')
    =C\cdot OPT(S').
\]
For sequences of at most one element, the
conclusion holds trivially.
\end{proof}

\section{Omitted Proofs from \Cref{sec:new-dst}} 
\begin{proof}[Proof of \Cref{cl:new-dst-minimum}]
We prove the claim by induction on $i$.
For $i=1$, the \dst\ is a sorted array, so the claim
is immediate.

Suppose $i>1$. If the \dst\ consists of a single
block with original section $O$, the claim follows
directly from the induction hypothesis applied to $O$.

Otherwise, write
\[
    S=O_1\mid T_2\mid O_2\mid O_3.
\]
By the Block Ordering invariant,
\[
    O_1<O_2<T_2,\qquad O_2<O_3.
\]
Thus, every element of $O_1$ is smaller than every
element outside $O_1$, so the minimum of $S$ lies
in $O_1$.

Since $O_1$ is an $(i-1)$-level \dst, the induction
hypothesis implies that its minimum occupies its
leftmost cell, which is also the leftmost cell of
$S$. This completes the induction.
\end{proof}

\begin{proof}[Proof of \Cref{cl:new-dst-maximum}]
We prove the claim by induction on $i$.
For $i=1$, the \dst\ is a sorted array, so the claim
is immediate.

Suppose $i>1$. If the \dst\ consists of a single
block with original section $O$, the claim follows
directly from the induction hypothesis applied to $O$.

Otherwise, write
\[
    S=O_1\mid T_2\mid O_2\mid O_3.
\]
By the Block Ordering invariant,
\[
    O_1<O_2<T_2,\qquad O_2<O_3.
\]
Thus, every element of $O_1$ and $O_2$ is smaller
than every element of $O_3$.

It remains to consider $T_2$. If $O_3$ is a
level-$1$ section, the ordering invariant gives
$T_2<O_3$. Otherwise, by the Block Sizes invariant,
$O_3$ consists of three blocks. Writing
\[
    O_3=O'_1\mid T'_2\mid O'_2\mid O'_3,
\]
the ordering invariant gives $T_2<O'_2$.
Since $O'_2$ is nonempty, the maximum element
of $S$ must belong to $O_3$.

By the induction hypothesis, the maximum of
$O_3$ occupies its rightmost cell, which is also
the rightmost cell of $S$. This completes the
induction.
\end{proof}

\begin{proof}[Proof of \Cref{cl:dst-cost}]
We prove the crossing bound by induction on $i$.

For $i=1$, the \dst\ is a sorted array and therefore
crosses any threshold at most once.

Suppose $i>1$. If $S$ consists of a single block,
the claim follows from the induction hypothesis.
Otherwise, write
\[
    S=O_1\mid T_2\mid O_2\mid O_3.
\]

Fix a threshold $z$. We first bound the crossings
within the four sections, and then account for the
crossings between them.

Since $O_1<O_2<O_3$, at most one of these three
original sections can contain elements on both sides
of $z$. Thus, if $T_2$ does not cross $z$, the
induction hypothesis bounds the total number of
crossings within the sections by $4(i-1)$.

Suppose instead that $T_2$ crosses $z$.
Since $O_1<O_2<T_2$, both $O_1$ and $O_2$ lie
entirely below $z$. The block ordering invariant
also ensures that $O_3$ crosses $z$ at most once.

Indeed, if $O_3$ is a level-$1$ section, then
$T_2<O_3$, so all its elements lie above $z$.
Otherwise, writing
\[
    O_3=O'_1\mid T'_2\mid O'_2\mid O'_3,
\]
the terminal ordering gives
\[
    O'_1<T_2<(T'_2\cup O'_2\cup O'_3).
\]
Hence $O_3$ consists of a prefix below $z$
followed by a suffix above $z$, and therefore
crosses $z$ at most once.

In either case, the total number of crossings
within the four sections is at most
$4(i-1)+1$.

Finally, there are at most three boundaries
between consecutive sections, each contributing
at most one additional crossing. Therefore,
\[
    \operatorname{cross}_z(S)\le 4(i-1)+1+3=4i.
\]

If all elements lie in an interval $I$, no
threshold outside $I$ is crossed. Integrating
the bound using \eqref{eq:cost-crossing} gives
\[
    \operatorname{cost}(S)
    =\int_I \operatorname{cross}_z(S)\,dz
    \le 4i|I|.\qedhere
\]
\end{proof}

\begin{proof}[Proof of \Cref{cl:add-max-invariants}]
Let $D$ be the \dst\ on which \addmax\ is invoked.
The operation fills its last cell with an element
$y$ larger than every element already stored in $D$.
This cell is also the last cell of every \dst\
along the chain of last original sections of $D$.
Since $y$ is larger than all elements in these
sections, by \Cref{cl:new-dst-maximum}, the insertion
is consistent with the position of the maximum
element at every recursive level.

We verify each invariant separately.

\begin{itemize}
    \item \textbf{Number and Structure of Blocks.}
    No section boundary or recursive structure is
    changed, so the invariant is preserved.

    \item \textbf{Base-Level Sorting.}
    Since $y$ is larger than all existing elements,
    filling the last cell of the level-$1$ section
    preserves its sorted order.

    \item \textbf{Block Sizes.}
    The operation fills an existing hole without
    changing cell ownership. Thus, all block sizes
    remain unchanged.

    \item \textbf{Block Ordering}.
    The new element is inserted along the chain
    of last original sections. At each three-block
    level, it is added to $O_3$, and since $y$ is
    larger than every existing element, the ordering
    $O_2<O_3$ remains valid.

    All other ordering constraints involve
    sections whose stored elements are unchanged,
    or require elements of the last original
    section to be larger than those of a preceding
    section. These constraints are therefore
    preserved as well.
\end{itemize}

Thus, \addmax\ preserves all structural invariants
within $D$.
\end{proof}

\begin{proof}[Proof of \Cref{cl:add-max-recourse}]
The operation \addmax only places $y$ in the
existing hole, without moving any other element.
Thus, its recourse is at most one.
\end{proof}

\begin{proof}[Proof of \Cref{cl:rebuild-invariants}]
We prove the claim by induction on the level of
the rebuilt section. For a level-$1$ section,
\rebuild\ produces a sorted array, so all invariants
hold.

Consider a level-$(i+1)$ section $D$, where $i\ge1$.
\begin{itemize}
   
\item \textbf{Block Sizes.}
We first establish the size bound needed for rebuilding.
Since higher-level size violations are handled first,
the block containing $D$ satisfies its size bound.
This block belongs to a level-$(i+2)$ section, so its
size is at most $12\cdot 2^{i+1}-1$.
Consequently,
\[
    |D|\le 24\cdot 2^i-1\le25\cdot 2^i-2.
\]

We now verify that rebuilding produces valid block sizes,
distinguishing the two rebuilding cases.

\emph{Repartition.}
If $|D|\le11\cdot 2^i$, we construct a single block,
whose original section contains fewer than
$12\cdot 2^i$ cells.

Otherwise, we construct three blocks.
The middle block contains exactly $2^i$ cells,
while the last and first blocks contain
\[
    L=\max\left\{
        8\cdot 2^i,
        \left\lceil\frac{|D|-2^i}{2}\right\rceil
    \right\},
    \qquad
    F=|D|-2^i-L
\]
cells, respectively.

Since $11\cdot 2^i<|D|\le25\cdot 2^i-2$,
these sizes satisfy
\[
    2^{i+1}\le F<12\cdot 2^i,
    \qquad
    8\cdot 2^i\le L<12\cdot 2^i.
\]
Hence all three blocks satisfy their required size bounds.

\emph{Cleanup.}
In this case, the existing three block sizes are preserved.
They already satisfy their bounds, since repartition
takes precedence whenever a size violation occurs.

In both cases, every recursive level-$i$ section
contains fewer than $12\cdot 2^i$ cells.
Since
\[
    12\cdot 2^i-1
    \le 25\cdot 2^{i-1}-2,
\]
the same capacity bound holds at the next recursive
level. The block-size invariants of these sections
therefore follow by induction.

\item \textbf{Number of Blocks.}
The rebuilding creates either a single block or
three blocks, as required. By induction, the
recursive sections also satisfy this invariant.

\item \textbf{Block Ordering.}
The stored elements are sorted before being
distributed among the new original sections.
Consequently, the original sections are ordered
by value. Since all transfer sections are empty,
every ordering constraint involving a transfer
section within $D$ holds trivially.
The recursive ordering invariants follow from
the induction hypothesis.

\item \textbf{Base-Level Sorting.}
Every level-$1$ section is reconstructed as a
sorted array, so this invariant follows directly
from the construction.

Finally, rebuilding preserves the total number
of cells and, if a hole is present, keeps it at
the last position of $D$. The hole is included
in the block-size calculations and does not
affect the ordering of stored elements.

\end{itemize}
Hence all internal structural invariants of
the rebuilt section are satisfied.

\end{proof}

\begin{proof}[Proof of \Cref{cl:rebuild-recourse}]
For $i=1$, the section contains at most $24$
cells, so rebuilding incurs $O(1)$ recourse.

For $i>1$, the capacity condition guarantees
that the section contains $O(2^i)$ cells.
Since \rebuild\ sorts and rebuilds the entire
section, moving each stored element at most once,
its recourse is at most the number of stored
elements, which is $O(2^i)$.
\end{proof}

\begin{proof}[Proof of \Cref{cl:swap-hole-invariants}]
The \swaphole\ operation transfers an empty cell
between the original and transfer sections of a
middle block. We verify the four structural
invariants separately.

\begin{itemize}
    \item \textbf{Number and Structure of Blocks.}
    The boundary change does not create or remove
    blocks or change their types. Any subsequent
    rebuilding preserves this invariant by
    \Cref{cl:rebuild-invariants}.

    \item \textbf{Base-Level Sorting.}
    Since no stored element moves during the
    boundary change, all level-$1$ sections
    remain sorted. Rebuilding also preserves
    this invariant by \Cref{cl:rebuild-invariants}.

    \item \textbf{Block Sizes.}
    The transferred cell changes the sizes of
    sections along the affected recursive paths,
    but the total size of the containing middle
    block remains unchanged.

    We examine the affected sections from highest
    level to lowest. Whenever a section requires
    repair, we rebuild either that section or
    its three-block parent, according to the
    rules of \swaphole.

    In the latter case, the parent was examined
    first and already satisfies its size bounds.
    Thus, \rebuild\ preserves its top-level
    block sizes. In either case,
    \Cref{cl:rebuild-invariants} guarantees that
    rebuilding restores the required size bounds
    within the rebuilt section.

    Repeating this procedure restores the
    Block Sizes invariant throughout the \dst.

    \item \textbf{Block Ordering.}
    The boundary change does not move any stored
    element, so all ordering constraints remain
    satisfied. By \Cref{cl:rebuild-invariants},
    rebuilding also restores the ordering
    constraints within the rebuilt section.

    The only difficulty arises when a section
    $S'$ requiring repair is the last original
    section of a three-block section
    \[
        S=O_1\mid T_2\mid O_2\mid S'.
    \]
    Rebuilding $S'$ may change the elements in
    its first original section, potentially
    violating their ordering with $T_2$.

    To avoid this, we rebuild $S$ instead.
    This empties $T_2$, eliminating the affected
    ordering constraint.

    We must also verify that rebuilding $S$
    preserves ordering constraints at higher
    levels. Suppose $S$ is itself the last
    original section of a higher-level
    three-block section
    \[
        \hat{S}=\hat{O}_1\mid\hat{T}_2
        \mid\hat{O}_2\mid S.
    \]
    When $\hat{T}_2$ is nonempty, the Block
    Ordering invariant requires
    \[
        O_1<\hat{T}_2<
        (T_2\cup O_2\cup S').
    \]
    Hence, it suffices to show that rebuilding
    $S$ preserves the elements belonging to $O_1$.

    Since $S$ was examined before $S'$, its
    middle original section $O_2$ is nonempty.
    Moreover, the hole is in $S'$ or outside
    $S$, so neither $O_1$ nor $O_2$ contains
    the hole. By the Block Ordering invariant,
    \[
        O_1<O_2,\qquad
        O_2<T_2,\qquad
        O_2<S'.
    \]
    Thus, $O_1$ contains precisely the smallest
    $|O_1|$ elements of $S$.

    Since the block sizes of $S$ are preserved
    during rebuilding, its first original
    section retains the same elements.
    Consequently, the ordering constraint
    involving $\hat{T}_2$ remains satisfied.

    Thus, rebuilding $S$ eliminates the ordering
    constraint involving $T_2$ without violating
    any ordering constraints at higher levels.
\end{itemize}

Therefore, all four structural invariants
are preserved.
\end{proof}

\begin{proof}[Proof of \Cref{cl:down-hole-invariants}]
We prove the claim by induction on $i$.
Consider an $i$-level \dst
\[
    S=O_1\mid T_2\mid O_2\mid O_3,
\]
with a hole at the end of $T_2$.

We verify each structural invariant separately.

\begin{itemize}
    \item \textbf{Number and Structure of Blocks.}
    Moving an element into the hole does not change
    the ownership of any cell. If $O_3$ is a
    level-$1$ section, the operation also leaves
    the recursive structure unchanged.

    Otherwise, the only change in cell ownership
    is performed by \swaphole$(O'_2,T'_2)$.
    By \Cref{cl:swap-hole-invariants}, this
    preserves the required block structure.
    The subsequent recursive call preserves it
    by induction.

    \item \textbf{Base-Level Sorting.}
    By \Cref{cl:add-max-invariants}, inserting
    the selected element into $T_2$ preserves
    its internal invariants.

    If $O_3$ is a level-$1$ section, removing
    its minimum and shifting the remaining
    elements preserves their sorted order.
    Otherwise, the remaining operations preserve
    Base-Level Sorting by
    \Cref{cl:swap-hole-invariants} and the
    induction hypothesis.

    \item \textbf{Block Sizes.}
    Moving an element into the hole does not
    change any block size, since the vacated
    cell becomes the new hole.

    If $O_3$ is a level-$1$ section, no further
    changes in cell ownership occur. Otherwise,
    \swaphole\ restores the size invariants
    of all affected sections by
    \Cref{cl:swap-hole-invariants}.
    The recursive call preserves these invariants
    by induction.

    \item \textbf{Block Ordering.}
    We verify that the element moved into $T_2$
    preserves the ordering constraints. The
    subsequent operations then preserve Block
    Ordering by the preceding claims and induction.

    First, suppose $O_3$ is a level-$1$ section.
    If $T_2$ is nonempty, the Block Ordering
    invariant gives $T_2<O_3$. Otherwise, we
    use $O_2<O_3$. Thus, moving $\min O_3$
    into $T_2$ preserves the required ordering
    between the sections.

    Now suppose $O_3$ consists of three blocks,
    and write
    \[
        O_3=O'_1\mid T'_2\mid O'_2\mid O'_3.
    \]
    If $T_2$ is nonempty, the Block Ordering
    invariant gives
    \[
        O'_1<T_2<O'_2.
    \]
    Hence, moving $\min O'_2$ into $T_2$
    preserves both inequalities: the inserted
    element is larger than every element of
    $O'_1$ and smaller than every element
    remaining in $O'_2$.

    If $T_2$ is empty, the same conclusion
    follows from $O'_1<O'_2$.

    Moreover, since $O_2<O_3$, the inserted
    element is larger than every element of
    $O_2$. Thus, the Block Ordering invariant
    of $S$ is also preserved.

    Consequently, the insertion satisfies
    all required ordering constraints.
    By \Cref{cl:add-max-invariants,cl:swap-hole-invariants},
    the subsequent operations preserve the
    affected ordering invariants, and the
    recursive call does so by induction.
\end{itemize}

It remains to verify that the hole reaches the
end of $S$.

If $O_3$ is a level-$1$ section, shifting its
remaining elements moves the hole directly
to the end of $S$.

Otherwise, after \swaphole$(O'_2,T'_2)$,
rebuilding may already have moved the hole
to the end of $O_3$ or an enclosing section,
in which case the operation is complete.
If not, the hole is at the end of $T'_2$,
and the induction hypothesis applies to $O_3$.

In either case, the hole reaches the last
position of $S$, and all structural invariants
are preserved.
\end{proof}

\begin{proof}[Proof of \Cref{cl:simpleslide-invariants}]
We prove the claim by induction on $i$.

For $i=1$, the \dst\ is a sorted array. Shifting its
elements one position to the left moves the hole to
the last position and preserves all invariants.

Assume now that $i>1$. If the \dst\ consists of a
single block with original section $O$, the claim
follows directly from the induction hypothesis
applied to $O$.

Otherwise, write
\[
    S=O_1\mid T_2\mid O_2\mid O_3.
\]
By the induction hypothesis, the recursive call to
\simpleslide($i-1$) moves the hole to the end of
$O_1$ while preserving its invariants.
We then move $x=\min O_2$ into the hole using
\addmax, apply \swaphole$(O_2,T_2)$, and finally
invoke \downhole($i$).

We verify that these operations preserve all
structural invariants.

\begin{itemize}
    \item \textbf{Number and Structure of Blocks.}
    The \addmax\ operation does not change the
    recursive structure. Any changes in cell ownership
    caused by \swaphole\ are repaired using \rebuild.
    The remaining operations preserve the invariant
    by the induction hypothesis and
    \Cref{cl:down-hole-invariants}.

    \item \textbf{Base-Level Sorting.}
    Since $O_1<O_2$, inserting $x=\min O_2$ at the
    end of $O_1$ preserves its internal invariants
    by \Cref{cl:add-max-invariants}.
    The remaining operations preserve base-level
    sorting by the induction hypothesis and
    \Cref{cl:swap-hole-invariants,cl:down-hole-invariants}.

    \item \textbf{Block Sizes.}
    The \addmax\ operation does not change cell
    ownership. Any size violations caused by
    \swaphole, including the possible exhaustion
    of $O_2$, are repaired using \rebuild.
    By \Cref{cl:rebuild-invariants}, these repairs
    restore the required size bounds.
    The remaining operations preserve the invariant
    by induction and \Cref{cl:down-hole-invariants}.

    \item \textbf{Block Ordering.}
    By the ordering invariants,
    \[
        O_1<O_2<T_2,\qquad O_2<O_3.
    \]
    Since $x=\min O_2$, moving $x$ into $O_1$
    preserves these relations: $x$ is larger than
    every element of $O_1$ and smaller than all
    remaining elements of $O_2$, as well as every
    element of $T_2$ and $O_3$.

    Moreover, \simpleslide\ is only applied when
    $S$ is not the last original section of a
    three-block parent. Thus, there is no
    higher-level transfer section whose ordering
    with $O_1$ could be violated.
    The operation preserves the set of elements
    in $S$, so the remaining ordering constraints
    involving sections outside $S$ are unchanged.

    Finally, \swaphole\ and \downhole\ preserve
    the ordering invariants by
    \Cref{cl:swap-hole-invariants,cl:down-hole-invariants}.
\end{itemize}

It remains to verify the position of the hole.
After moving $x$ into $O_1$, the hole is at the
beginning of $O_2$. The \swaphole\ operation moves
it to the end of $T_2$, and \downhole($i$) moves it
to the last position of $S$.
If rebuilding has already moved the hole to the
end of $S$ or an enclosing section, the remaining
operations terminate.

Thus, \simpleslide($i$) preserves all structural
invariants and moves the hole to the last position,
completing the induction.
\end{proof}

\begin{proof}[Proof of \Cref{cl:slide-invariants}]
We prove the claim by induction on $i$.

For $i=1$, the \dst\ is a sorted array. Shifting its
elements one position to the left moves the hole
to the last position and preserves all invariants.

Assume now that $i>1$. If the \dst\ consists of a
single block with original section $O$, the claim
follows from the induction hypothesis applied to $O$.

Otherwise, write
\[
    S=O_1\mid T_2\mid O_2\mid O_3.
\]
We first apply \simpleslide($i-1$) to $O_1$.
By \Cref{cl:simpleslide-invariants}, this moves the
hole to the end of $O_1$ while preserving its
invariants.

If $S$ is not the last original section of a
three-block parent with a nonempty transfer section,
the remaining operations are those of \simpleslide,
and correctness follows from
\Cref{cl:simpleslide-invariants}.

It remains to consider the exceptional case, where
\[
    \widehat S
    =\widehat O_1\mid\widehat T_2
    \mid\widehat O_2\mid S,
\]
and $\widehat T_2$ is nonempty.
We move $\min\widehat T_2$ into $O_1$ using
\addmax, apply \simpleslide($i$) to $\widehat T_2$,
and then move $\min O_2$ into $\widehat T_2$
using \addmax.
Finally, we apply \swaphole$(O_2,T_2)$ and
\downhole($i$) to $S$.

We verify that these operations preserve all
structural invariants.

\begin{itemize}
    \item \textbf{Number and Structure of Blocks.}
    The \addmax\ operations do not change the
    recursive structure. The calls to \simpleslide,
    \swaphole, and \downhole\ preserve this invariant
    by \Cref{cl:simpleslide-invariants,cl:swap-hole-invariants,cl:down-hole-invariants},
    including any rebuilding they trigger.

    \item \textbf{Base-Level Sorting.}
    Since $O_1<\widehat T_2$, inserting
    $\min\widehat T_2$ at the end of $O_1$
    preserves its internal invariants by
    \Cref{cl:add-max-invariants}.
    Similarly, since $\widehat T_2<O_2$,
    inserting $\min O_2$ at the end of
    $\widehat T_2$ preserves its internal
    invariants.

    All other operations preserve base-level
    sorting by the correctness of \simpleslide,
    \swaphole, and \downhole.

    \item \textbf{Block Sizes.}
    The \addmax\ operations do not change cell
    ownership and hence do not affect block sizes.
    Any violations caused by \swaphole, including
    the possible exhaustion of $O_2$, are repaired
    using \rebuild.
    By \Cref{cl:rebuild-invariants}, these repairs
    restore the required size bounds.
    The remaining operations preserve this
    invariant by the preceding correctness claims.

    \item \textbf{Block Ordering.}
    The ordering invariants of $S$ and its parent
    imply
    \[
        O_1<\widehat T_2<O_2<T_2,
        \qquad O_2<O_3,
    \]
    as well as
    \[
        \widehat O_2<\widehat T_2,
        \qquad \widehat O_2<S.
    \]

    First, moving $\min\widehat T_2$ into $O_1$
    preserves $O_1<\widehat T_2$, since the
    transferred element is smaller than every
    remaining element of $\widehat T_2$.
    It also preserves $O_1<O_2$, since
    $\widehat T_2<O_2$.

    The subsequent call to \simpleslide\ on
    $\widehat T_2$ preserves its set of stored
    elements and therefore does not affect
    these ordering relations.

    Next, moving $\min O_2$ into $\widehat T_2$
    preserves $\widehat T_2<O_2$, since the
    transferred element is larger than every
    remaining element of $\widehat T_2$ and
    smaller than every remaining element of $O_2$.
    Moreover, since $O_2<T_2$ and $O_2<O_3$,
    the inserted element is also smaller than
    every element of $T_2$ and $O_3$.
    Thus, all ordering constraints between
    these sections are preserved.

    The ordering constraints
    $\widehat O_2<\widehat T_2$ and
    $\widehat O_2<S$ also remain valid, since
    $\widehat O_2$ was smaller than every
    element of both sections before the transfers.
    Furthermore, the transfers preserve the
    multiset of elements in $\widehat S$, while
    leaving $\widehat O_1$ and $\widehat O_2$
    unchanged. Hence, ordering constraints
    involving higher-level sections are preserved.

    Finally, \swaphole\ and \downhole\ preserve
    the ordering invariants by
    \Cref{cl:swap-hole-invariants,cl:down-hole-invariants}.
\end{itemize}

It remains to verify the position of the hole.
After moving $\min\widehat T_2$ into $O_1$,
the hole is at the beginning of $\widehat T_2$.
The call to \simpleslide($i$) moves it to the
end of $\widehat T_2$, after which moving
$\min O_2$ leaves the hole at the beginning
of $O_2$.

The \swaphole\ operation moves this hole to
the end of $T_2$, and \downhole($i$) moves it
to the last position of $S$.
As usual, if rebuilding has already moved
the hole to the end of $S$ or an enclosing
section, the remaining operations terminate.

Thus, \slide($i$) preserves all structural
invariants and moves the hole to the last
position of the \dst, completing the induction.
\end{proof}

\begin{proof}[Proof of \Cref{cl:route-hole-invariants}]
We prove the claim by induction on $i$.

For $i=1$, the \dst\ is a sorted array.
Shifting the elements following the hole one
position to the left preserves its sorted order
and moves the hole to the last position.

Assume now that $i>1$. If the \dst\ consists
of a single block with original section $O$,
the claim follows from the induction hypothesis
applied to $O$.

Otherwise, write
\[
    S=O_1\mid T_2\mid O_2\mid O_3,
\]
and let $A$ be the section containing the hole.
By the induction hypothesis, the initial call
to \routehole($i-1$) moves the hole to the end
of $A$ while preserving all invariants.

We verify that the remaining operations preserve
the four structural invariants.

\begin{itemize}
    \item \textbf{Number and Structure of Blocks.}
    The \addmax\ operations do not change the
    recursive structure. Any changes in cell
    ownership caused by \swaphole\ are repaired
    using \rebuild. The remaining operations
    preserve this invariant by
    \Cref{cl:slide-invariants,cl:simpleslide-invariants,cl:down-hole-invariants}.

    \item \textbf{Base-Level Sorting.}
    The only new element movements not already
    covered by the preceding correctness claims
    occur when $A=O_2$. In this case, we move
    either $\min O_3$ or $\min T_2$ into the
    hole at the end of $O_2$.
    Since
    \[
        O_2<O_3,\qquad O_2<T_2,
    \]
    the inserted element is larger than every
    remaining element of $O_2$.
    Thus, \addmax\ preserves its internal
    invariants by \Cref{cl:add-max-invariants}.
    All subsequent operations preserve
    base-level sorting by their correctness
    claims.

    \item \textbf{Block Sizes.}
    Moving an element into the hole does not
    change cell ownership, since its previous
    cell becomes the new hole. Any size
    violations caused by \swaphole, including
    the possible exhaustion of a middle original
    section, are handled by \rebuild.
    By \Cref{cl:swap-hole-invariants,cl:rebuild-invariants}, these repairs
    restore the required size bounds.
    The remaining operations preserve the
    invariant by their correctness claims.

    \item \textbf{Block Ordering.}
    We distinguish the section $A$ containing
    the original hole.

    If $A=O_3$, no further operations are
    required. If $A=T_2$, we apply \downhole,
    which preserves all ordering invariants by
    \Cref{cl:down-hole-invariants}.

    If $A=O_1$, we perform the remaining steps
    of \slide. These steps require only that
    the hole is at the end of $O_1$ and that
    the ordering invariants hold, both of which
    follow from the induction hypothesis.
    Thus, correctness follows from the same
    argument as \Cref{cl:slide-invariants},
    including the ordering constraints involving
    a higher-level transfer section.

    It remains to consider $A=O_2$.
    We distinguish two cases according to
    the element moved into the hole.

    \begin{enumerate}
        \item If $T_2$ is empty or
        $\min O_3<\min T_2$, we move
        $x=\min O_3$ into $O_2$.
        Since $O_2<O_3$, the new element
        is smaller than every remaining
        element of $O_3$, preserving
        $O_2<O_3$.
        Moreover, if $T_2$ is nonempty,
        then $x<\min T_2$, so
        $O_2<T_2$ is also preserved.

        The subsequent call to \slide\
        on $O_3$ preserves the remaining
        ordering constraints, including
        those involving $T_2$, by
        \Cref{cl:slide-invariants}.

        \item Otherwise, $T_2$ is nonempty
        and $\min T_2<\min O_3$.
        We first observe that $O_3$ must
        be a level-$1$ section. Indeed,
        otherwise $O_3$ consists of three
        blocks, and writing
        \[
            O_3=O_1'\mid T_2'\mid O_2'\mid O_3',
        \]
        the ordering invariant
        $O_1'<T_2$ would imply
        $\min O_3<\min T_2$, a
        contradiction.

        Since $O_3$ is a level-$1$
        section, the terminal ordering
        invariant gives $T_2<O_3$.
        We move $x=\min T_2$ into $O_2$.
        The relations
        \[
            O_2<x<T_2\setminus\{x\}<O_3
        \]
        show that all ordering constraints
        between these sections are preserved.

        We then apply \simpleslide\ to
        $T_2$ and \downhole\ to $S$.
        By \Cref{cl:simpleslide-invariants,cl:down-hole-invariants}, these
        operations preserve all ordering
        invariants.
    \end{enumerate}

    Finally, we verify the ordering
    constraints involving sections outside
    $S$. The case $A=O_1$ is already
    covered by \Cref{cl:slide-invariants}.
    In the other cases, the operations
    preserve the multiset of elements in
    $S$ and do not modify $O_1$.

    In particular, if $S$ is the last
    original section of a three-block
    parent with nonempty transfer section
    $\widehat T_2$, the additional ordering
    constraint is
    \[
        O_1<\widehat T_2<O_2.
    \]
    In the case $A=O_2$, the elements
    moved into $O_2$ come from $T_2$
    or $O_3$, both of which lie above
    $\widehat T_2$ by the ordering
    constraints before the hole was created.
    Thus, this constraint is preserved.
    All other external ordering constraints
    are preserved as well.
\end{itemize}

It remains to verify the position of the hole.
After the initial recursive call, the hole is
at the end of $A$.

If $A=O_3$, it has already reached the end
of $S$. If $A=T_2$, \downhole\ moves it
to the end of $S$. If $A=O_1$, the remaining
steps of \slide\ complete the movement.

If $A=O_2$, moving the minimum of $O_3$
or $T_2$ leaves the hole at the beginning
of the corresponding section, by
\Cref{cl:new-dst-minimum}.
In the first case, \slide\ moves the hole
to the end of $O_3$, and hence of $S$.
In the second case, \simpleslide\ moves
the hole to the end of $T_2$, after which
\downhole\ moves it to the end of $S$.

As usual, if rebuilding has already moved
the hole to the end of $S$ or an enclosing
section, the remaining operations terminate.

Thus, \routehole($i$) preserves all
structural invariants and moves the hole
to the last position of the \dst,
completing the induction.
\end{proof}

\section{Omitted Proofs from \Cref{sec:fully-dynamic-lowcost}} 

\begin{proof}[Proof of \Cref{cl:resort-invariants}]
We first verify that every \resort\ operation
can reconstruct the corresponding \adst\
as a single \dst\ of the same level.

Consider a level-$j$ auxiliary \adst\ $A_j$,
where $j<d$.
Whenever \resort($j+1$) or a higher-level
reconstruction is performed, $A_j$ becomes empty.
Before the next such reconstruction, at most
$2^j$ updates occur, each increasing the number
of cells in $A_j$ by at most one.
Thus, even during an update,
\[
    |A_j|\le 2^j.
\]
The entire level-$d$ \adst\ contains $n\le 2^d$
cells, so the same bound holds at the top level.

Consequently, whenever \resort($i$) is invoked,
it reconstructs at most $2^i$ cells.
The recursive construction used by \rebuild\
can accommodate this size, as established in
\Cref{cl:rebuild-invariants}.
Indeed, the rebuilding rule constructs a valid
level-$i$ \dst\ from any collection of at most
$24\cdot 2^{i-1}$ cells, which is more than
sufficient.

We now verify the structural invariants.

\begin{itemize}
    \item \textbf{DST Invariants.}
    The elements are sorted and reconstructed
    into a clean level-$i$ \dst, with all
    transfer sections empty.
    By \Cref{cl:rebuild-invariants}, the new
    \dst\ satisfies the Number of Blocks,
    Block Sizes, Block Ordering, and Base-Level
    Sorting invariants.

    All components outside the rebuilt \adst\
    remain unchanged. Since no ordering is
    required between different components,
    their invariants remain satisfied.

    \item \textbf{Suffix Sizes.}
    The reconstruction leaves the level-$(i-1)$
    \adst\ empty, so all its lower-level
    auxiliary \adst s are also empty.

    The total number of cells in the rebuilt
    level-$i$ \adst\ remains unchanged.
    Therefore, the sizes of all enclosing
    auxiliary \adst s are unchanged as well,
    preserving their bounds.
\end{itemize}

Thus, \resort\ maintains all structural
invariants of the \adst.
\end{proof}

\begin{proof}[Proof of \Cref{cl:adst-swap-hole-invariants}]
We verify the Suffix Sizes invariant and the
internal \dst\ invariants separately.

\begin{itemize}
    \item \textbf{Suffix Sizes.}
    Consider a level-$j$ auxiliary \adst\ $A_j$,
    where $j<d$.
    Whenever \resort($j+1$) or a higher-level
    reconstruction is performed, $A_j$ becomes empty.

    By the \resort\ schedule, at most $2^j-1$
    completed updates occur before the next update
    that triggers such a reconstruction.
    Each update can transfer at most one cell into
    $A_j$, since the hole crosses its left boundary
    at most once.

    Thus, during the current update, including
    the temporary hole, we have
    \[
        |A_j|\le (2^j-1)+1=2^j.
    \]
    Rebuilding within a component does not change
    the total number of cells in $A_j$.
    Hence the Suffix Sizes invariant is preserved.

    \item \textbf{DST Invariants.}
    Transferring the hole changes cell ownership
    without moving any stored element, so the
    ordering and base-level sorting invariants
    are preserved. Any resulting structural
    violations are repaired independently within
    the two components.

    For rebuilt sections not containing the hole,
    correctness follows from
    \Cref{cl:swap-hole-invariants,cl:rebuild-invariants}.
    In the receiving component $D'$, every section
    containing the hole lies along its chain of first
    original sections, with the hole in its first cell.
    In particular, such a section is never the last
    original section of a three-block parent.

    We rebuild these sections by treating the hole
    as the smallest element. This leaves the hole
    in the first cell, while sorting the stored
    elements and emptying the transfer sections.
    The same block-size construction as in \rebuild\
    restores the structural invariants. Moreover,
    since the set of stored elements in each rebuilt
    section is unchanged, the ordering constraints
    with surrounding sections remain satisfied.

    Finally, the Suffix Sizes bound guarantees
    sufficient capacity for rebuilding the root
    sections. Since the \adst\ imposes no ordering
    constraints between distinct components,
    no additional conditions are required.
\end{itemize}

Thus, all structural invariants of the \adst\
are preserved, and the hole remains in the first
cell of $D'$.
\end{proof}

\begin{proof}[Proof of \Cref{cl:rebuild-amortized}]
We charge the recourse incurred by \rebuild\ to
changes in cell ownership caused by \swaphole.
Consider a repair triggered by a level-$(i+1)$
section $S$, and let $q=2^i$.
The repair rebuilds $S$, or its parent if $S$
is the last original section of a three-block
parent. We include all recursive rebuilding
in the cost of this repair.

\emph{Cost of rebuilding.}
The section $S$ contains at most $24q$ cells,
by the size bound of its enclosing block, or
by the capacity bound if $S$ is a root.
If we rebuild its parent instead, the parent's
first and last blocks each contain fewer than
$24q$ cells, and its middle block contains
$2q$ cells. Thus, the repair incurs at
most $50q$ recourse.

\emph{Changes required for rebuilding.}
We show that each repair requires at least
$q/2$ preceding changes in cell ownership,
distinguishing the two rebuilding conditions.

\begin{itemize}
    \item \textbf{Block-size violation.}
    Whenever new block sizes are chosen, the
    rebuilding rule produces either a single
    block of size at most $11q$, or three blocks
    satisfying
    \[
        2q\le |O_1|\le 23q/2,
        \qquad
        8q\le |O_3|\le 23q/2.
    \]
    The permitted bounds are $|S|<12q$ for
    a single block, and
    \[
        q\le |O_1|<12q,\qquad
        7q\le |O_3|<12q
    \]
    for three blocks.

    For a single block, reaching the upper bound
    of $12q$ requires at least $q$ increases in
    size. For three blocks, violating either
    lower bound requires more than $q$ decreases,
    while reaching the upper bound of $12q$
    requires at least $q/2$ increases.
    Since each change in cell ownership changes
    a block size by at most one, any violation
    requires at least $q/2$ such changes since
    the block sizes were last chosen.

    \item \textbf{Empty original section of the
    middle block.}
    If $O_2$ becomes empty, then $|T_2|=q$,
    since the middle block contains exactly
    $q$ cells. As $T_2$ was empty after its
    previous reconstruction, at least $q$
    cells must have entered it since then.
\end{itemize}

Thus, each repair costs at most $50q$ and can
be charged to at least $q/2$ preceding ownership
changes, assigning at most $100$ recourse to each.

\emph{Charging.}
It remains to show that each ownership change is
charged only $O(d)$ times. We charge each repair to
the section $S$ that triggered it, even if its parent
is rebuilt instead. Fix such a section $S$.

For block-size violations, we charge the repair
to ownership changes since the block sizes of $S$
were last chosen. Each such repair chooses a new
block division, so successive violations are
charged to disjoint sets of changes. Rebuilding
$S$ without changing its block sizes does not
reset this accounting.

For an empty middle original section, we charge
the repair to transfers into $T_2$ since it was
last emptied. Since rebuilding empties $T_2$,
successive violations of this type are also
charged to disjoint sets of changes.

Rebuilding an ancestor may reconstruct $S$ as
well, but this incurs no separate charge, since
its cost is included in the ancestor's repair.
If the reconstruction chooses new block sizes
for $S$, the accounting for size violations
restarts; otherwise, it is preserved.
The accounting for an empty middle section
restarts whenever $T_2$ is emptied.

Thus, each ownership change is charged at most
once for each of the two violation types in
a given section. Since each \swaphole\ changes
cell ownership along at most two recursive paths,
it affects only $O(1)$ sections at each of the
$d$ levels. Consequently, each \swaphole\
receives $O(d)$ charge, giving a total rebuilding
recourse of $O(dM)$.

\end{proof}

\begin{proof}[Proof of \Cref{cl:adst-delete-invariants}]
Let $D_j$ be the component containing the deleted
element $x$. After removing $x$, we apply
\routehole($j$), which moves the resulting hole
to the last cell of $D_j$ while preserving its
structural invariants, by
\Cref{cl:route-hole-invariants}.

We then process the remaining components from
left to right. At each step, \swaphole\ transfers
ownership of the empty cell from the current
component to the next one, where it becomes the
first cell. By \Cref{cl:adst-swap-hole-invariants},
this preserves the structural invariants of the
participating components. We then apply \slide\
to move the hole to the last cell of the next
component, preserving its invariants by
\Cref{cl:slide-invariants}.
Repeating this procedure brings the hole to the
last cell of $D_1$, which is also the last
position of the \adst.

We now verify the structural invariants.

\begin{itemize}
    \item \textbf{Number of Blocks.}
    By \Cref{cl:route-hole-invariants,cl:adst-swap-hole-invariants,cl:slide-invariants},
    the operations preserve the recursive
    structure of each participating \dst,
    using \rebuild\ when necessary.

    \item \textbf{Block Sizes.}
    Transferring the hole between components
    changes their numbers of cells.
    By the same correctness claims, any resulting
    violations of the Block Sizes invariant are
    repaired using \rebuild, whose correctness
    follows from \Cref{cl:rebuild-invariants}.

    \item \textbf{Block Ordering.}
    The ordering constraints within each component
    are preserved by the operations above.
    Since the \adst\ imposes no ordering between
    different components, no additional constraints
    are required.

    \item \textbf{Base-Level Sorting.}
    By the correctness of \routehole, \swaphole,
    and \slide, every level-$1$ section remains
    sorted. In particular, the final component
    $D_1=A_1$ remains sorted.

    \item \textbf{Suffix Sizes.}
    Recall that $A_k$ is the suffix consisting
    of components $D_k,\ldots,D_1$, and that
    $|A_k|$ counts all cells, including holes.

    During the deletion, the only operation
    that can increase $|A_k|$ is the transfer
    of the hole from $D_{k+1}$ into $D_k$.
    This transfer adds exactly one cell to $A_k$.
    Since the hole moves through the components
    from left to right, this boundary is crossed
    at most once. Thus, each deletion increases
    $|A_k|$ by at most one.

    It remains to bound the accumulated increase.
    Initially, $A_k$ is empty. Moreover, each
    execution of \resort$(k+1)$, or of a
    higher-level \resort, reconstructs the
    corresponding structure with $A_k$ empty.
    By the rebuilding schedule, at most $2^k$
    deletion--insertion updates occur between
    consecutive such reconstructions.

    Since each deletion increases $|A_k|$ by
    at most one, while the subsequent insertion
    does not change cell ownership, we obtain
    \[
        |A_k|\le 2^k.
    \]
    This bound also holds during the deletion
    immediately preceding a scheduled \resort.
\end{itemize}

Consequently, all structural invariants are
preserved, and the hole is at the last cell
of $A_1$, ready for the subsequent insertion.
\end{proof}

\begin{proof}[Proof of \Cref{cl:adst-insertion-invariants}]
By \Cref{cl:adst-delete-invariants}, the preceding
deletion preserves the structural invariants and
leaves a hole at the last position of the \adst,
inside $A_1$.

The \insertadst\ operation inserts the new element
in sorted order into $A_1$, using this existing hole.
Since $A_1$ is a level-$1$ \dst, this preserves
the Base-Level Sorting invariant.

Moreover, the insertion does not change the number
of cells or their ownership by sections.
In particular, the sizes of all blocks and
auxiliary \adsts\ remain unchanged, since holes
are counted toward their sizes.
All other components are unaffected.
Thus, the Number of Blocks, Block Sizes,
Block Ordering, and Suffix Sizes invariants
remain satisfied.

Finally, any scheduled \resort\ is performed
after the insertion and preserves the invariants
by \Cref{cl:resort-invariants}.
\end{proof}


\end{document}